\documentclass[twocolumn,showpacs,preprintnumbers,amsmath,amssymb,amsfonts,final,a4paper,aps,citeautoscript,footinbib,pra,superscriptaddress]{revtex4-2}
\usepackage{braket}
\usepackage[pdftex]{graphicx,color}
\usepackage{amssymb,amsmath,amsfonts}
\graphicspath{{./figures/}}
\usepackage[latin1]{inputenc}
\usepackage{mathrsfs}
\usepackage{epstopdf}
\usepackage{times}
\usepackage{float}
\usepackage[normalem]{ulem}
\usepackage[sort&compress]{natbib}
\usepackage[colorlinks,bookmarks=true,citecolor=blue,linkcolor=blue,urlcolor=blue, breaklinks=true]{hyperref}
\newcommand{\eq}[1]{\begin{equation}{#1}\end{equation}}

\newcommand{\sun}[1]{\mathfrak{su}([1])}

\newcommand{\Ua}{\Uparrow}

\newcommand{\Da}{\Downarrow}

\newcommand{\ua}{\uparrow}
\newcommand{\da}{\downarrow}
\newcommand{\calM}{\mathcal{M}}
\newcommand{\gr}[4]{\begin{figure}\centering\includegraphics[width=#1cm]{#2}\caption{#3}\label{#4}\end{figure}}
\newcommand{\ds}{\displaystyle}
\begin{document}
%%%%%%%%%%%%%%%%%%%%%%%%%%%%%%%%%%%%%%%%%%%%%%%%%%%%%%%
% Use the \preprint command to place your local institutional report
% number in the upper righthand corner of the title page in preprint mode.
% Multiple \preprint commands are allowed.
% Use the 'preprintnumbers' class option to override journal defaults
% to display numbers if necessary
%\preprint{}
%%%%%%%%%%%% Title of paper %%%%%%%%%%%%%%%%%%%%%%%%%%%%%%%%%%%%
%\title{Electric and Magnetic Properties of Higher-Spin Kondo Lattice Models in Strong-Coupling Regions}
\title{Electric and Magnetic Properties of Higher-Spin Kondo-Heisenberg Models at Strong Coupling
}
%%%%%%%%%%%%%%%%%%%%%%%%%%%%%%%%%%%%%%%%%%%%%%%%%%%%%%%
% repeat the \author .. \affiliation  etc. as needed
% \email, \thanks, \homepage, \altaffiliation all apply to the current
% author. Explanatory text should go in the []'s, actual e-mail
% address or url should go in the {}'s for \email and \homepage.
% Please use the appropriate macro foreach each type of information

% \affiliation command applies to all authors since the last
% \affiliation command. The \affiliation command should follow the
% other information
% \affiliation can be followed by \email, \homepage, \thanks as well.
\author{Riku Masui}
\affiliation{Division of Physics and Astronomy, Graduate School of Science, Kyoto University, Kyoto 606-8502, Japan}
\affiliation{Center for Gravitational Physics and Quantum Information (CGPQI), Yukawa Institute for Theoretical Physics, Kyoto University, Kyoto 606-8502, Japan}
%\email[]{Your e-mail address}
%\homepage[]{Your web page}
%\thanks{}
%\altaffiliation{}
\author{Keisuke Totsuka}
\affiliation{Center for Gravitational Physics and Quantum Information (CGPQI), Yukawa Institute for Theoretical Physics, Kyoto University, Kyoto 606-8502, Japan}
%Collaboration name if desired (requires use of superscriptaddress
%option in \documentclass). \noaffiliation is required (may also be
%used with the \author command).
%\collaboration can be followed by \email, \homepage, \thanks as well.
%\collaboration{}
%\noaffiliation

\date{\today}

\begin{abstract}
We study higher-spin ($S \geq 1$) generalization of the one-dimensional Kondo-Heisenberg model, 
in which the local spin-$S$ moments of the Kondo lattice model interact with each other via the antiferromagnetic
Heisenberg interaction ($J_{\text{H}}$), by analytical and numerical methods.  
The strong-coupling (i.e., large Kondo-coupling) expansion maps out an insulating phase at half-filling 
whose magnetic correlation depends on the parity of $2S$ as well as a ferromagnetic metallic phase which dominates 
the strong-coupling region at generic fillings.   Then, we carried out the Density-Matrix Renormalization Group (DMRG) simulations 
for $S=1$ to closely investigate the phase structure at large but finite Kondo coupling.  At half-filling, the Kondo coupling and 
$J_{\text{H}}$ do not compete and the insulating spin-gapless phase is stable, 
while the competition of the two leads to a stepwise collapse of the strong-coupling ferromagnetism 
via an intervening dimerized insulating phase with power-law spin correlation at quarter-filling.   
\end{abstract}

% insert suggested keywords - APS authors don't need to do this
%\keywords{}

%\maketitle must follow title, authors, abstract, and keywords
\maketitle

% body of paper here - Use proper section commands
% References should be done using the \cite, \ref, and \label commands
%%%%%%%%%%%%%%%%%%%%%%%%%%%%%%%%%%%%%%%%%%%%%%%%%%%%%%%%%%%%%%
\section{Introduction}
%%%%%%%%%%%%%%%%%%%%%%%%%%%%%%%%%%%%%%%%%%%%%%%%%%%%%%%%%%%%%%
Heavy fermion systems are typical examples of the strongly correlated electron systems \cite{Fazekas-book-99,Coleman-book-15}, 
where the interaction among electrons plays 
crucial roles.  In heavy fermion systems, the interaction among electrons results in quasi-particles with large effective mass, 
and realizes a variety of ground states depending on materials.
One of the standard minimal theoretical models of heavy fermion system is the Kondo lattice (KL) model 
(see, e.g., Refs.~\cite{Tsunetsugu_review,Gulacsi-review-04}, for reviews),  
in which tight-biding electrons interact with localized spins through the exchange interaction.  
Historically, the Kondo lattice model is derived from the Kondo model, where a single magnetic impurity exists in conduction electron system.  
In the ground state of the Kondo model, the conduction electron strongly couples to the impurity   
thereby screening its magnetic moment by forming the spin-singlet state (the Kondo singlet) \cite{Yoshida-66}.  
On the other hand, when there are many magnetic impurities, the competition between the long-range 
spin-spin interaction mediated by the conduction electrons (the RKKY interaction \cite{RK,K,Y}) and the Kondo screening 
is expected.   The resulting global phase structure is summarized in the celebrated Doniach phase diagram \cite{Doniach-77}.     
The minimal lattice model that allows us to study the competition between the Kondo screening and the formation of magnetic order 
is the Kondo lattice model whose Hamiltonian is given by \cite{Doniach-77}:
\eq{
\begin{split}
\hat{H}_{\text{KL}} & = -t\sum_{i,\alpha} \left(c_{i,\alpha}^\dagger c_{i+1,\alpha} +\text{H.C.} \right)
+J_{\text{K}} \sum_{i}\vec{s}_i  \cdot \vec{S}_i      \\
& =: \hat{H}_{\text{e}} + \hat{H}_{\text{K}}   \; .
\end{split}
\label{klH}
}
In Eq.~\eqref{klH}, $c_{i,\alpha}$ ($c_{i,\alpha}^\dagger$) denotes the annihilation (creation) operator of the conduction electron 
with spin $\alpha =\ua,\da$ at site $i$. 
%So, the first term of Eq.~\eqref{klH} stands for the nearest-neighbor hopping of the conduction electrons.  
The first term $\hat{H}_{\text{e}}$ is the kinetic energy (the hoping term) of the conduction electron,  
while the second describes the exchange interaction between the spin 
$\vec{s}_i = c^{\dagger}_{i,\alpha} [\vec{\sigma}]_{\alpha\beta} c_{i,\beta} /2$ of the conduction electron (the symbols $\vec{\sigma}$ 
denote the Pauli matrices and the summation over repeated indices is implied) 
and the localized spin $\vec{S}_i$ (spin-$S$) at the same site, which is known as the Kondo coupling.  
Since the localized spins $\vec{S}_i$ in the Kondo lattice model \eqref{klH} originate from the spin degree of freedom 
of $d$ or $f$-electrons, the case $S=1/2$ has been mainly studied \cite{Tsunetsugu_review,Coleman-G-T-97}.  
It has been also proposed that the model \eqref{klH} itself can be quantum-simulate in a well-controlled manner using alkaline-earth-like  
fermionic cold atoms (e.g., ${}^{171}\text{Yb}$) loaded on optical lattices \cite{Gorshkov-et-al-10,Riegger-et-al-KLM-18}.   

Recently, the possibility that the $S=1$ Kondo lattice model with a uniaxial anisotropy under a transverse magnetic field can describe the coexistence of ferromagnetism and superconductivity 
in materials like URhGe has been pointed out \cite{Suzuki_Hattori1,Suzuki_Hattori2}.  
This motivates us to study the Kondo lattice model 
with spin $S$ larger than 1/2 and broaden the range of materials to which the Kondo lattice model can apply \footnote{%
\textcolor[rgb]{1,0,0}{
If $S=1$ or higher,  the ground state multiplet of the corresponding f-ion must be four-degenerate by spin-orbit coupling. This degeneracy will be lifted under the crystalline field, so the realization of the magnitude $S=1$ has the delicate problem.} }. 
Another interesting aspect of considering the higher-spin ($S$) generalization is that, in one dimension, 
the magnetically-ordered region of the Doniach phase diagram may exhibit intrinsically different properties depending on,  
e.g., the parity of $2S$.   

On top of considering higher-spin cases, we shall incorporate below the {\em direct} interaction between the adjacent localized spins.  
Here, by ``direct" we mean that the spin-spin interaction is not mediated by the conduction electrons.  
In order to incorporate these two generalizations to the Kondo lattice model, 
we consider the following spin-$S$ Kondo-Heisenberg (KH) Hamiltonian \cite{Sikkema-A-W-97,Zachar-01,Zachar-T-01}:
%\begin{widetext}
\begin{equation}
\begin{split}
\hat{H}_{\text{KH}} =&  - t\sum_{i,\alpha} \left(c_{i,\alpha}^\dagger c_{i+1,\alpha} +\rm{H.C.} \right)
+J_{\text{K}} \sum_{i}\vec{s}_i  \cdot \vec{S}_i   \\
& + J_{\text{H}} \sum_{i} \vec{S}_{i} \cdot \vec{S}_{i+1}  \\
=: & \hat{H}_{\text{e}} + \hat{H}_{\text{K}} + \hat{H}_{\text{H}}   \; .
\end{split}
\label{khH}
\end{equation}
%\end{widetext}
The first two terms $\hat{H}_{\text{e}}$ and $\hat{H}_{\text{K}}$ are common to the two models 
$\hat{H}_{\text{KL}}$ \eqref{klH} and $\hat{H}_{\text{KH}}$ \eqref{khH}.     
The last term $\hat{H}_{\text{H}}$ ($J_{\text{H}}>0$) is the direct antiferromagnetic interaction 
between the adjacent localized spins mentioned above. 
Physically, this interaction corresponds to the superexchange interaction among $f$ electrons which would arise  
when small hopping of the $f$-electrons is taken into account.   

In the case $S=1/2$ and $0<n<1$ of the KL model \eqref{klH}, the electron motion favors ferromagnetism when $J_{\text{K}}$ 
is sufficiently large \cite{kondo_lowdendity_FM,Sigrist-T-U-R-92}.  
One of the important effects of the $J_\text{H}$ term is to suppress this ferromagnetic ground state and stabilize  
the paramagnetic one \cite{Moukouri-C-96}. Another interesting effect is that moderate $J_\text{H}$ term can open 
the spin gap even away from half-filling $0<n<1$ \cite{Berg-F-K-10}. 
The KH model \eqref{khH} with the localized spin-1/2 has been studied in the context of, e.g., the uranium-based heavy-fermion superconductors \cite{Thalmeier2002}, the pair density wave in superconducting state of $\text{La}_{2-x}\text{Ba}_x\text{CuO}_4$ \cite{Berg-F-K-10}, and 
the topological Kondo insulators \cite{topological_kondo_insulator,topological_kondo_insulator2,topological_kondo_insulator3}.  
The model $\hat{H}_{\text{KH}}$ with $S=1/2$ is also used as a simplest model that may describe the interplay between two different orders  
in a certain class of organic compounds [e.g., $\text{(Per)}_{2}\text{Pt}\text{(mnt)}_{2}$] 
in which the systems consist of partially-filled metallic part and half-filled insulating 
one \cite{Henriques-et-al-84,Bourbonnais-et-al-91,Green-et-al-11}.  

Yet another motivation to study the KH model $\hat{H}_{\text{KH}}$ is related to the physics of open quantum systems.  
Instead of viewing it as a generalization of the Kondo lattice model \eqref{klH}, 
we can think of the KH model \eqref{khH} as the spin-$S$ Heisenberg chain ($\hat{H}_{\text{H}}$) 
coupled to the environment ($\hat{H}_{\text{e}}$) of the conduction electrons through the Kondo coupling.  
In fact, the ground state of the spin-$S$ Heisenberg chain is known to be deeply connected to topology \cite{Haldane_conjecture,Haldane_conjecture-2} 
and is quite interesting in its own right.   
For example, the gapped ground state of the $S=1$ Heisenberg chain is one of the typical examples of  
the symmetry-protected topological (SPT) phases \cite{Gu-W-09,spt},
which can be used as the resource states of the measurement-based quantum computation \cite{Else-S-B-D-12}.   
%it is known to be characterized by non-trivial string order parameter \cite{string_order,Garcia-W-S-V-C-08}, 
%which differentiates the SPT phase from other (topologically) trivial phase.   
This motivates us to study the effects of coupling non-trivial (topological) many-body states hosted in the localized spin system 
to a gapless environment (i.e., the conduction electrons).  
The investigation of the robustness of the SPT states against perturbation from the environment through the KH model \eqref{khH}  
would be a very important theme also from the quantum-computational point of view and will be discussed elsewhere.   
As the first step toward the understanding of the physics of the genralized KH model \eqref{khH}, we study in this paper its phase structure 
in the region of strong Kondo coupling where we can determine the ground-state properties accurately (sometimes rigorously). 

The organization of the rest of the paper is as follows.  
In Sec.~\ref{sec:strong-coupling}, we derive the low-energy effective Hamiltonian in the strong-coupling region 
(i.e., $J_{\text{K}} \gg t, J_{\text{H}}$) both at half-filling and away from half-filling, which gives an important insight into the structure 
of the phase diagram.  In particular, we will show that, in the strong-coupling region,  
the ferromagnetic phase is generically stabilized (except at half-filling) through a mechanism similar to the double-exchange interaction 
and that this tendency competes with antiferromagnetism stabilized by the direct antiferromagnetic interaction $J_{\text{H}}$. 

In Sec. \ref{sec:numerical_results}, in order to investigate this competition between the ferromagnetism and the $J_{\text{H}}$-induced antiferromagnetism, 
we carry out numerical density-matrix-renormalization-group (DMRG) simulations \cite{White1,White2,SCHOLLWOCK} 
combined with the sine-square-deformation (SSD) technique \cite{SSD,SSD_chargegap,SSD_nishino} for the special case of $S=1$ 
to find that the competition indeed stabilizes a new dimerized (i.e., bond-centered) phase with power-law spin correlation and a finite charge gap. 
We summarize the main results in Section \ref{sec:conclusion}, and some technical details including the proof of ferromagnetism 
are presented in the appendices.   
%%%%%%%%%%%%%%%%%%%%%%%%%%%%%%%%%%%%%%%%%%%%%%%%%%%%%%%%%%%%%
\section{Strong-coupling effective Hamiltonian}
\label{sec:strong-coupling}
%%%%%%%%%%%%%%%%%%%%%%%%%%%%%%%%%%%%%%%%%%%%%%%%%%%%%%%%%%%%%
In this section, we carry out the perturbation theory from the strong-coupling limit ($J_{\text{K}} \to \infty$) 
to derive the low-energy effective Hamiltonian that enables us to map out the strong-coupling phases.  
Since $\hat{H}_{\text{KH}}$ can commute with the total electron number 
$N_{\text{e}}=\sum_{i,\alpha} c_{i,\alpha}^\dagger c_{i,\alpha} = \sum_{i} n_{i}$,  
the electron density $n=N_{\text{e}}/L$ (with $L$ being the system size) of the conduction electrons 
is a conserved quantum number to be fixed.  
Moreover, since the particle-hole transformation $c_{i,\alpha} \leftrightarrow c_{i,\alpha}^{\dagger}$  
maps the KH model at filling $n$ onto the same model at filling $2-n$ as in the KL model \cite{Tsunetsugu_review},  
we can safely restrict ourselves to $n\leq 1$ without the loss of generality.  
Another important conserved quantum number is: 
$T^{z}_{\text{tot}} = \sum_{i} T^{z}_{i} = \sum_{i} (s^z_i + S^z_i)$.  
Throughout this paper, we reserve the notation $\vec{T}_{i}$ to denote the composite spin on each site:
\begin{equation}
\vec{T}_{i} := \vec{s}_{i} + \vec{S}_{i}   \; .
\label{total_spin_operator}
\end{equation}
To be specific, unless otherwise stated, we set $S=1$ in what follows, although the generalization to arbitrary $S$ is straightforward.  
Some of the generalizations are discussed in appendices.
%%%%%%%%%%%%%%%%%%%%%%%%%%%%%%%%%%%%%%%%%%%%%%%%%%%%%%%%%%%%%
\subsection{Half-filling ($n=1$)} 
\label{halffilling_case}
%%%%%%%%%%%%%%%%%%%%%%%%%%%%%%%%%%%%%%%%%%%%%%%%%%%%%%%%%%%%%
\subsubsection{Strong-coupling ground state}
%%%%%%%%%%%%%%%%%%%%%%%%%%%%%%%%%%%%%%%%%%%%%%%%%%%%%%%%%%%%%
At half-filling $n=1$, the number of conduction electrons $N_{\text{e}}$ equals to the number of the sites $L$. 
In the strong-coupling limit $J_{\text{K}} \to \infty$, where we can ignore the other two terms $\hat{H}_{\text{e}}$ and $\hat{H}_{\text{H}}$,  
we can find the ground state of $\hat{H}_{\text{K}}$ by minimizing 
the Kondo coupling $J_{\text{K}} \vec{s}_i  \cdot \vec{S}_i$ site by site; 
the ground state has no doubly-occupied or vacant sites, and at each site the spin-1/2 from a conduction electron and the localized spin-1 
form a doublet:  
\begin{equation}
\begin{split}
& \ket{\Ua}_i := \sqrt{\frac{2}{3}} 
\Ket{
\begin{array}{c}
\da \\
1
\end{array}
}_i
-
\sqrt{\frac{1}{3}} \Ket{
\begin{array}{c}
\ua \\
0
\end{array}
}_{i,} \\
\label{doublet}
\\
& \ket{\Da}_i :=
\sqrt{\frac{2}{3}} \Ket{
\begin{array}{c}
\ua \\
-1
\end{array}
}_i
-
\sqrt{\frac{1}{3}} \Ket{
\begin{array}{c}
\da \\
0
\end{array}
}_{i}   \;  . 
\end{split}
\end{equation}
On the right-hand sides of (\ref{doublet}), we have introduced the symbols 
${\tiny \Ket{
\begin{array}{c}
\alpha \\
S^z
\end{array}
}_i}$ to denote the tensor-product state $\ket{\alpha}_{i,c\text{-electron}} \otimes \ket{S^z}_{i,\text{local spin}}$ 
with $\ket{\alpha}$ being one of the four electronic states $|\text{emp}\rangle=|0\rangle$, $|\!\!\uparrow\rangle = c^{\dagger}_{i,\ua}|0\rangle$, 
$|\!\!\downarrow\rangle=c^{\dagger}_{i,\da}|0\rangle$, and 
$|\!\! \uparrow\downarrow\rangle= c^{\dagger}_{i,\ua}c^{\dagger}_{i,\da}|0\rangle$, and $S^z=\pm 1,0$.    
From now on, we call this effective spin-1/2 state as the {\em Kondo doublet}.  
In addition, as the Kondo doublets at the individual sites do not interact with each other in this limit,  
the ground state of the entire system is $2^L$-fold degenerate; all the possible tensor-products of these local Kondo doublet states 
$\otimes_{i} | A \rangle_{i}$ ($A=\Ua,\Da$) span the basis of 
the huge ground-state eigenspace $\mathcal{H}_{\text{hf}}$ in the strong-coupling limit.
%%%%%%%%%%%%%%%%%%%%%%%%%%%%%%%%%%%%%%%%%%%%%%%%%%%%%%%%%%%%%%%%%%
\subsubsection{Perturbation theory from strong-coupling limit}
\label{Perturbation_theory_from_strong-coupling_limit}
%%%%%%%%%%%%%%%%%%%%%%%%%%%%%%%%%%%%%%%%%%%%%%%%%%%%%%%%%%%%%%%%%%
Now, we consider the parameter region where the Kondo coupling is finite but still much larger than the other two terms, 
i.e. where the hopping term and the Heisenberg term can be viewed as the small perturbation.  

%\noindent%
\underline{\it (i) Second-order perturbation in the hopping $t$.}
At half-filling, the first-order perturbation of the hopping term is prohibited because the application of the hopping 
to the unperturbed ground state always gives rise to the states with exactly 
one pair of a doubly-occupied and a vacant (``emp'') sites which is out of the ground-state subspace: 
\begin{eqnarray}
\left\{
\begin{array}{l}
c_{i+1,\ua}^\dagger c_{i,\ua} \ket{\Ua}_i \otimes \ket{\Ua}_{i+1} =-\frac{\sqrt{2}}{3} 
\Ket{
\begin{array}{c}
\text{emp} \\
0
\end{array}
}_i
\otimes
\Ket{
\begin{array}{c}
\ua \da \\
1
\end{array}
}_{i+1} \; ,
\\
c_{i+1,\ua}^\dagger c_{i,\ua} \ket{\Ua}_i \otimes \ket{\Da}_{i+1} =\frac{1}{3}
\Ket{
\begin{array}{c}
\text{emp} \\
0
\end{array}
}_i
\otimes
\Ket{
\begin{array}{c}
\ua \da \\
0
\end{array}
}_{i+1}\; ,
\\
c_{i+1,\ua}^\dagger c_{i,\ua} \ket{\Da}_i \otimes \ket{\Ua}_{i+1} =\frac{2}{3} 
\Ket{
\begin{array}{c}
\text{emp} \\
-1
\end{array}
}_i
\otimes
\Ket{
\begin{array}{c}
\ua \da \\
1
\end{array}
}_{i+1}\; ,
\\
c_{i+1,\ua}^\dagger c_{i,\ua} \ket{\Da}_i \otimes \ket{\Da}_{i+1} =-\frac{\sqrt{2}}{3}
\Ket{
\begin{array}{c}
\text{emp} \\
-1
\end{array}
}_i
\otimes
\Ket{
\begin{array}{c}
\ua \da \\
0
\end{array}
}_{i+1} \; .
\\
\end{array}
\right. 
\end{eqnarray}
Here, the sign of the right-hand side comes from the definition of the doubly-occupied state $\ket{\ua \da}_{i}=c_{i,\ua}^\dagger c_{i,\da}^\dagger \ket{\mathrm{emp}}_{i}$.
Therefore, we need to go to the second-order perturbation in the hopping $t$ to find the effective interaction among the Kondo doublets.  

As is illustrated in Fig.~\ref{2nd_order_hopping2}, 
the second-order process consists of (i) hopping from $i$ to $i+1$ (from $i+1$ to $i$) that excites $\hat{H}_{\text{K}}$  
to the intermediate state with 
exactly one pair of a doubly-occupied and a vacant sites, 
and (ii) hopping from $i+1$ to $i$ (from $i$ to $i+1$).
Therefore, the second-order processes induce the following transitions among the four states $(\ket{\Ua \Ua} \ket{\Ua \Da} \ket{\Da \Ua} \ket{\Da \Da})$ 
of the neighboring Kondo doublets $(i,i+1)$: 
%%%%%%%%%%%%%%%%%%%%%%%%%%%%%%%%%%%%%%%%%%%%%%%%%%%%%%%%%%%%%%%
\begin{equation}
\begin{split}
& P_{\text{G.S.}} 
\left(\sum_{\alpha=\ua,\da} c_{i,\alpha}^\dagger c_{i+1,\alpha} + \text{H.C.} \right) \\
& \qquad  \times  \frac{1}{E_{\text{G.S.}}-\hat{H}_{\text{K}} } 
\left(\sum_{\alpha} c_{i+1,\alpha}^\dagger c_{i,\alpha} + \text{H.C.} \right) P_{\text{G.S.}}  \\
& =
-\frac{t^2}{9J_{\text{K}}}
\left(
\begin{array}{cccc}
\ds 4 & 0 &0 &0 \\
0&\ds 5 &\ds -1 &0\\
0&\ds -1 &\ds 5 &0\\
0&0&0&\ds 4
\end{array}
\right) \; , 
\end{split}
\end{equation}
where $E_{\text{G.S.}} =- J_{\text{K}} L$ is the ground-state energy of $\hat{H}_{\text{K}}$ 
and $P_{\text{G.S.}}$ is the projector onto the ground-state subspace.    
%%%%%%%%%%%%%%%%%%%%%%%%%%%%%%%%%%%%%%%%%%%%%%%%%%%%%%%%%%%%%%%
In deriving the above, we have used the fact that we can substitute the denominator $E_{\text{G.S.}}-\hat{H}_{\text{K}}$ on the left-hand side with 
the constant $-2J_{\text{K}}$ because any allowed intermediate states have exactly one pair of a doubly-occupied and a vacant sites 
each of which contributes the energy cost $J_{\text{K}}$.  
Similar effective interactions arise from any neighboring doublet pairs $(i,i+1)$, and 
we finally obtain the following effective antiferromagnetic spin exchange among the Kondo doublets: 
\eq{\sum_{i}\left( \frac{2t^2}{9J_K} \vec{D}_i\cdot \vec{D}_{i+1} -\frac{t^2}{2J_K} \right)  \label{eqn:eff-Ham-half-filling-S1} \; .}
Here, $\vec{D}_i$ denotes the spin-1/2 operator for the Kondo doublet at site $i$.  
%%%%%%%%%%%%%%%% FIG 1    %%%%%%%%%%%%%%%%%%%%%%%%%%%%%%%%%%%%%
\gr{8}{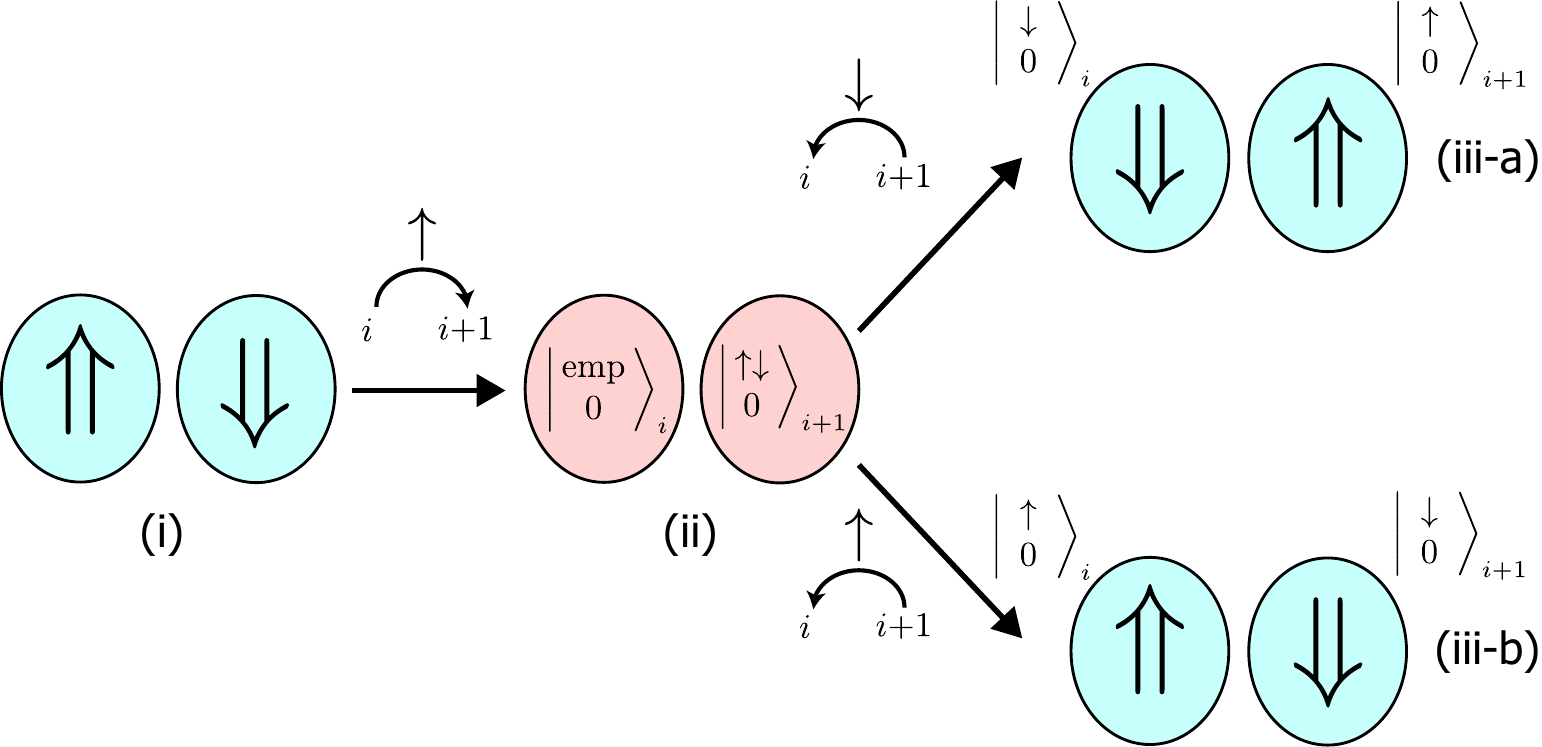}{A process of second-order perturbation of hopping term. (i)One of neighboring sites 
in the non-perturbed ground state. Thick arrow in an oval descibes a Kondo doublet in the strong-coupling limit.  
(ii) shows a possible first-order process. This picture describes the state after an electron with up-spin hops from the left site 
to the right one. This state is not included in vastly degenerated ground states.  
(iii)Second-order process. We can take two possible ways to back to the eigenspace which is spanned by non-perturbed ground states.}
{2nd_order_hopping2}
%%%%%%%%%%%%%%%%%%%%%%%%%%%%%%%%%%%%%%%%%%%%%%%%%%%%%%%%%%

%%%%%%%%%%%%%%%%%%%%%%%%%%%%%%%%%%%%%%%%%%%%%%%%%%%%%%%%%%
%\noindent%
\underline{\it (ii) First-order perturbation in the Heisenberg interaction $J_{\text{H}}$.} 
On top of the second-order kinetic exchange, there is the contribution from the Heisenberg term $\hat{H}_{\text{H}}$.  
As the Heisenberg interaction $J_{\text{H}}$ does not change the electronic state, it can generate first-order processes 
within the half-filled ground-state subspace $\mathcal{H}_{\text{hf}}$. 
In the basis \eqref{doublet} spanning $\mathcal{H}_{\text{hf}}$, the matrix elements of the localized spin-1 operators $\vec{S}_{i}$ are:
\begin{align}
\begin{split}
& \bra{\Ua}_i S_i^z \ket{\Ua}_i = 2/3 = 4/3 \bra{\Ua}_i D_i^z \ket{\Ua}_i   \\
& \bra{\Ua}_i S_i^z \ket{\Da}_i=\bra{\Da}_i S_i^z \ket{\Ua}_i =0  \\ 
& \bra{\Da}_i S_i^z \ket{\Da}_i = - 2/3 = 4/3 \bra{\Da}_i D_i^z \ket{\Da}_i   \; ,
\end{split}
\label{doublet_base_sz}
\\
\begin{split}
& \bra{\Ua}_i S_i^+ \ket{\Ua}_i = \bra{\Da}_i S_i^+ \ket{\Da}_i = 0\\
& \bra{\Ua}_i S_i^+ \ket{\Da}_i= - 4/3   \\
& \bra{\Da}_i S_i^+ \ket{\Ua}_i= 0   \; ,
\end{split}
\label{doublet_base_sp}
\end{align}
which means that the localized spin $\vec{S}_{i}$ projected onto $\mathcal{H}_{\text{hf}}$ is given by 
$\vec{S}_{i} \xrightarrow{\mathcal{H}_{\text{hf}}} (4/3) \vec{D}_{i}$ [see Eq.~\eqref{eqn:localized-spin-projected-gen-S} for 
the expression for general $S$].   
Therefore, the first-order degenerate perturbation of the Heisenberg term $\hat{H}_{\text{H}}$ gives 
the following effective antiferromagnetic spin exchange among the neighboring Kondo doublets:
\eq{
\begin{split}
& J_{\text{H}} \sum_i \vec{S}_i\cdot \vec{S}_{i+1}  
\xrightarrow{\mathcal{H}_{\text{hf}}} 
\frac{16}{9}J_{\text{H}} \sum_i \vec{D}_i\cdot \vec{D}_{i+1} \; .
\end{split}
\label{eqn:1st-order-JH-S1}
}

Combining Eqs.~\eqref{eqn:eff-Ham-half-filling-S1} and \eqref{eqn:1st-order-JH-S1}, 
we obtain the following spin-1/2 antiferromagnetic Heisenberg model for the Kondo doublets 
as the strong-coupling effective Hamiltonian at half-filling $n=1$:
\eq{
H^{(n=1)}_{\text{eff}}
=\left( \frac{2t^2}{9J_{\text{K}}}+\frac{16}{9}J_{\text{H}} \right) \sum_{i=1}^{L}  \vec{D}_i \cdot \vec{D}_{i+1} 
+ E_{0} \; .  
\label{effHam}
}
[with the constant $E_{0}$ given by $-(t^{2}/2J_{\text{K}}+J_{\text{K}})L$].  
Therefore, the ground state of the half-filled KH model in the strong-coupling region exhibits the quasi-long-range antiferromagnetic 
order with gapless spinon excitations (a spin Luttinger liquid) \cite{Giamarchi}.  
Moreover, as adding an electron or changing the electron configuration in the ground state costs a finite energy ($\sim J_\text{K}$), 
the ground state is expected to be insulating.  
Hence, we conclude that the ground state of strongly-coupled spin-1 KH chain at half-filling is an insulator 
with power-law antiferromagnetic spin-spin correlation regardless of the value of $J_{\text{H}}(\ll J_{\text{K}})$ \cite{Giamarchi}.    
This is in stark contrast to the spin-gapped insulating ground state (the Kondo insulator) found in the $S=1/2$ 
KL model \cite{kondo_spingap,Yu-W-93}.  

In general, the spin sector of spin-$S$ half-filled KH chain is described effectively by the antiferromagnetic spin-$(S-1/2)$ Heinseberg model 
for the partially-screened moments (see Appendix \ref{Appendix_derivation_2nd_order_hopping} for the details):
\begin{equation}
\begin{split}
& H_{\text{eff}}^{(n=1)} \\
& = \left\{ \frac{4t^{2}}{(2S+1)^{2}(S+1) J_{\text{K}} } + \left( \frac{2(S+1)}{2S+1} \right)^{2} J_{\text{H}} \right\} 
\sum_{i} \vec{\mathfrak{S}}_{i}  {\cdot} \vec{\mathfrak{S}}_{i+1}   
\end{split}
\label{eqn:eff-Ham-half-filling-gen-S} 
\end{equation} 
where $\vec{\mathfrak{S}}_{i}$ is the effective spin-$(S{-}1/2)$ operator which replaces the doublet $\vec{D}_{i}$ in the case of $S=1$.  
The ground state of the above effective Hamiltonian depends 
on the value of $S$ \cite{Haldane_conjecture,Haldane_conjecture-2}; 
when $2S$ is even, the insulating ground state has gapless spin excitations, while the ground state is fully gapped 
(i.e., both the charge and spin gaps are finite) when $2S$ is odd.   
This conclusion is consistent with that of a field-theory argument \cite{Tsvelik_kondo,Tsvelik-Y-19}.  
In Sec.~\ref{sec:numerics-at-HF}, we will numerically check this prediction for $S=1$ 
by increasing $J_\text{H}$ up to $\sim J_\text{K}$ while keeping $J_\text{K} \gg t$.
%%%%%%%%%%%%%%%%%%%%%%%%%%%%%%%%%%%%%%%%%%%%%%%%%%%%%%%%%%%%%%%
\subsection{Other filling ($n<1$)}
\label{sec:less-than-HF}
%%%%%%%%%%%%%%%%%%%%%%%%%%%%%%%%%%%%%%%%%%%%%%%%%%%%%%%%%%%%%%%
Now let us consider the filling less than half-filling, i.e., $N_{\text{e}} < L$.   
In this case, the strong-coupling ground state is $_L \text{C}_{N_{\text{e}}} \times 2^{N_{\text{e}}} \times 3^{L-N_{\text{e}}}$-fold degenerate.  
This degree of degeneracy includes the value $2^{L}$ at half-filling as a spacial case $N_{\text{e}}=L$. 
Unlike at half-filling, there exist some sites without conduction electrons 
(since $n<1$, doubly-occupied sites are not allowed in the strong-coupling ground state), and electrons can move even in the limit 
$J_{\text{K}} \to \infty$.    
As we will see, this difference dramatically changes the magnetism.  

In deriving the effective Hamiltonian, we first note that the Kondo doublets carry the spin degrees of freedom in contrast to the case 
of spin-1/2 KL model \cite{Tsunetsugu_review},  
where the spin degrees of freedom are quenched at the sites occupied by the conduction electrons by forming the Kondo singlets.   
For these reasons, 
at $n < 1$, the electron motion contributes to the magnetism already at the first-order in $t$.   
The first-order effective Hamiltonian reads:
%\begin{widetext}
\eq{
\begin{split}
H^{(n<1)}_{\text{eff}}= &
-t\sum_{i} \biggl\{  \hat{d}^{\dagger}_{i+1} \hat{d}_{i} \,  f^{(S=1)}_{i\to i+1} ( \vec{D}_i{\cdot} \vec{S}_{i+1} ) \, 
\hat{n}_{\text{d},i} (1-\hat{n}_{\text{d},i+1})  \\
&+  \hat{d}^{\dagger}_{i} \hat{d}_{i+1} \,  f^{(S=1)}_{i+1 \to i} ( \vec{S}_i{\cdot} \vec{D}_{i+1} )  \, 
(1-\hat{n}_{\text{d},i}) \hat{n}_{\text{d},i+1}   \biggr\}  \; , 
\end{split}
\label{hopping1st_effectiveH} 
}
%\end{widetext}
where $\hat{n}_{\text{d},i}~(=0,1)$ denotes the number of the Kondo doublets at site $i$, which are created (annihilated) by the fermionic operator 
$\hat{d}^\dagger_i$ ($\hat{d}_i$), and the effective spin-dependent hopping amplitudes of the doublets are given by:
\begin{equation}
\begin{split}
& f^{(S=1)}_{i\to i+1} ( \vec{D}_i{\cdot} \vec{S}_{i+1} ) =  (2/3)  \vec{D}_i \cdot \vec{S}_{i+1}  + 1/3   \\
& f^{(S=1)}_{i+1 \to i} ( \vec{S}_i{\cdot} \vec{D}_{i+1} ) =  (2/3)  \vec{S}_i \cdot \vec{D}_{i+1}  + 1/3  
\end{split}
\label{eqn:eff-int-S-1}
\end{equation}   
[see Eq.~\eqref{eqn:polynomial-spin-S} for the expression for general $S$].  
The derivation of the above equation \eqref{hopping1st_effectiveH} and the generalization to the arbitrary spin-$S~(\geq 1)$ are given 
in Appendix \ref{Appendix_derivation_1st_order_hopping}.  

The amplitude of the doublet hopping $\hat{d}^{\dagger}_{i\pm 1} \hat{d}_{i}$ in the Hamiltonian \eqref{hopping1st_effectiveH} 
takes its maximal value ($-2t/3$) when a doublet ($D=1/2$) and the localized spin ($S=1$) on the adjacent site are coupled ferromagnetically, which suggests a ferromagnetic 
ground state similar to that of the double-exchange model \cite{Zener-51,Anderson-H-55,deGennes-60,Kubo-82}.  
In fact, as is discussed in Appendix \ref{sec:ferro-proof}, exploiting the non-positivity and the indecomposability of the effective Hamiltonian \eqref{hopping1st_effectiveH}, we can rigorously show that the ground state of the effective Hamiltonian 
\eqref{hopping1st_effectiveH} is unique (up to trivial degeneracy associated with the rotational symmetry) 
and ferromagnetic with the maximal total spin $S_{\text{tot}} = L - N_{\text{e}}/2$ for $1 \leq N_{\text{e}} \leq L-1$.    
%%%%%%%%%%%%%%%%%%%%%
Hence, the ground state of the spin-1 KL model in the strong-coupling region is ferromagnetic for generic filling $0 < n < 1$ (and for $1 < n < 2$ 
by the particle-hole symmetry) [in fact, the statement can be generalized to arbitrary $S \geq 1$ in which 
the maximal total spin is $S_{\text{tot}} = L S - N_{\text{e}}/2$ ; see Appendix \ref{sec:ferro-proof}].   
This is consistent with the recent numerical observation for the spin-1 KL model \cite{Suzuki_Hattori1,Suzuki_Hattori2}.  
The ferromagnetic phase in the large-$J_{\text{K}}$ region is reminiscent of the situation in the spin-1/2 KL model 
\cite{kondo_lowdendity_FM,Sigrist-T-U-R-92,McCulloch-J-R-G-02,Peters-K-12},  
but the way how the hopping of conduction electrons causes ferromagnetism is different 
from each other; 
the mechanism of ferromagnetic ground state of spin-1 KL model \eqref{klH} rather resembles the double-exchange interaction
first-order in $t$, while, in the latter case, the ferromagnetism occurs  
through the second-order ($\propto t^{2}$) effective interactions \cite{Sigrist-T-U-R-92}.  
The ferromagnetic-metal phase found in the strong-coupling region persists down to $J_{\text{K}}\to 0$ 
at least in the low-density ($n\to 0$) limit as in the $S=1/2$ case \cite{boundary_FM_PM}.   
In fact, it is straightforward to generalize the proof in Ref.~\cite{kondo_lowdendity_FM} to $S \geq 1$ to show that 
the ground state of the {\em single-electron} spin-$S$ KL model is ferromagnetic.    

Now let us consider the effects of the Heisenberg term $\hat{H}_{\text{H}}$.  
By the same argument as that leading to \eqref{eqn:1st-order-JH-S1}, 
we see that the projection $\vec{\widetilde{S}}_{i}$ of the localized spin $\vec{S}_{i}$ onto the ground-state subspace is given by:
\begin{equation}
\vec{\widetilde{S}}_{i} = 
\begin{cases}
\vec{S}_{i} & \text{when site-$i$ is unoccuied} \\
(4/3) \vec{D}_{i}  & \text{when site-$i$ is occuied} \; .
\end{cases}
\label{eqn:localized-spin-projected}
\end{equation}
Then, the first-order perturbation in $J_{\text{H}}$ results in 
the following antiferromagnetic spin-spin exchange on the adjacent spins (either $S=1$ or $S=1/2$ depending 
on how the individual sites are occupied by the conduction electrons):
\begin{equation}
J_{\text{H}} \sum_{i} \vec{\widetilde{S}}_{i}  \cdot \vec{\widetilde{S}}_{i+1}   \; .
\label{eqn:J_H-int-away-from-HF}
\end{equation}
Therefore, in  the strong-coupling region of the KH model ($J_\text{H} > 0$), 
the ferromagnetic order found above for the KL model 
may be destabilized by the antiferromagnetic interaction generated by the Heisenberg term $J_{\text{H}}$.   
A rough estimate of the critical value of $J_{\text{H}}$ may be obtained by comparing 
the spin-dependent hopping amplitude \eqref{eqn:eff-int-S-1} and the projected Heisenberg interaction \eqref{eqn:J_H-int-away-from-HF}: 
$J_{\text{H}}^{\text{c}}/t \sim 1/2$.  A more precise calculation for the two-site system shows that the ferromagnetic ground state ends at 
$J_{\text{H}}/t= 1/6$.   

Note that the effects of the Heisenberg term $\hat{H}_{\text{H}}$ is very different for $n=1$ (half-filling) and $n <1$; 
in the former, $\hat{H}_{\text{H}}$ stabilizes the antiferromagnetic correlation in the insulating phase, while 
it competes with the hopping-assisted (double-exchange) ferromagnetism in the latter.   
We will closely investigate this competition in the next section.

%%%%%%%%%%%%%%%%%%%%%%%%%%%%%%%%%%%%%%%%%%%%%%%%%%%%%%%%%%%%%%%%%%%
\section{Numerical results for Kondo-Heisenberg model}
\label{sec:numerical_results}
%%%%%%%%%%%%%%%%%%%%%%%%%%%%%%%%%%%%%%%%%%%%%%%%%%%%%%%%%%%%%%%%%%%
In this section, we report the numerical results for a particular case of the spin-1 localized moments.  
To obtain the ground state of the $S=1$ KH model \eqref{khH}, 
we carried out density-matrix renormalization group (DMRG) simulations 
using an open source library ITenosr \cite{itensor} for the DMRG algorithm.  
In addition, in some DMRG simulations, 
we adopted the sine-square-deformed Hamiltonian \cite{SSD,SSD_chargegap} in order to reduce   
the effects of the open boundary condition, in which the DMRG algorithm  works better \cite{White1,White2}.  
Specifically, we simulated the following Hamiltonian instead of the original one \eqref{khH}:
%\begin{widetext}
\begin{subequations}
\begin{equation}
\begin{split}
& H_{\text{KH,SSD}}  \\
& =   - t\sum_{i=1,\alpha}^{i=L-1}f_1(i) \left( c^{\dagger}_{i,\alpha}c_{i,\alpha}+\text{H.C.} \right) 
 -\mu\sum_{i=1,\alpha}^{i=L}f_0(i)c^{\dagger}_{i,\alpha}c_{i,\alpha} \\
&  \phantom{=}   
+J_{\text{K}}\sum_{i=1}^{i=L} f_{0}(i)\vec{s}_i \cdot \vec{S}_i 
+ J_{\text{H}} \sum_{i=1}^{i=L-1}f_1(i)\vec{S}_i \cdot \vec{S}_{i+1}  \; ,
\end{split}
\label{SSD_Hamiltonian}
\end{equation}
where the deformation functions are given by:
\begin{equation}
f_l(i) = \text{sin}^2 \left[ \frac{\pi}{L} \left( i+\frac{l-1}{2} \right) \right] 
\quad (l=0,1) 
\end{equation}
\end{subequations}
and $l=0$ ($l=1$) is used for the on-site (on-bond) operators.  
%\end{widetext}
With the Sine-Square Deformation~(SSD) \cite{GendiarK-N-SSD-09}, the vicinity of the center of a finite system well approximates the bulk 
of the infinite system.  In our simulations, we considered finite systems of the sizes $L \leq 100$ under open boundary condition,  
and set the block sizes $m \leq 720$.  In all the cases, we found that the truncation errors were less than $\sim 10^{-7}$.  
Throughout this section, we set $t=1$ as the unit of energy.
%\label{sec:numerics-at-HF}
%%%%%%%%%%%%%%%%%%%%%%%%%%%%%%%%%%%%%%%%%%%%%%%%%%%%%%%%%%%%%%%%%%%
%%%%%%%%%%%%%%%%%%%%%%%%%%%%%%%%%%%%%%%%%%%%%%%%%%%%%%%%%%%%%%%%%%%
\subsection{half-filling}
\label{sec:numerics-at-HF}
\begin{figure}[h]
\begin{center}
\includegraphics[width=\columnwidth,clip]{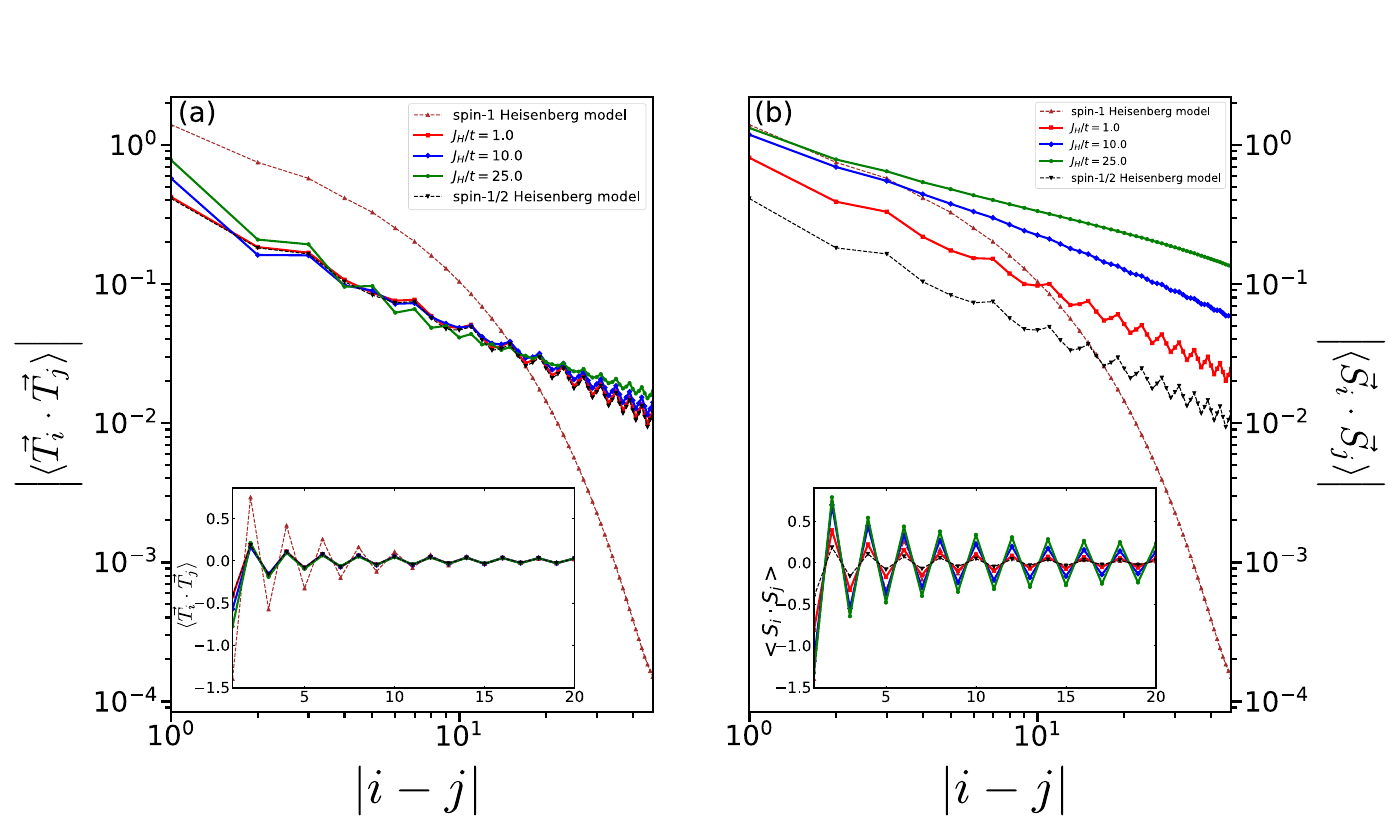}
\caption{Correlation functions (log-log plots) of (a) the total spin 
$\braket{\vec{T}_i \cdot \vec{T}_{j} }$ := $\braket{(\vec{s}_i+\vec{S}_i) \cdot (\vec{s}_{j}+\vec{S}_{j})}$ 
and (b) the $S=1$ localized spins $\braket{\vec{S}_i \cdot \vec{S}_{j}}$ 
in the ground state of the {\em uniform} (i.e., without SSD) spin-1 KH model \eqref{khH} at half-filling.  
Both are calculated at the fixed $J_\text{K}/t=10.0$ for varying $J_\text{H}/t$.  
For comparison, the spin-spin correlation functions of the $S=1/2,~1$ Heisenberg models are also plotted (dashed curves).  
The zoom-up of the short-range part is shown in the insets which clearly indicates the antiferromagnetic nature of the correlation. 
\label{half_filling_spin_corr_long_range_log_log}}
\end{center}
\end{figure}

%%%%%%%%%%%%%%%%%%%%%%%%%%%%%%%%%%%%%%%%%%%%%%%%%%%%%%%%%%%%%%%%%%%
In Sec.~\ref{halffilling_case}, we have studied the ground state in the strong-coupling limit ($J_\text{K} \gg t, J_{\text{H}}$), 
where the perturbation theory in $t$ and $J_{\text{H}}$ predicts that the ground state is insulating; the low-energy physics is described, 
when $S=1$, by the spin-$1/2$ Heisenberg model \eqref{effHam}, which indicates the power-law antiferromagnetic spin-spin correlation.    
To check this for increasing $J_{\text{H}}$ ($0 \leq J_{\text{H}} \lesssim J_{\text{K}}$), 
we numerically investigate in this section the ground-state spin-spin correlation at half-filling.   
Specifically, we fixed $J_\text{K}/t=10$, and increased $J_\text{H}/t$ from the Kondo-lattice limit ($J_\text{H}=0$) 
up to $J_\text{H} \sim J_\text{K}$ to calculate the spin correlation between distant sites for each $J_\text{H}/t$.  
In doing so, we used the uniform (i.e., undeformed) Hamiltonian \eqref{khH}.  
The results for the correlation functions of (a) the composite spins $\langle \vec{T}_{i}{\cdot} \vec{T}_{j} \rangle$ 
[$\vec{T}_{i}$ is defined in \eqref{total_spin_operator}]  
and (b) the localized spins $\langle \vec{S}_{i} {\cdot} \vec{S}_{j} \rangle$, as well as those for the spin-1 and 1/2 Heisenberg models, 
are plotted in Fig.~\ref{half_filling_spin_corr_long_range_log_log}.    

%%%%%%%%%%%%%%%%%%%     FIG       %%%%%%%%%%%%%%%%%%%%%%%%%%%%%%%%%%%%%%%%%%
%%%%%%%%%%%%%%%%%%%%%%%%%%%%%%%%%%%%%%%%%%%%%%%%%%%%%%%%%%
%%%%%%%%%%%%%%%%%%%%%%%%%%%%%%%%%%%%%%%%%%%%%%%%%%%%%%%%%%
%%%%%%%%%%%%%%%%%%%%%%%%%%%%%%%%%%%%%%%%%%%%%%%%%%%%%%%%%%%%%%%%%%%

From the numerical results, we can first read off that the composite-spin correlation function 
$\langle \vec{T}_{i}{\cdot} \vec{T}_{j} \rangle$ essentially coincides with the ordinary spin-spin correlation 
of the spin-1/2 Heisenberg chain (shown by the dashed line), up to fairly large $J_\text{H}/t$ 
[see Fig.~\ref{half_filling_spin_corr_long_range_log_log}(a)].   
As the system is insulating and in the strong-coupling region, 
this quasi-long-range antiferromagnetic correlation is not attributed to the RKKY interaction which requires metallicity 
and is valid in the weak-coupling regime.    
Rather, this implies that the strong-coupling picture discussed in Sec.~\ref{halffilling_case} 
remains valid even for large $J_{\text{H}}(\sim J_{\text{K}})$, suggesting that 
the Kondo doublets, which are well-defined when $J_{\text{K}} \gg t, J_{\text{H}}$, are rather robust 
against the interaction ($J_{\text{H}}$) among the localized spins.   

We can also confirm this persistence of the Kondo doublets 
by the results shown in Fig.~\ref{half_filling_spin_corr_long_range_log_log}(b).    
According to the strong-coupling argument in Sec.~\ref{halffilling_case}, 
the correlation function of the localized $S=1$ spins in the KH model \eqref{khH} 
should behave like that of the spin-1/2 Heisenberg chain \eqref{effHam}:   
\begin{equation}
\langle \vec{S}_{i} {\cdot} \vec{S}_{j}\rangle_{\text{KH}} \xrightarrow{\mathcal{H}_{\text{hf}}}  
(4/3)^{2} \langle \vec{D}_{i} {\cdot} \vec{D}_{j} \rangle_{\text{Heisenberg}}   \; .
\end{equation}
In fact, the plots in Fig.~\ref{half_filling_spin_corr_long_range_log_log}(b) clearly show that 
the localized-spin correlation function $\langle \vec{S}_{i} {\cdot} \vec{S}_{j}\rangle_{\text{KH}}$ and the spin-spin correlation function 
of the spin-1/2 Heisenberg chain (dashed line) behave similarly \footnote{%
Almost parallel shifts of the four curves in Fig.~\ref{half_filling_spin_corr_long_range_log_log}(b) suggest that 
$\langle \vec{S}_{i} {\cdot} \vec{S}_{j}\rangle$ and $\langle \vec{D}_{i} {\cdot} \vec{D}_{j} \rangle$ 
differ only by numerical factors.} even when $J_{\text{H}}$ is fairly large;  
the correlation function $\langle \vec{S}_{i} {\cdot} \vec{S}_{j}\rangle_{\text{KH}}$ exhibits behavior essentially different from the short-range 
(i.e., exponentially-decaying) correlation 
in the spin-1 Heisenberg chain which describes the physics of the localized spins when $J_{\text{K}}=0$.   
All these suggest that the exchange interaction $J_{\text{H}}$ does not really interfere with the antiferromagnetism 
stabilized by the motion of the conduction electrons, and that the Kondo-doublet physics dominates a wide range of 
the parameter space (i.e., $0 \leq J_{\text{H}} \lesssim J_{\text{K}}$) at half-filling.    
%%%%%%%%%%%%%%%%%%%%%%%%%%%%%%%%%%%%%%%%%%%%%%%%%%%%%%%%%%%%%%%%%%%
\subsection{Away from half-filling}
\label{sec:numerics-at-QF}
%%%%%%%%%%%%%%%%%%%%%%%%%%%%%%%%%%%%%%%%%%%%%%%%%%%%%%%%%%%%%%%%%%%
In the last section, we have seen that the inclusion of antiferromagnetic $J_{\text{H}}$ does not essentially affect the insulating ground state 
with gapless antiferromagnetic spinon excitations at half-filling found in the strong-coupling limit.  
Away from half-filling (i.e., $0 < n < 1$), on the other hand, the situation is very different.  In fact, the strong-coupling argument tells us that 
the electron hopping tends to stabilize the ferromagnetic ground state which may be eventually destabilized by the antiferromagnetic 
interaction \eqref{eqn:J_H-int-away-from-HF}.  
In this section, we consider various ground-state phases resulting from the competition between the kinetic-energy-driven ferromagnetism 
and the $J_{\text{H}}$-induced antiferromagnetism.   
Specifically, we fix $J_{\text{K}}$ large, i.e., $J_{\text{K}}/t =5.0$ and numerically investigate the stability of the ferromagnetic order 
found in the KL-model limit $J_\text{H} =0$ against the antiferromagnetic interaction $J_{\text{H}}$.   
As long as we know from the preliminary calculations, the quarter-filling case $n=1/2$ seems most interesting, 
and we mainly focus on the case with $n=1/2$ in this subsection unless otherwise stated.   
%%%%%%%%%%%%%%%%%%%%%%%%%%%%%%%%%%%%%%%%%%%%%%%%%%%%%%%%%%%%%%%%%%%
\subsubsection{Magnetic properties}
\begin{figure}[h]
\begin{center}
\includegraphics[width=\columnwidth,clip]{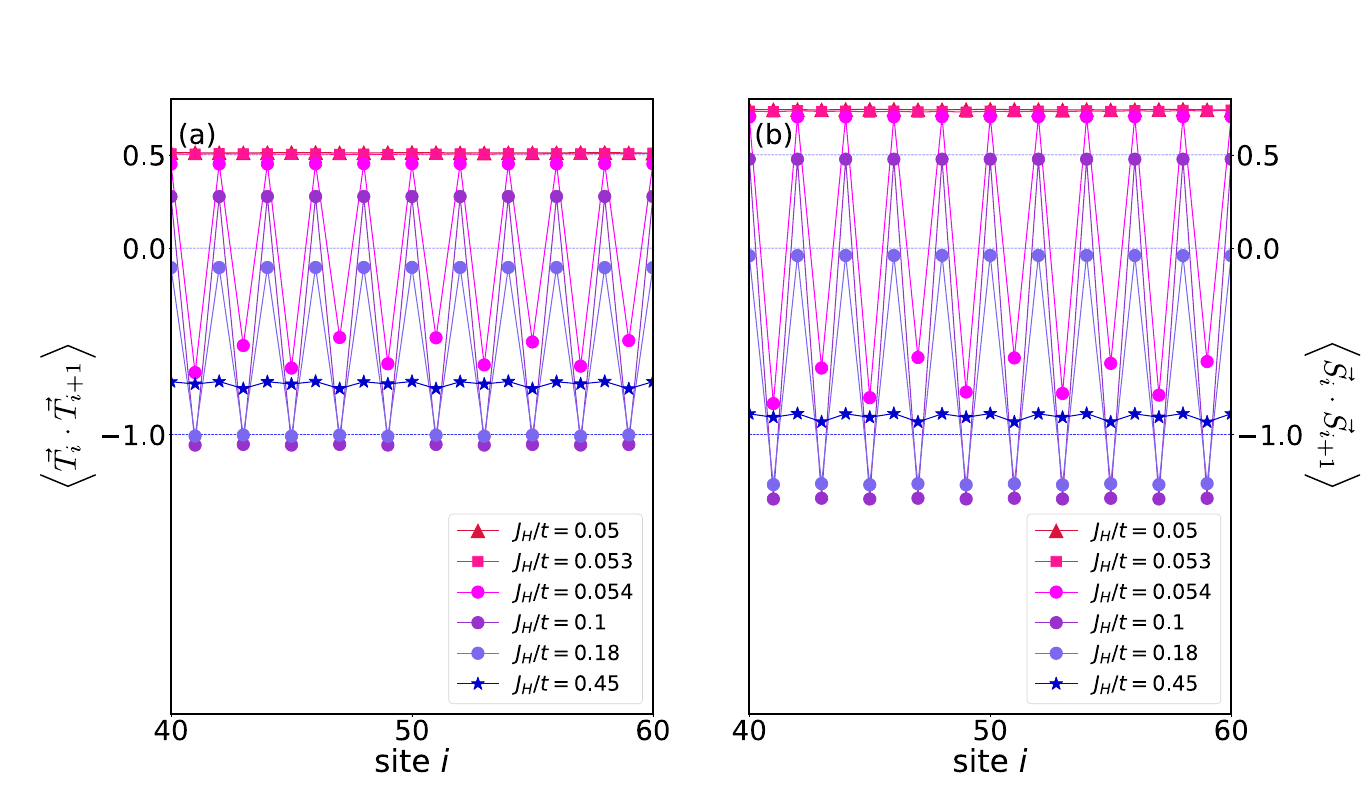}
\caption{
The nearest-neighbor spin-spin correlation (a) $\braket{\vec{T}_i \cdot \vec{T}_{i+1}}$ 
and (b) $\braket{\vec{S}_i \cdot \vec{S}_{i+1}}$
in the ground state of strongly coupled spin-1 KH model at quarter-filling. $\braket{\vec{T}_i \cdot \vec{T}_{i+1}}$ and $\braket{\vec{S}_i \cdot \vec{S}_{i+1}}$ are calculated for fixed $J_\text{K}/t=5.0$ and $L=100$. 
\label{q_spin1KH_SS_corr_jh2}}
\end{center}
\end{figure}

%%%%%%%%%%%%%%%%%%%%%%%%%%%%%%%%%%%%%%%%%%%%%%%%%%%%%%%%%%%%%%%%%%%
To investigate how the magnetic properties change as the direct antiferromagnetic interaction $J_{\text{H}}$ is increased,  
we calculated the correlation function of the total spin $\vec{T}_i (=\vec{s}_{i} + \vec{S}_{i})$ at each site as the probe. 
First, we show in Fig.~\ref{q_spin1KH_SS_corr_jh2}(a) and (b) the nearest-neighbor spin-spin correlation functions  
$\braket{\vec{T}_{i} \cdot \vec{T}_{i+1}}$ [(a)] and $\braket{\vec{S}_{i} \cdot \vec{S}_{i+1}}$ [(b)] 
between neighboring sites for various $J_{\text{H}}/t$ at a fixed Kondo coupling $J_\text{K}/t=5.0$.    

It can be seen from Fig.~\ref{q_spin1KH_SS_corr_jh2}(a) that for sufficiently weak $J_\text{H}$, 
i.e., $0\leq J_\text{H}/t \lesssim 0.053$, the value of the neighboring spin correlation 
$\braket{\vec{T}_i \cdot \vec{T}_{i+1}}$ is site($i$)-independent and takes a positive constant value (i.e. ferromagnetic) 
regardless of the value of $J_{\text{H}}$, while for $J_\text{H}/t \gtrsim 0.054$, it is alternating between two values.  
A similar behavior was observed for the localized spins as well [see Fig.~\ref{q_spin1KH_SS_corr_jh2}(b)].  
The period-two behavior in the bond-centered operators $\braket{\vec{T}_i \cdot \vec{T}_{i+1}}$ 
and $\braket{\vec{S}_{i} \cdot \vec{S}_{i+1}}$ suggests that the localized spins are dimerized 
for $J_\text{H}/t \gtrsim 0.054$.   

If we further increase $J_{\text{H}}$, the clear period-2 behavior disappears at around $J_{\text{H}}/t \sim 0.45$ and 
both $\braket{\vec{T}_i \cdot \vec{T}_{i+1}}$ and $\braket{\vec{S}_i \cdot \vec{S}_{i+1}}$ become negatively constant, 
which means that, when $J_{\text{H}}/t \gtrsim 0.45$, the dimerized phase is taken over by a new phase in which 
short-range antiferromagnetic correlation develops [see Figs.~\ref{phase_qfill}(a)-(d)].  
This is consistent with that the system asymptotically approach the spin-1 Heisenberg model, which exhibits 
short-range antiferromagnetic correlation, if we increase $J_{\text{H}}$ to a large value with $J_\text{K}/t$ fixed.   
In contrast to the naive expectation based on the energetic argument in Sec.~\ref{sec:less-than-HF}, the ferromagnetic phase 
yields first to the dimerized one at much smaller value of $J_{\text{H}}$ before the antiferromagnetic tendency due to $J_{\text{H}}$ finally wins.  

\begin{figure}[h]
\begin{center}
\includegraphics[width=\columnwidth,clip]{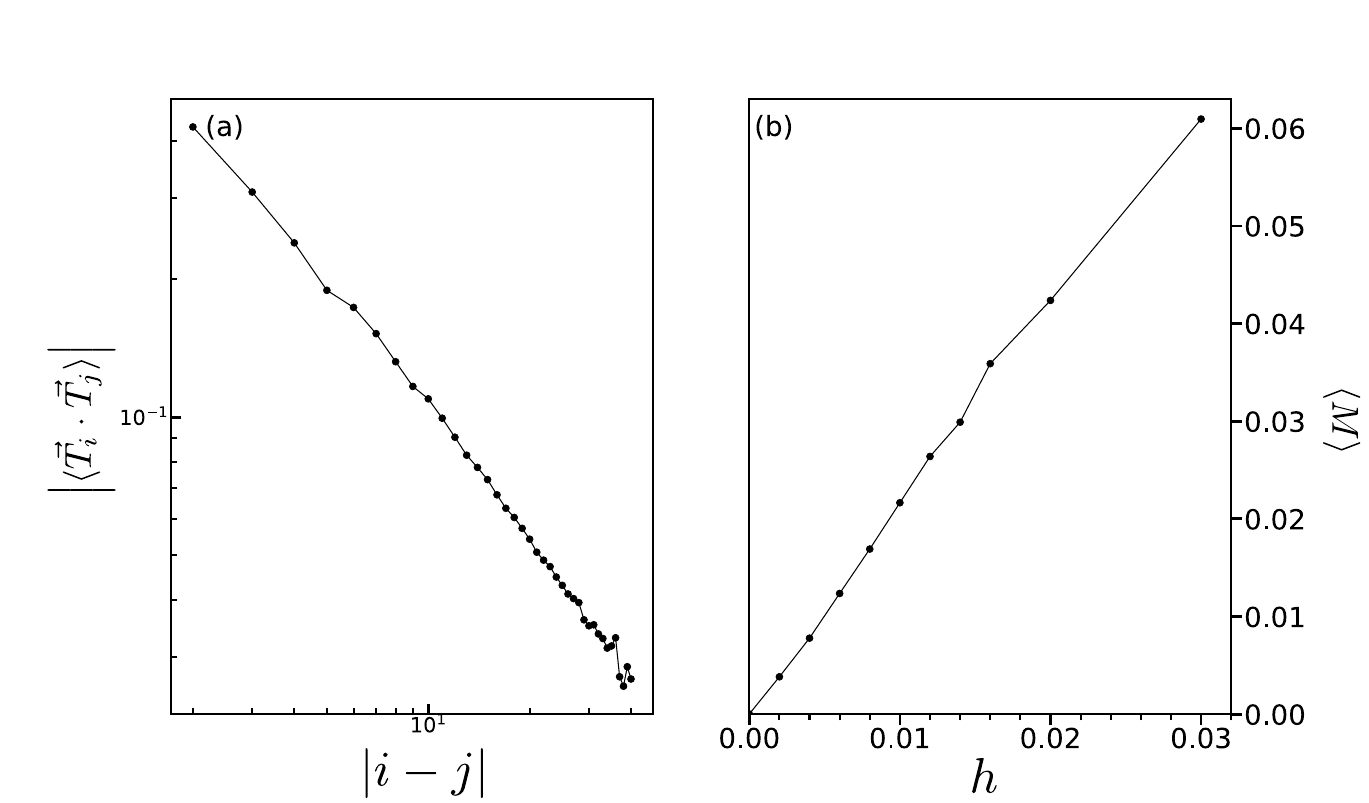}
\caption{
(a) Spin-spin correlation $\braket{\vec{T}_i \cdot \vec{T}_{j}}$ and (b) the ground-state magnetization process 
of the quarter-filled spin-1 KH model in the dimerized phase: $(J_\text{K}/t, J_\text{H}/t)=(5.0, 0.1)$. 
The correlation function seems to decay as $| \braket{\vec{T}_i \cdot \vec{T}_{j}} | \sim |i - j|^{-1}$.  
In obtaining (b), the SSD Hamiltonian \eqref{SSD_Hamiltonian} was used. 
\label{gapless_dimer}}
\end{center}
\end{figure}

Despite the usual lore that the spin dimerization is accompanied by a finite spin gap, 
the intermediate ``dimerized'' phase found above in fact has quasi-long-range antiferromagnetic correlation, i.e.,  
the correlation function $\braket{\vec{S}_i \cdot \vec{S}_j}$ exhibits power-law decay [see Fig.~\ref{gapless_dimer} (a)], 
indicating a vanishing spin gap \footnote{%
Although in critical isotropic spin systems, the spin-spin correlation function is generically 
expected to behaves like $(\ln |i-j|)^{1/2}/|i-j|$ \cite{Giamarchi} except at the fine-tuned points, 
we did not find such logarithmic corrections in our simulations.  
We do not know whether this absence of the logarithmic correction is explained by  
some effective long-range spin-spin interactions generated by the electron motion or not.}.   
To check whether the spin gap vanishes, we calculated the magnetization $\langle M \rangle = \sum_{i} T^{z}_{i}/L$ 
with increasing external magnetic field $h$ (in the $z$-direction).  
To this end, we used the SSD Hamiltonian \eqref{SSD_Hamiltonian} with the (deformed) Zeeman term 
$-h \sum_{i} f_0(i)(S^z_i +s^z_i )$ added.   The results are shown in Fig.~\ref{gapless_dimer}(b).  
The linear increase of the magnetization $\langle M \rangle \propto h$ ($h \ll 1$) strongly suggests that the spin gap indeed vanishes 
in the dimerized phase.    

\begin{figure}[h]
\begin{center}
\includegraphics[width=\columnwidth,clip]{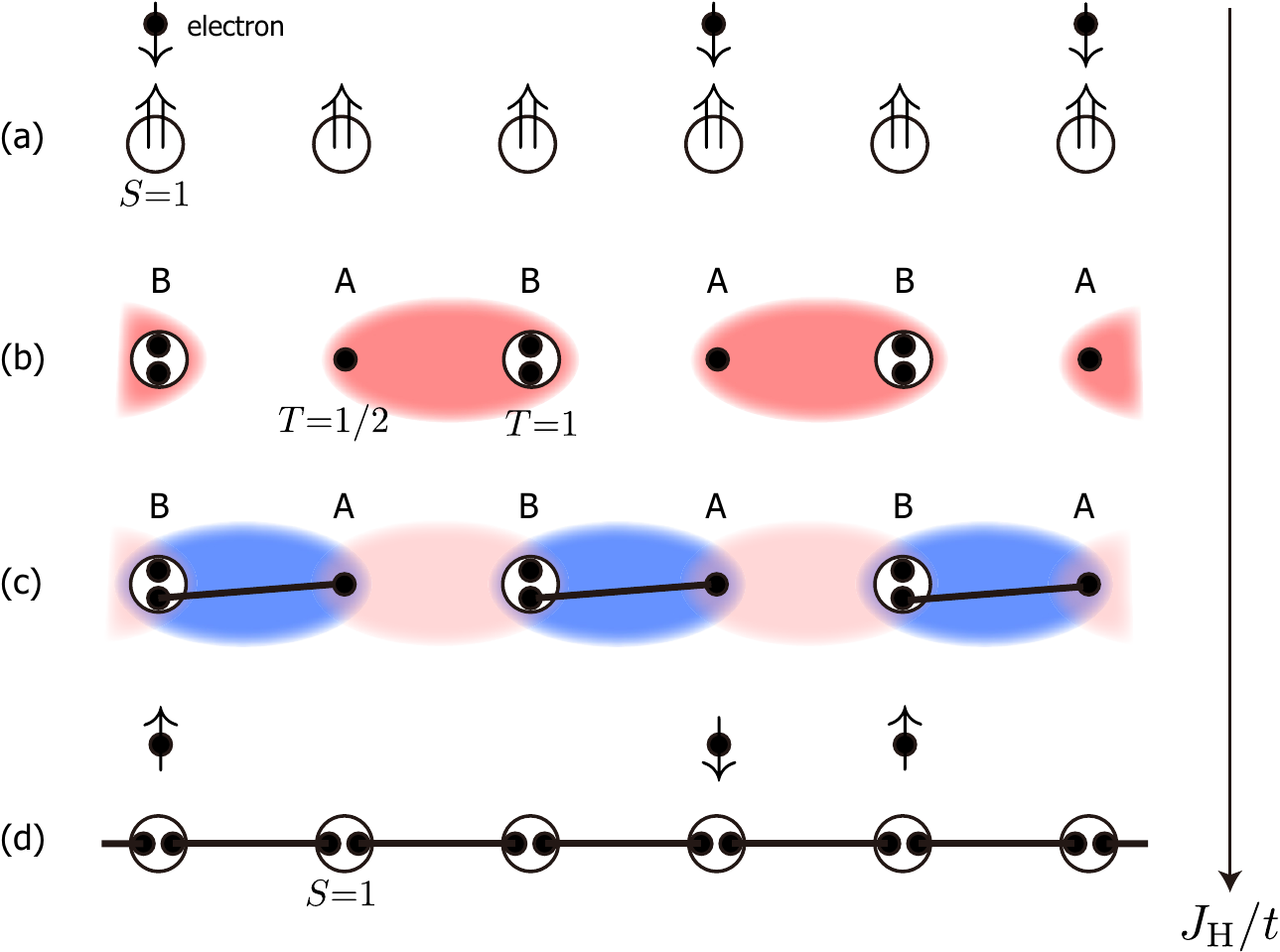}
\caption{
Schematic pictures illustrating how magnetic property changes as $J_{\text{H}}/t$ increases.  
(a) ferromagnetic phase, (b) dimerized phase for small $J_{\text{H}}$ in which spin-3/2s (i.e., local ferromagnetic correlation between  
$T=1/2$ and $T=1$) are formed on A-B bonds (red ovals), (c) dimerized phase 
for larger $J_{\text{H}}$ where doublets develop on B-A bonds (blue ovals), and (d) metallic phase with short-range antiferromagnetic correlation. 
Note that there is no transition between (b) and (c).  
\label{process_of_breaking_ferromagnetism}}
\end{center}
\end{figure}

All these properties of the dimerized phase may be best understood 
in the strong-coupling limit in which the system is described only in terms of spin-1/2 (the Kondo doublets $\vec{D}_{i}$) 
and the unscreened localized spin-1 ($\vec{S}_{i}$) [see Eq.~\eqref{hopping1st_effectiveH}].   
Let us consider the situation where $J_{\text{H}}$ is much larger than $t$ and we can neglect the the order-$t$ perturbation 
\eqref{hopping1st_effectiveH}.  
Depending on the configurations, $\braket{\vec{T}_i \cdot \vec{T}_{i+1}}$ is given by 
\begin{equation}
\braket{\vec{T}_i \cdot \vec{T}_{i+1}} 
= 
\begin{cases}
\braket{\vec{S}_i \cdot \vec{S}_{i+1}}  & \text{when $(T_{i},T_{i{+}1})=(1,1)$}  \\
\braket{\vec{S}_i \cdot \vec{D}_{i+1}}  & \text{when $(T_{i},T_{i{+}1})=(1,1/2)$}  \\
\braket{\vec{D}_i \cdot \vec{S}_{i+1}}  & \text{when $(T_{i},T_{i{+}1})=(1/2,1)$}  \\
\braket{\vec{D}_i \cdot \vec{D}_{i+1}}  & \text{when $(T_{i},T_{i{+}1})=(1/2,1/2)$}  \; .
\end{cases}
\end{equation}
Similarly, for the (projected) localized spins [see Eq.~\eqref{eqn:localized-spin-projected}], we have:
\begin{equation}
\begin{split}
& \braket{\vec{\widetilde{S}}_i \cdot \vec{\widetilde{S}}_{i+1}}  \\
& = 
\begin{cases}
\braket{\vec{S}_i \cdot \vec{S}_{i+1}}  & \text{when $(T_{i},T_{i{+}1})=(1,1)$}  \\
(4/3) \braket{\vec{S}_i \cdot \vec{D}_{i+1}}  & \text{when $(T_{i},T_{i{+}1})=(1,1/2)$}  \\
(4/3) \braket{\vec{D}_i \cdot \vec{S}_{i+1}}  & \text{when $(T_{i},T_{i{+}1})=(1/2,1)$}  \\
(16/9) \braket{\vec{D}_i \cdot \vec{D}_{i+1}}  & \text{when $(T_{i},T_{i{+}1})=(1/2,1/2)$}  \; .
\end{cases}
\end{split}
\end{equation}
The value $\braket{\vec{T}_i \cdot \vec{T}_{i+1}}=1/2$ means that neighboring $S=1/2$ and $S=1$ form spin-3/2 pairs 
for, e.g., $J_{\text{H}}=0.054$ [see Fig.~\ref{process_of_breaking_ferromagnetism}(b)].  
The expected value $\braket{\vec{\widetilde{S}}_i \cdot \vec{\widetilde{S}}_{i+1}} = 2/3$ 
is consistent with the numerical results in Fig.~\ref{q_spin1KH_SS_corr_jh2}(b).   
On the other hand, the value $\braket{\vec{T}_i \cdot \vec{T}_{i+1}}=-1$ allows the two possibilities 
$(T_{i},T_{i{+}1})=(1,1)$ and $(1/2,1)$ [or $(1,1/2)]$.  
However, looking at the value $\braket{\vec{S}_i \cdot \vec{S}_{i+1}} \approx - 4/3$ [see Fig.~\ref{q_spin1KH_SS_corr_jh2}(b)], 
we may conclude that the second realizes for, e.g., $J_{\text{H}}/t \gtrsim 0.1$ and that the spin pairs $(1/2,1)$ form doublets [Fig.~\ref{process_of_breaking_ferromagnetism}(c)].  

From these observations, the following picture emerges. 
First we note that this dimerized state is in fact insulating as will be shown in the next subsection, which allows us to treat the spin-$1/2$ and $1$ 
(there are equal numbers of them at quarter-filling) as immobile.  The numerical results indicate that these spin-$1/2$ and $1$ alternate in the insulating dimerized phase.  
For small values of $J_{\text{H}}/t~(\gtrsim 0.054)$, local ferromagnetic correlation still remains and 
magnetism is described by the {\em preformed} quartets on the A-B bonds 
[see the red ovals in Fig.~\ref{process_of_breaking_ferromagnetism}(b)].  The weak fluctuations among these 
quartets may be captured by the spin-$3/2$ Heisenberg chain which eventually leads to a gapless collective singlet ground state.  
For larger values of $J_{\text{H}}/t~(\gtrsim 0.1)$, on top of the ferromagnetic correlation on the A-B bonds, 
short-range antiferromagnetic correlation develops on the B-A bonds, and doublets 
are formed on these bonds [see the blue ovals Fig.~\ref{process_of_breaking_ferromagnetism}(c)].   
Note that, in contrast to the usual spin-singlet dimerization, these two different kinds of correlation do {\em not} interfere with each other, and 
the state shown in Fig.~\ref{process_of_breaking_ferromagnetism}(b) smoothly crosses over 
to Fig.~\ref{process_of_breaking_ferromagnetism}(c).  Again, the fluctuations among these doublets may be taken into account by 
the spin-1/2 Heisenberg chain, which exhibits power-law antiferromagnetic correlation.  
A remark is in order here about the nature of ``antiferromagnetic'' correlation.  As the effective ``spin''-1/2s are formed on dimers, 
the $\pi$-oscillating correlation in the effective model translates to the period-4 oscillation in the original model. 
In fact, we numerically observed such power-law decaying period-4 behavior in the spin-spin correlation $\braket{\vec{T}_i \cdot \vec{T}_{j}}$ 
at $J_\text{H}/t= 0.1$.    
%%%
%%%%%%%%%%%%%%%%     FIG 3  %%%%%%%%%%%%%%%%%%%%%%%%%%%%%%%%%%%%%%%%%%
%%%%%%%%%%%%%%%%%%%%%%%%%%%%%%%%%%%%%%%%%%%%%%%%%%%%%%%%%%
%%%%%%%%%%%%%%%%%%%%%%%%%%%%%%%%%%%%%%%%%%%%%%%%%%%%%%%%%%
%%%%%%%%%%%%%%%%%%%%%%%%%%%%%%%%%%%%%%%%%%%%%%%%%%%%%%%%%%%%%%%%%%%

%%%%%%%%%%%%%%%%%%%     FIG  4   %%%%%%%%%%%%%%%%%%%%%%%%%%%%%%%%%%%%%%%%%%
%%%%%%%%%%%%%%%%%%%%%%%%%%%%%%%%%%%%%%%%%%%%%%%%%%%%%%%%%%
%%%%%%%%%%%%%%%%%%%%%%%%%%%%%%%%%%%%%%%%%%%%%%%%%%%%%%%%%%
%%%%%%%%%%%%%%%%%%%%%%%%%%%%%%%%%%%%%%%%%%%%%%%%%%%%%%%%%%%%%%%%%%%

\begin{figure}[h]
\begin{center}
\includegraphics[width=\columnwidth,clip]{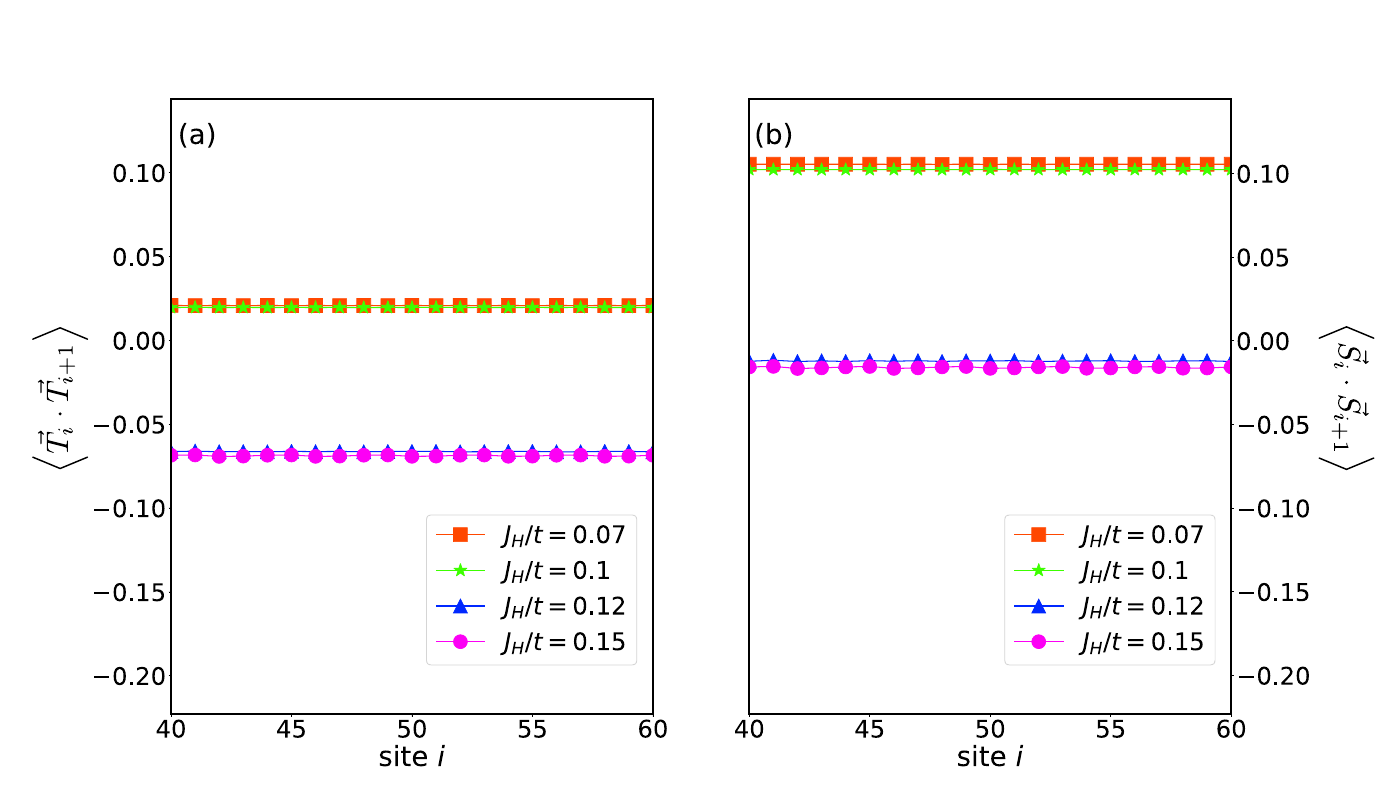}
\caption{
The same correlation functions as in Fig.~\ref{q_spin1KH_SS_corr_jh2} for $S=1/2$.  Again $J_\text{K}/t=5.0$ was used.  
Note that no alternating (period-2) behavior is observed for $S=1/2$.  
\label{halfspin_KH_q_filling_spin_corr_neighbor}}
\end{center}
\end{figure}

A similar ``gapless dimerized phase'' has also been found recently in the quarter-filled Kondo lattice model ($S=1/2$, $J_{\text{H}}=0$) 
at {\em small} Kondo coupling $J_\text{K}/t =0.6$ \cite{dimer2}.  
However, we would like to stress here that the above dimerized phase found in the strong-coupling region does not exist 
in the $S=1/2$ KH model at quarter-filling.  In fact, as is seen in Fig.~\ref{halfspin_KH_q_filling_spin_corr_neighbor}(a) and (b),  
a similar intermediate spin-dimerized state is absent in the $S=1/2$ case, 
and instead there seems to be a jump at $J_{\text{H}}/t \sim 0.1$ 
in both $\braket{\vec{T}_i \cdot \vec{T}_{i+1}}$ and $\braket{\vec{S}_{i} \cdot \vec{S}_{i+1}}$ 
from a positive value to a negative one.  
This sudden suppression of ferromagnetim by $J_\text{H}$ in the $S=1/2$ case is consistent 
with the analytic argument in Ref.~\cite{Moukouri-C-96}.  
%This is in stark contrast to the situation in the spin-1 KH model.  
Therefore, the existence of this intermediate gapless dimerized phase is one of the unique properties of the $S=1$ KH model at quarter-filling.  
In Fig.~\ref{process_of_breaking_ferromagnetism}, we illustrate how the ferromagnetic order is lost 
via the dimerized phase as we increase the interaction $J_{\text{H}}$ among the localized spins. 

%%%%%%%%%%%%%%%%%%%     FIG  5   %%%%%%%%%%%%%%%%%%%%%%%%%%%%%%%%%%%%%%%%%%
%%%%%%%%%%%%%%%%%%%%%%%%%%%%%%%%%%%%%%%%%%%%%%%%%%%%%%%%%%
%%%%%%%%%%%%%%%%%%%%%%%%%%%%%%%%%%%%%%%%%%%%%%%%%%%%%%%%%%
%%%%%%%%%%%%%%%%%%%%%%%%%%%%%%%%%%%%%%%%%%%%%%%%%%%%%%%%%%%%%%%%%%%

%%%%%%%%%%%%%%%%%%%     FIG       %%%%%%%%%%%%%%%%%%%%%%%%%%%%%%%%%%%%%%%%%%
%%%%%%%%%%%%%%%%%%%%%%%%%%%%%%%%%%%%%%%%%%%%%%%%%%%%%%%%%%
%%%%%%%%%%%%%%%%%%%%%%%%%%%%%%%%%%%%%%%%%%%%%%%%%%%%%%%%%%
%%%%%%%%%%%%%%%%%%%%%%%%%%%%%%%%%%%%%%%%%%%%%%%%%%%%%%%%%%%%%%%%%%%

%%%%%%%%%%%%%%%%%%%%%%%%%%%%%%%%%%%%%%%%%%%%%%%%%%%%%%%%%%%%%%%
\subsubsection{Electrical properties}
\label{sec:electronic-properties}
%%%%%%%%%%%%%%%%%%%%%%%%%%%%%%%%%%%%%%%%%%%%%%%%%%%%%%%%%%%%%%%

In the previous subsection, we have investigated the spin correlation $\braket{T_i \cdot T_{i+1}}$ and 
$\braket{S_i \cdot S_{i+1}}$ and concluded that the competition between the hopping-induced ferromagnetism and 
the direct antiferromagnetic interaction $J_{\text{H}}$ leads to two quantum phase transitions; one at $J_{\text{H}}/t \approx 0.054$ 
from the ferromagnetic phase to the new {\em gapless} dimerized phase, 
and another at $J_{\text{H}}/t \approx 0.45$ from the dimerized phase to a non-dimerized one with short-range antiferromagnetic correlation.  
In this subsection, we investigate the phase structure and the quantum phase transition(s) 
from the viewpoint of the electrical transport.  

\begin{figure}[ht]
\begin{center}
\includegraphics[width=\columnwidth,clip]{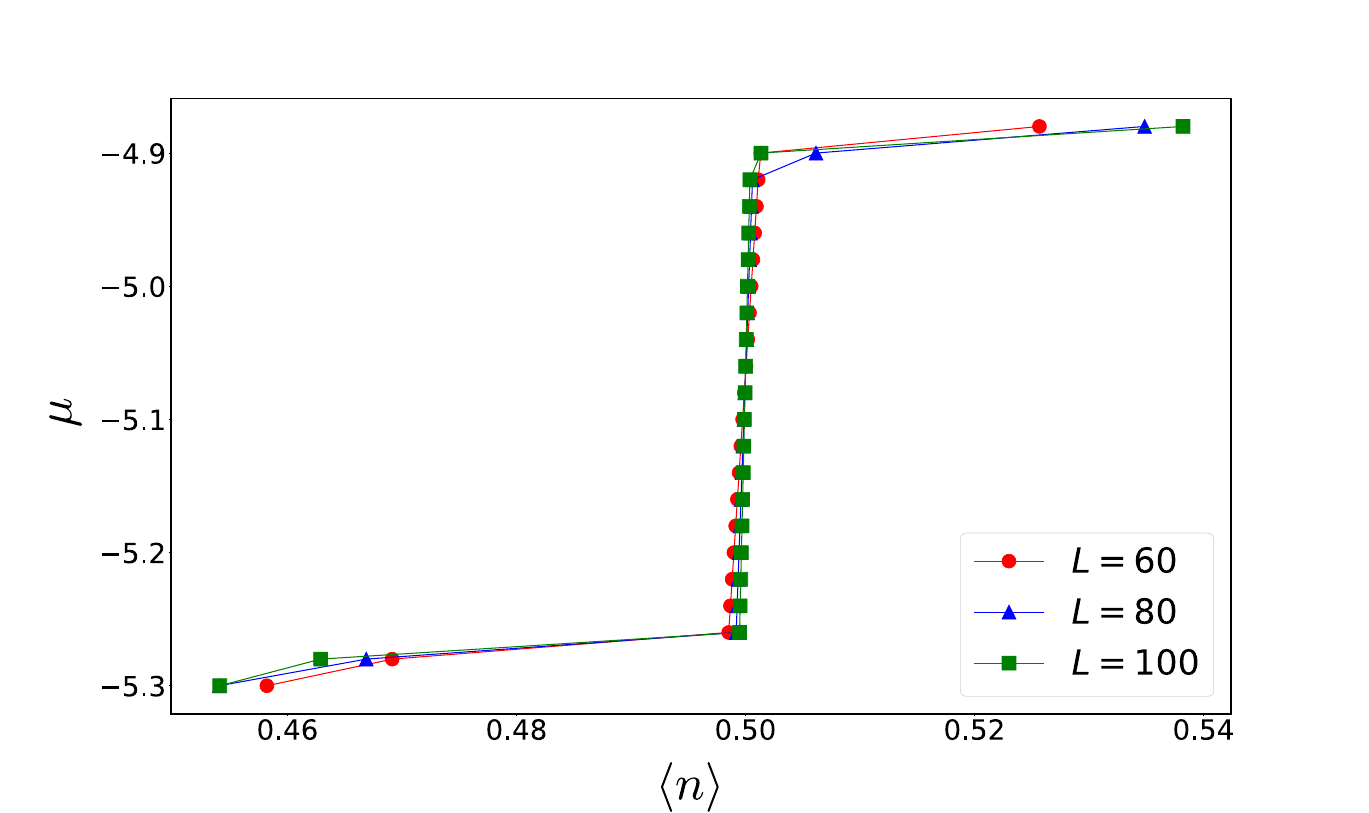}
\caption{
A typical behavior of $\mu$ as a function of the electron density $\braket{n}$ around $n=1/2$
for various system sizes $L=60,80,100$. The clear jump at $n=1/2$ indicates 
a finite charge gap at quarter-filling and the absence of the charge gap for other fillings (around $n=1/2$).  
$(J_{\text{K}}/t,J_{\text{H}}/t) =(5.0,0.1)$ is used.}\label{typical_charge_gap}
\end{center}
\end{figure}

To this end, we calculated the charge gap $\Delta_{\text{c}}$ for various $J_\text{H}/t$ with the Kondo coupling $J_\text{K}/t$ fixed.   
In doing so, we first fixed the chemical potential $\mu$ and calculated the bulk electron density $\braket{n(\mu)}$ 
by averaging the {\em local} density $\braket{n_i}$ around the center of the system.  
In Fig.~\ref{typical_charge_gap}, we plot the electron density $\braket{n(\mu)}$ obtained at $(J_{\text{K}}/t,J_{\text{H}}/t)=(5.0,0.1)$  
for several system sizes $L$.  There is a clear jump in $\mu$ {\em only} at $\braket{n}=0.5$, which suggests that 
a finite charge gap opens at quarter-filling and that the system is metallic for other densities around $n=1/2$.   
This seems consistent with the field-theoretical prediction in Ref.~\cite{Tsvelik-Y-19} which concludes that a metallic phase with collinear 
spin fluctuations ({\em collinear metal}) occupies the region around $n=1/2$ (when $J_{\text{H}}=0$).  
Also the size-dependence seems to be relatively small.  
The charge gap $\Delta_{\text{c}}$ is obtained by appropriately extrapolating the width of the jump to $L \to \infty$ \cite{SSD_chargegap}.   

\begin{figure}[ht]
\begin{center}
\includegraphics[width=\columnwidth,clip]{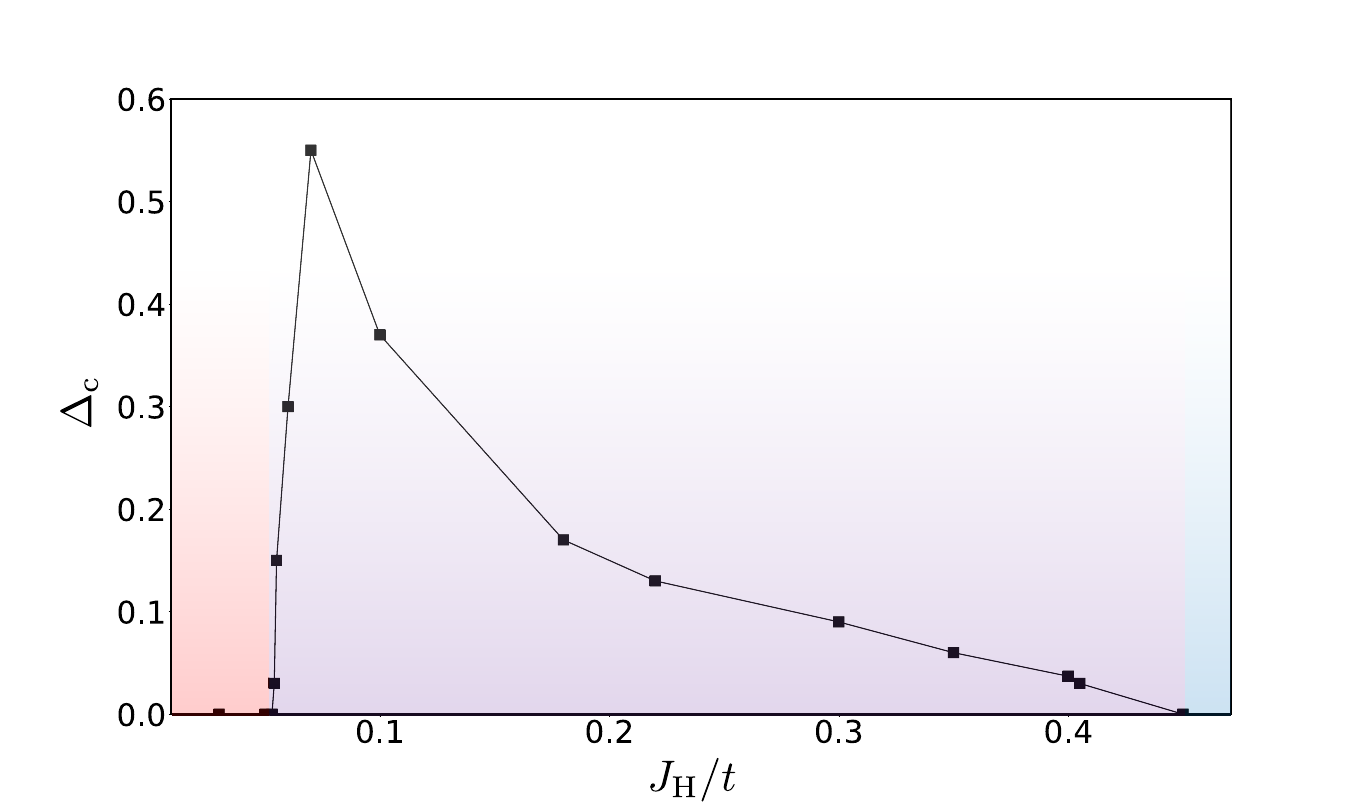}
\caption{
Relation between $J_{\text{H}}/t$ and the charge gap $\Delta_{\text{c}}$ for fixed $J_{\text{K}}/t=5.0$.  
The charge gap is obtained by extrapolating the finite-size values (which are given by the jump in the $n$-$\mu$ plot; 
see Fig.~\ref{typical_charge_gap}).  
Note that the charge gap $\Delta_{\text{c}}$ is finite only in the dimerized phase ($0.054 < J_{\text{H}}/t \lesssim 0.45$).  
\label{jh_vs_charge_gap}}
\end{center}
\end{figure}

\begin{figure}[ht]
\begin{center}
\includegraphics[width=\columnwidth,clip]{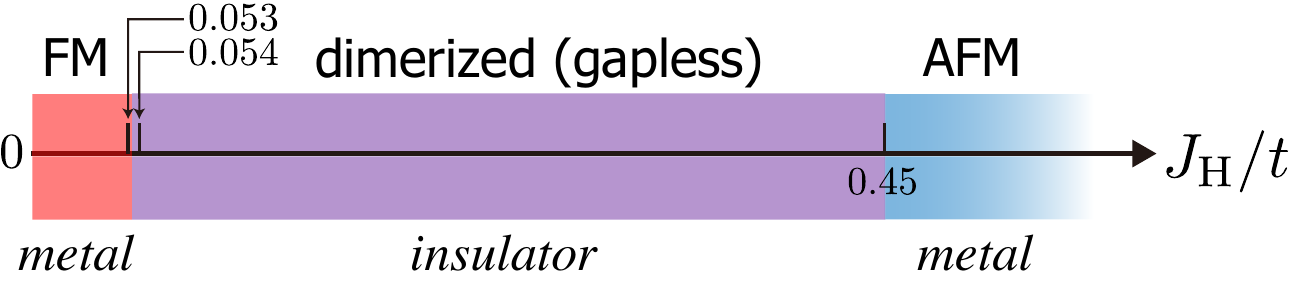}
\caption{
Phase diagram at quarter-filling ($n=1/2$) along the $J_{\text{H}}/t$ axis with fixed $J_{\text{K}}/t=5.0$.  
The magnetic and electrical properties of the phases are determined by $\braket{T_{i} \cdot T_{i+1}}$ ($\braket{S_{i} \cdot S_{i+1}}$) 
and the charge gap $\Delta_{\text{c}}$.  In the dimerized phase, the spin-spin correlation is power-law decaying with period-4 oscillation. 
\label{phase_qfill}}
\end{center}
\end{figure}

The value of the charge gap $\Delta_{\text{c}}$ at $n=1/2$ (quarter-filling) obtained in this way 
is shown as a function of $J_\text{H}/t$ ($J_\text{K}/t=5.0$ is fixed) in Fig.~\ref{jh_vs_charge_gap}.  
It clrearly shows that for $0\leq J_\text{H} \lesssim 0.053$ the ground state is metallic with a vanishing charge gap,  
while for $J_\text{H} \gtrsim 0.054$ the ground state is an insulator.  
Moreover, Fig.~\ref{jh_vs_charge_gap} shows that, after attaining a maximum at around $J_{\text{H}} \sim 0.07$,  
the charge gap $\Delta_{\text{c}}$ decreases monotonically until it vanishes at around $J_{\text{H}}/t =0.45$.   
Combining all these with the results of the last subsection, we conclude that 
the region where the system has a finite charge gap matches that of the dimerized phase.  
To put it another way, the two magnetic quantum phase transitions into and out of the dimerized phase 
(at $J_{\text{H}}/t\approx 0.054$ and $J_{\text{H}}/t\approx 0.45$), and the metal-insulator transitions found here occur simultaneously.   
The final phase diagram at quarter-filling along $J_H/t$ axis is shown in Fig.~\ref{phase_qfill}.   
Note that the third phase (``AFM'') is determined only by the order parameter $\braket{\vec{T}_i \cdot \vec{T}_{i+1}}$ and the charge gap $\Delta_{\text{c}}$, 
and the precise characterization, e.g., in the light of the heavy Luttinger liquid \cite{Khait-A-H-S-A-18} is yet to be done.  

The mechanism of this dimerization-induced metal-insulator transitions at quarter-filling is an intriguing question. 
One may naively expect that magnetic dimerization somehow induces the modulation of the hopping amplitude 
thereby halving the Brillouin zone and leading to a Mott-insulating state in the {\em half-filled} bonding band \cite{hubbard_qfill_dimer}.  
To clarify this point, we measured the hopping amplitude $\left\langle \sum_{\alpha}c^{\dagger}_{i\alpha} c_{i{+}1,\alpha} + \text{h.c.} \right\rangle$ 
in the dimerized phase to find no sign of alternation.   Therefore, this appealing scenario does not seem to work in our situation.  

However, once we assume the spin-dimerization in the local moments, a combination of bosonization and a mean-field-like argument 
seems to reasonably explain the opening of the charge gap at quarter-filling.  
When the spin correlation $\langle \vec{S}_{i} \cdot \vec{S}_{i+1} \rangle$ exhibits alternation, second-order perturbation in $J_{\text{K}}$ 
induces the following effective interaction among the conduction electrons \cite{Xavier-P-M-A-03}:
\begin{equation}
(-1)^{i} g_{\text{d}} ( \vec{s}_{i} \cdot \vec{s}_{i+1}  )
\end{equation}
where the coupling constant $g_{\text{d}}$ is proportional to the amplitude of the spin-dimerization.  
Then, it is straightforward to treat the above interaction in the framework of bosonization \cite{Giamarchi}, and we see that, at $n=1/2$,  
the charge sector of the conduction electron acquires the interaction $\cos (\sqrt{8} \phi_{\rho})$ whose scaling dimension is 
$2/K_{\rho}$ (with $\phi_{\rho}$ and $K_{\rho}$ respectively being the charge boson field and the corresponding Luttinger-liquid parameter) \footnote{%
We follow the convention of Ref.~\cite{Giamarchi}.  Precisely, we have one more interaction of the form 
$\cos(\sqrt{8}\phi_{\rho})\cos(\sqrt{8}\phi_{\sigma})$ with the scaling dimension $2+2/K_{\rho} >2$.  This is irrelevant and 
we can safely drop it.}, which, when $K_{\rho} >1$, opens the charge gap.   Note that the period-2 component of 
the hopping amplitude $\sin (\sqrt{8} \phi_{\rho} )$ has a zero expectation value consistent with the above numerical observation.   
Although this argument seems reasonable, the spin-dimerization and the opening of the charge gap actually occur hand in hand, and 
a clear explanation of the mechanism of the magnetic dimerization still remains to be an important open question.   
%%%%%%%%%%%%%% FIG 4 %%%%%%%%%%%%%%%%%%%%%%%%%%%%%%%%%%%%%%%%%%
%%%%%%%%%%%%%% FIG 5 %%%%%%%%%%%%%%%%%%%%%%%%%%%%%%%%%%%%%%%%%%
%%%%%%%%%%%%%%%%%%%%%%%%%%%%%%%%%%%%%%%%%%%%%%%%%%%%%%%%%%
%%%%%%%%%%%%%%%%%%%%%%%%%%%%%%%%%%%%%%%%%%%%%%%%%%%%%%%

%%%%%%%%%%%%%% FIG 5 %%%%%%%%%%%%%%%%%%%%%%%%%%%%%%%%%%%%%%%%%%
%%%%%%%%%%%%%%%%%%%%%%%%%%%%%%%%%%%%%%%%%%%%%%%%%%%%%%%%%%
%%%%%%%%%%%%%%%%%%%%%%%%%%%%%%%%%%%%%%%%%%%%%%%%%%%%%%%

%%%%%%%%%%%%%%%%%%%%%%%%%%%%%%%%%%%%%%%%%%%%%%%%%%%%%%%%%%%%%%
\section{Summary and discussion}
\label{sec:conclusion}
\begin{figure}[h]
\begin{center}
\includegraphics[width=\columnwidth,clip]{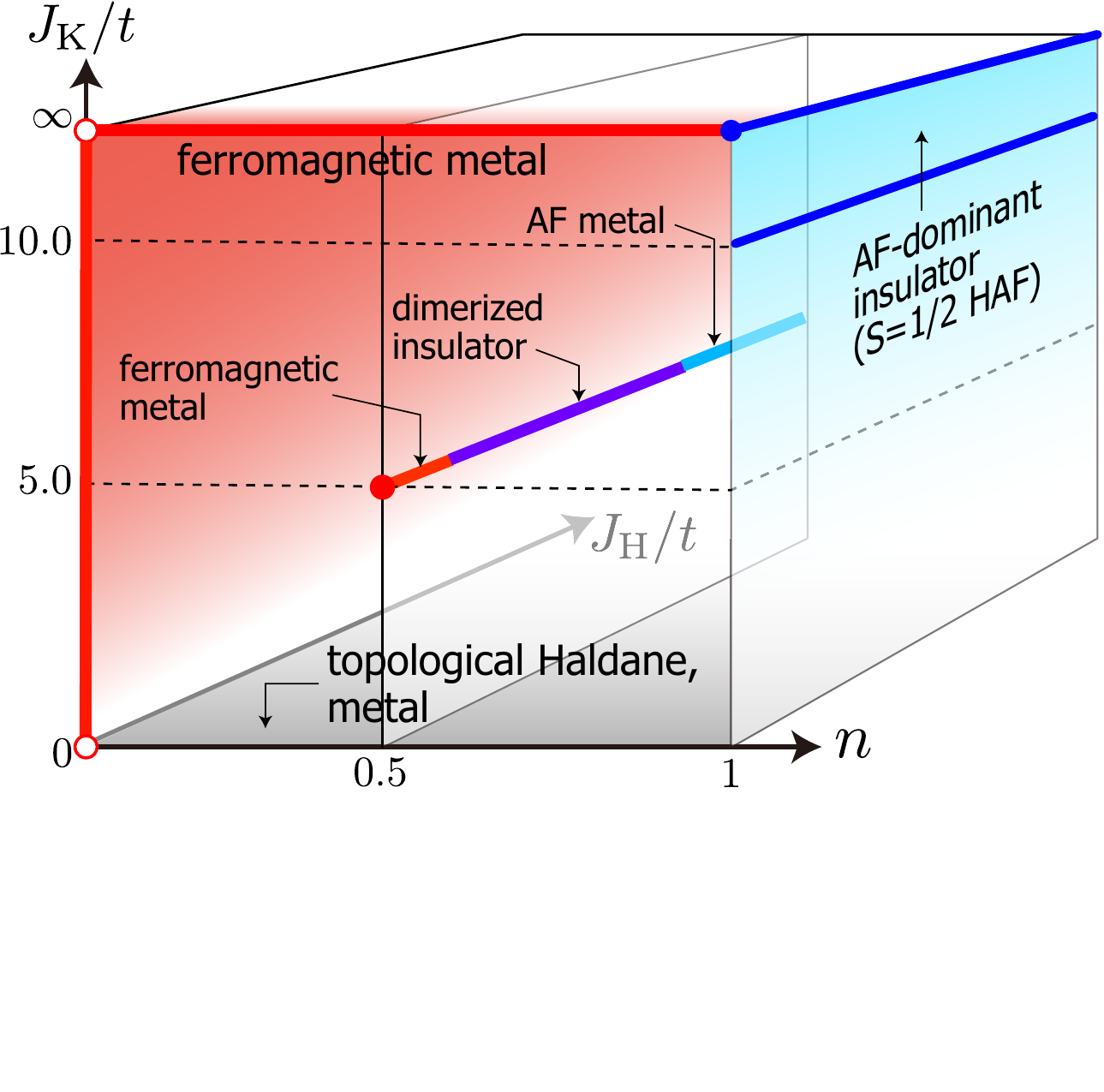}
\caption{
Schematic phase diagram of the spin-1 Kondo-Heisenberg model obtained from strong-coupling expansions and numerical simulations.  
The parameter regions studied in this paper are shown by the thick lines.  
On the line $J_{\text{K}}=\infty$, $0 < n <1$, the system is rigorously shown to be ferromagnetic metal.  
On the plane $n=1$ (half-filling), an insulating phase with power-law antiferromagnetic spin correlation is stable.  
At quarter-filling ($n=1/2$), we found at least three different phases at $J_{\text{K}}/t =5.0$: (i) ferromagnetic metal, (ii) dimerized insulator 
with power-law period-4 spin correlation, and (iii) metallic phase with antiferromagnetic correlation.  
\label{3D_phase_diagram}}
\end{center}
\end{figure}

%%%%%%%%%%%%%%%%%%%%%%%%%%%%%%%%%%%%%%%%%%%%%%%%%%%%%%%%%%%%%%
In this paper, we investigated the ground-state phases of the spin-$S$ Kondo-Heisenberg model in one dimension by means of 
analytical calculations in the limit of strong Kondo coupling ($J_{\text{K}}$) and the numerical DMRG simulations for $S=1$.  
The main results are summarized schematically in Fig.~\ref{3D_phase_diagram}.  
First, we derived the strong-coupling effective Hamiltonian both for and away from half-filling to obtain the insight into 
the global phase structure.   
At half-filling, the charge gap of the order of $J_{\text{K}}$ opens and the magnetic sector is described solely by the partially screened local 
moments [with spin-$(S{-}1/2)$], whose dynamics is governed by the antiferromagnetic Heisenberg model.  
The resulting physics of the magnetic sector depends on the parity of $2S$; when $2S$ is odd, 
the half-filled ground state is a spin-gapped insulator while when $2S$ is even the system is an insulator with power-law 
antiferromagnetic correlation (i.e., a spin Luttinger liquid).     As far as the direct interaction $J_{\text{H}}$ is much smaller than $J_{\text{K}}$, 
the two do not compete with each other and the only effect of $J_{\text{H}}$ is to renormalize the effective antiferromagnetic interaction 
among the partially screened local moments.     
Away from half-filling, on the other hand, we can prove that the system (at $J_{\text{H}}=0$) in strong coupling is generically in the ferromagnetic 
metallic phase (see Fig.~\ref{3D_phase_diagram}) in which the unscreened moments (spin-$S$) and the (partially) screened ones [spin-$(S{-}1/2)$] 
form a collective ferromagnetic state.   
Now this ferromagnetic state is challenged by the direct antiferromagnetic interaction $J_{\text{H}}$ among the local moments.   

To substantiate these expectations quantitatively for large but finite $J_{\text{K}}$, 
we carried out the DMRG simulations combined with the SSD method for the case of $S=1$.  
At half-filling ($n=1$), the spin-spin correlation indeed exhibits a power-law antiferromagnetic behavior, which agrees very well with that of 
the spin-1/2 ($S{-}1/2=1/2$ here) Heisenberg chain up to fairly large values of $J_{\text{H}} \, (\gtrsim J_{\text{K}})$.  
This implies that the picture of the insulating phase 
with correlated Kondo-doublets, which is established in the perturbative regime (i.e., $J_{\text{H}} \ll J_{\text{K}}$), 
in fact extends over a wide range of the parameter space (see ``AF-dominant insulator'' in Fig.~\ref{3D_phase_diagram}).   
Combining this with the results of the weak-coupling approach \cite{Tsvelik_kondo,Tsvelik-Y-19}, we expect that the AF-dominant insulator 
persists all the way down to small $J_{\text{K}}$.   

At quarter-filling ($n=1/2$) where ferromagnetism and antiferromagnetic $J_{\text{H}}$ compete with each other, 
the phase diagram is much richer (Figs.~\ref{phase_qfill} and \ref{3D_phase_diagram}).   
The ferromagnetic metal which is found for rather small $J_{\text{H}}$ is destabilized by increasing $J_{\text{H}}$ 
and yields to a dimerized insulating phase with period-4 power-law spin-spin correlation (labeled as ``dimerized insulator'' 
in Fig.~\ref{3D_phase_diagram}).  The critical value of $J_{\text{H}}$ is much smaller than we expect from the strong-coupling effective Hamiltonian. 
We also characterized the magnetic structure in the dimerized phase by a simple phenomenological argument. 
If $J_{\text{H}}$ is further increased, we encounted another quantum phase transition at $J_{\text{H}}/t \approx 0.45$ 
where the system becomes metallic again.   It remains open to understand how magnetic dimerization is stabilized by $J_{\text{H}}$ and  
opens a charge gap.   Perhaps direct simulations for the large-$J_{\text{K}}$ effective Hamiltonian \eqref{hopping1st_effectiveH}  
might give an important hint.  
Also, as already noted in section \ref{sec:electronic-properties}, 
the third phase with short-range antiferromagnetic correlation (``AF metal'') is determined only 
by the behavior of $\braket{\vec{T}_i \cdot \vec{T}_{i+1} }$ 
and the absence of the charge gap, and the full characterization of it is an important future problem.     

In this paper, we have focused on the strong-coupling phases of the spin-1 KH chain with small $J_\text{H}$.  
On the other hand, when the spin-1 moments are replaced with spin-1/2s, 
the model at weak oupling $J_\text{k} \ll J_\text{H}$ is known to possess co-existing CDW and superconducting orders \cite{Berg-F-K-10}, 
and, when inter-chain couplings are turned on, it even exhibits a topologically nontrivial ground state \cite{Tsvelik-16}.   
These facts hint at a possibility that, in the weak-coupling region, our spin-1 KH model might have a rich phase structure.   
Therefore, it is also an important future problem to study whether this is the case or not for $S > 1/2$.  
%%%%%%%%%%% FIG6 %%%%%%%%%%%%%%%%%%%%%%%%%%%%%%%%%%%%%%%%%%%%%%%%
%%%%%%%%%%%%%%%%%%%%%%%%%%%%%%%%%%%%%%%%%%%%%%%%%%%%%%%%%%%%%%

%%%%%%%%%%%%%%%%%%%%%%%%%%%%%%%%%%%%%%%%%%%%%%%%%%%%%%%%%%%%%%
\section*{Acknowledgements}
The authors would like to thank C.~Hotta for helpful discussions on the SSD. 
One of the authors (R.M.) is supported by JST, the establishment of university fellowships towards the creation of science technology innovation, 
Grant No. JPMJFS2123. 
K.T. is supported in part by JSPS KAKENHI Grant No. 18K03455 and No. 21K03401.  

%%%%%%%%%%%%%%%%%%%%%%%%%%%%%%%%%%%%%%%%%%%%%%%%%%%%%%%%%%%%%%
\appendix
%%%%%%%%%%%%%%%%%%%%%%%%%%%%%%%%%%%%%%%%%%%%%%%%%%%%%%%%%%%%%%
\section{Effective Hamiltonian of spin sector of spin-$S$ KH chain}
\label{Appendix_derivation_2nd_order_hopping}
In this section, we generalize our observation in Sec.~\ref{halffilling_case} that the strong-coupling effective Hamiltonian 
for the spin-1 Kondo lattice model at half-filling is given by the spin-$1/2$ Heisenberg model to the case of localized spin-$S$.  
%let us consider the second-order perturbation of the hopping term for arbitrary spin-$S$ half-filled KH model in the strong-coupling region.  
As already mentioned in Sec.~\ref{Perturbation_theory_from_strong-coupling_limit}, 
the low-energy effective Hamiltonian is the spin-$(S{-}1/2)$ antiferromagnetic Heisenberg model.

Basically, we follow the strategy in Sec.~\ref{halffilling_case}.  
The $4(2S+1)$ possible states at site-$i$ are shown in Table~\ref{tab:local-states} together with the Kondo energy and 
the degeneracy.  
In the strong coupling limit ($J_\text{K} \to \infty$) at half-filling (i.e, one electron per site $n_{i}=1$), the conduction electron and 
spin-$S$ localized spin at each site form $2S$-fold degenerate $S_\text{tot}=(S-1/2)$ states, in which the Kondo energy is given by: 
$J_\text{K} \vec{s}_{i} \cdot \vec{S}_{i} = - J_\text{K}(S+1)/2$.  
%is given by:
%\eq{
%J_\text{K} \vec{s}_{i} \cdot \vec{S}_{i} \Ket{S_\text{tot}=S- 1/2 }_{i} 
%= - J_\text{K}(S+1)/2 \Ket{S_\text{tot}=S-1/2 }_{i} \; . 
%}
The remaining $(2S+2)$ states with $S_\text{tot}=(S+1/2)$ have higher energy $J_\text{K}S/2$.   

On the other hand, the action of the creation operators is locally written as:
\begin{subequations}
\begin{align}
& c_{i\ua}^\dagger  = 
\sum_{\mathcal{M}=-S}^{S} \left\{ 
\Ket{
\begin{array}{c}
\scriptsize{ \ua } \\
\tiny{ \mathcal{M}  }
\end{array}
}_i
\Bra{
\begin{array}{c}
\text{emp} \\
\mathcal{M}
\end{array}
}_i
+ 
\Ket{
\begin{array}{c}
\ua \da \\
\mathcal{M}
\end{array}
}_i
\Bra{
\begin{array}{c}
\da \\
\mathcal{M}
\end{array}
}_i  
\right\}
\label{eqn:Hubbard-op-1}
\\
& c_{i\da}^\dagger = \sum_{\mathcal{M}=-S}^{S} 
\Ket{
\begin{array}{c}
\da \\
\mathcal{M}
\end{array}
}_i
\Bra{
\begin{array}{c}
\text{emp} \\
\mathcal{M}
\end{array}
}_i
-
\sum_{\mathcal{M}=-S}^{S}
\Ket{
\begin{array}{c}
\ua \da \\
\mathcal{M}
\end{array}
}_i
\Bra{
\begin{array}{c}
\ua \\
\mathcal{M}
\end{array}
}_i  \; ,
\label{eqn:Hubbard-op-2}
\end{align}
\end{subequations}
where we have introduced the notations 
${\tiny \Ket{
\begin{array}{c}
\alpha \\
\mathcal{M}
\end{array}
}_i}$ similar to those used in Sec.~\ref{sec:strong-coupling}, and 
the minus sign in the second equation comes from the definition of doubly-occupied state 
as $\ket{\ua \da}_i=c^\dagger_{i,\ua} c^\dagger_{i,\da} \ket{\text{emp}}_i$.     
Let us rewrite these operator in the basis where $S_{\text{tot}}$ is diagonal. 
To this end, we note that the relevant states are explicitly written as
%%%%%
%\begin{widetext}
\begin{subequations}
\begin{align}
\begin{split}
& \Ket{ S+\frac{1}{2}; \mathcal{M}+\frac{1}{2}}_{i}  \\
& =  \sqrt{\frac{S+\mathcal{M}+1}{2S+1} } 
\Ket{
\begin{array}{c}
\ua\\
\mathcal{M}
\end{array}
}_{i} 
+ \sqrt{\frac{S-\mathcal{M}}{2S+1} }
\Ket{
\begin{array}{c}
\da \\
\mathcal{M}+1
\end{array}
}_{i}   
 \end{split}
 %%%%%%%%%%%%%%%%%%%%%
 \\
\begin{split} 
& \Ket{S-\frac{1}{2}; \mathcal{M}+\frac{1}{2}}_{i}  \\
& = \sqrt{ \frac{S-\mathcal{M}}{2S+1} }
\Ket{
\begin{array}{c}
\ua\\
\mathcal{M}
\end{array}
}_{i} -\sqrt{ \frac{S +\mathcal{M}+1}{2S+1} } 
\Ket{
\begin{array}{c}
\da \\
\mathcal{M}+1
\end{array}
}_{i}  
\\
& (\mathcal{M}=-S,\cdots,S-1) \; .
\end{split}
\end{align}
\end{subequations}
%\end{widetext}
By inverting these equations, we can express ${\tiny \Ket{ \begin{array}{c} \ua/\da \\  \mathcal{M} \end{array}}_{i} }$ in terms of 
${\tiny \Ket{ S \pm \frac{1}{2}; \mathcal{M}+\frac{1}{2}}_{i} }$.  
Plugging those expressions into Eqs.~\eqref{eqn:Hubbard-op-1} and 
\eqref{eqn:Hubbard-op-2}, and dropping the states with $S_{\text{tot}} =S+1/2$, 
we obtain the expressions of $c^\dagger_{i,\ua/\da}$ and $c_{i,\ua/\da}$ projected onto the $S_{\text{tot}}=S{-}1/2$ states: 
\begin{subequations}
\begin{equation}
\begin{split}
& \mathcal{P}_{\text{d}} c^\dagger_{i,\ua} \mathcal{P}_{S-1/2} 
= - \sum_{\mathcal{M}=-S}^{S} \sqrt{\frac{S+\mathcal{M}}{2S+1}} \Ket{
\begin{array}{c}
\ua \da \\
\mathcal{M}
\end{array}
}
\Bra{S-\frac{1}{2};\mathcal{M}-\frac{1}{2}} 
\\
& \mathcal{P}_{\text{e}} c_{i,\ua} \mathcal{P}_{S-1/2}  
= \sum_{\mathcal{M}=-S}^{S} \sqrt{\frac{S-\mathcal{M}}{2S+1}} \Ket{
\begin{array}{c}
\text{emp} \\
\mathcal{M}
\end{array}
}
\Bra{S-\frac{1}{2};\mathcal{M}+\frac{1}{2}} 
\end{split}
\end{equation}
and 
\begin{equation}
\begin{split}
& \mathcal{P}_{\text{d}} c^\dagger_{i,\da} \mathcal{P}_{S-1/2} =  \sqrt{\frac{S-\mathcal{M}}{2S+1}} \Ket{
\begin{array}{c}
\ua \da \\
\mathcal{M}
\end{array}
}
\Bra{S-\frac{1}{2};\mathcal{M}+\frac{1}{2}}   \\
& \mathcal{P}_{\text{e}} c_{i,\da} \mathcal{P}_{S-1/2} = - \sqrt{\frac{S+\mathcal{M}}{2S+1}} 
\Ket{
\begin{array}{c}
\text{emp} \\
\mathcal{M}
\end{array}
}
\Bra{S-\frac{1}{2};\mathcal{M}-\frac{1}{2}} \; ,
\end{split}
\end{equation}
\end{subequations}
where $\mathcal{P}_{S-1/2}$ projects the state st site-$i$ onto the space of total spin $S-1/2$, and 
$\mathcal{P}_{\text{d}}$ and $\mathcal{P}_{\text{e}}$ respectively are the projectors onto the doubly-occupied and empty states.  
Therefore, the nearest-neighbor hopping on the two adjacent $\ket{S-1/2}$'s are:
%\begin{widetext}
\begin{subequations}
\begin{equation}
\begin{split}
& \mathcal{P}_{\text{d}}^{i} \mathcal{P}_{\text{e}}^{i+1} c^\dagger_{i,\ua} c_{i+1,\ua}\mathcal{P}_{S-1/2}^{i} \mathcal{P}_{S-1/2}^{i+1}  \\
& =\frac{1}{2S+1}  \displaystyle{\sum_{\mathcal{M}_{i}}} \displaystyle{\sum_{\mathcal{M}_{i+1}}}  
\sqrt{S+\mathcal{M}_i} \sqrt{S-\mathcal{M}_{i+1}} 
\\
&\times
\Ket{
\begin{array}{c}
\ua \da \\
\mathcal{M}_{i}
\end{array}
}
\Ket{
\begin{array}{c}
\text{emp} \\
\mathcal{M}_{i+1}
\end{array}
}
\Bra{S-\frac{1}{2};\mathcal{M}_{i}-\frac{1}{2}} 
\Bra{S-\frac{1}{2};\mathcal{M}_{i+1}+\frac{1}{2}} 
\end{split}
\end{equation}
%%%%%%%%%%%%%%%%%%%%%%%%%%%%%%%
\begin{equation}
\begin{split}
& \mathcal{P}_{\text{d}}^{i} \mathcal{P}_{\text{e}}^{i+1}  c^\dagger_{i,\da} c_{i+1,\da}\mathcal{P}_{S-1/2}^{i}\mathcal{P}_{S-1/2}^{i+1}  \\
& = -\frac{1}{2S+1} \displaystyle{\sum_{\mathcal{M}_{i}}} \displaystyle{\sum_{\mathcal{M}_{i+1}}}  \sqrt{S-\mathcal{M}_i} \sqrt{S+\mathcal{M}_{i+1}}
\\
&\times
\Ket{
\begin{array}{c}
\ua \da \\
\mathcal{M}_{i}
\end{array}
}
\Ket{
\begin{array}{c}
\text{emp} \\
\mathcal{M}_{i+1}
\end{array}
}
\Bra{S-\frac{1}{2};\mathcal{M}_{i}+\frac{1}{2}} 
\Bra{S-\frac{1}{2};\mathcal{M}_{i+1}-\frac{1}{2}} 
\end{split}
\end{equation}
\end{subequations}
The expressions for $c^\dagger_{i+1,\ua} c_{i,\ua}$ and $c^\dagger_{i+1,\da} c_{i,\da}$ are obtained from the above by 
interchanging $i \leftrightarrow i+1$.  
Similarly, we can write down 
$\mathcal{P}_{S-1/2}^{i} \mathcal{P}_{S-1/2}^{i+1} c^\dagger_{i,\ua} c_{i+1,\ua} \mathcal{P}_{\text{e}}^{i} \mathcal{P}_{\text{d}}^{i+1}$, etc. 
%\end{widetext}

\begin{figure}[h]
\begin{center}
\includegraphics[width=\columnwidth,clip]{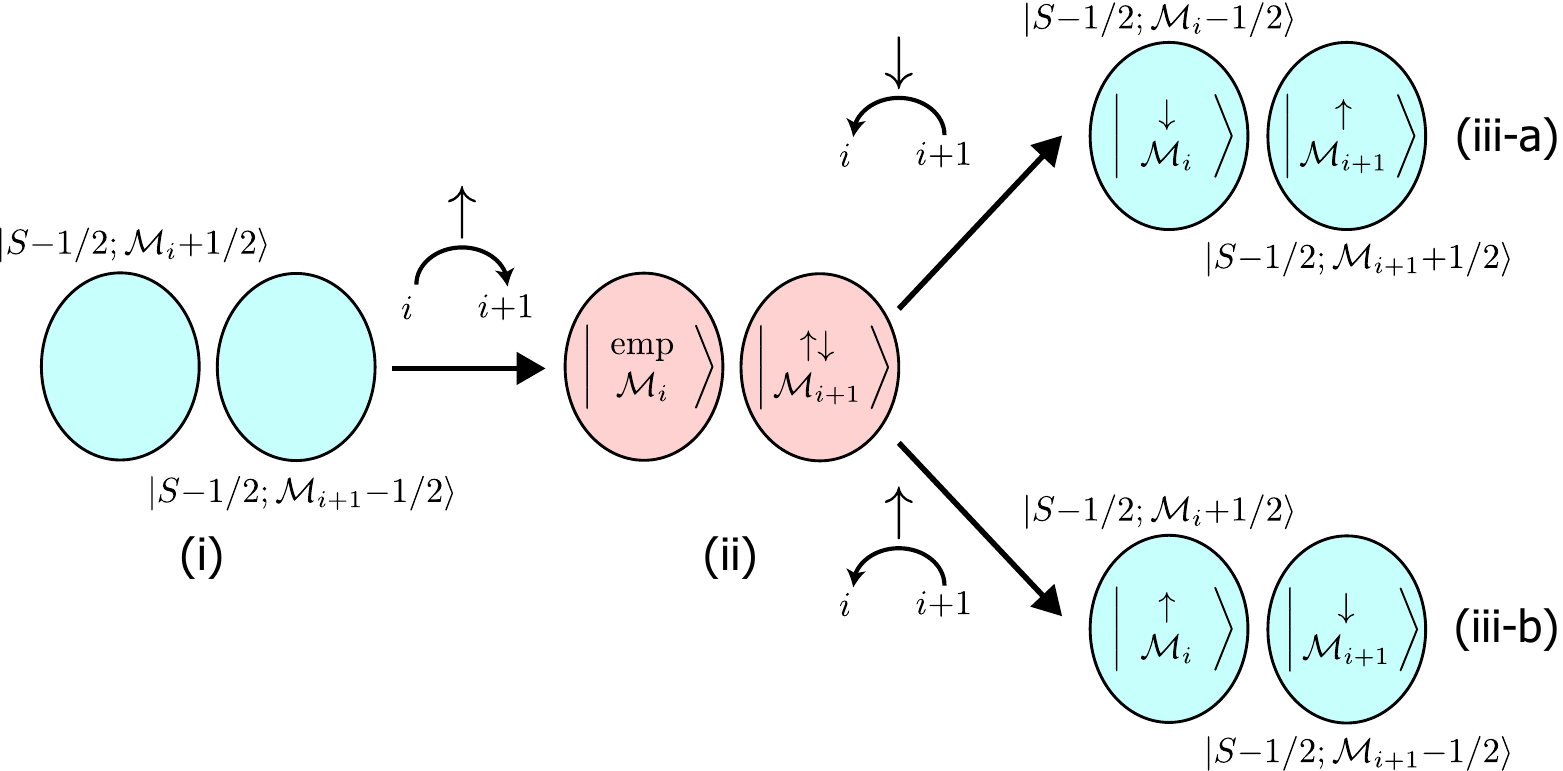}
\caption{
Typical second-order processes in $t$ for generic spin-$S$. 
(i) A pair of adjacent sites in the unperturbed ($t=0$) ground state. 
(ii) Electron ($\uparrow$ here) hopping from the site $i$ to $i+1$ generates a pair of sites in excited states.  
(iii) Second hopping back to the site $i$ returns the state to the ground-state subspace.  
Depending on the spin of the electron going back, different final states are obtained [(iii-a) and (iii-b)].  
\label{spinS_2nd_order_hopping}}
\end{center}
\end{figure}

%\begin{widetext} 
Combining all these, we can calculate the matrix elements of the second-order processes shown in Fig.~\ref{spinS_2nd_order_hopping}.  
For example, the matrix element of the process $\text{(i)}\to\text{(ii)} \to \text{(iii-a)}$ is:
\begin{equation}
\begin{split}
& \frac{(-t)^2}{-J_{\text{K}}(S+1) }  \\
& \times \mathcal{P}_{S-1/2}^{i} \mathcal{P}_{S-1/2}^{i+1} c^\dagger_{i,\da} c_{i+1,\da} 
\mathcal{P}^{i}_{\text{e}} \mathcal{P}^{i+1}_{\text{d}} c^\dagger_{i+1,\ua} c_{i,\ua} \mathcal{P}_{S-1/2}^{i} \mathcal{P}_{S-1/2}^{i+1}  \\
&= \frac{t^2}{J_{\text{K}}(S+1)(2S+1)^2 } \mathfrak{S}^{-}_{i}\mathfrak{S}^{+}_{i+1} \; , 
\end{split}
\end{equation}
where $\vec{\mathfrak{S}}$ are the spin-$(S{-}1/2)$ operators and we have used 
\[
\begin{split}
\mathfrak{S}^{-}_{i} 
=&  \sum_{\calM=-(S{-}1/2)}^{S{-}1/2} \sqrt{\left\{(S{-}1/2)+\calM_{i}\right\}\left\{ (S{-}1/2)-\calM_{i}+1 \right\}}   \\
& \times  |S{-}1/2; \calM_{i}- 1 \rangle_{2} \langle S{-}1/2; \calM_{i} |_{2} \; , 
\end{split}
\]
etc. in obtaining the final expression.  
Similarly, the process $\text{(i)}\to\text{(ii)} \to \text{(iii-b)}$ gives the diagonal term:
\begin{equation}
\begin{split}
& \frac{t^2}{J_\text{K}(S+1)(2S+1)^2} \left\{ 
\mathfrak{S}^{z}_{i} \mathfrak{S}^{z}_{i+1} 
- \frac{2S+1}{2} \left( \mathfrak{S}^{z}_{i} - \mathfrak{S}^{z}_{i+1} \right) \right\}  \\
& -\frac{t^2}{4J_\text{K}(S+1)} \; . 
\end{split}
\end{equation}
If we collect all the possible processes, the terms proportional to $\left( \mathfrak{S}^{z}_{i} - \mathfrak{S}^{z}_{i+1} \right)$ cancel out and 
we obtain the following effective Hamiltonian:
\begin{equation}
H_{\text{eff}} = \frac{4t^2}{(2S+1)^2 (S+1)J_{\text{K}}} \sum_{i} \vec{\mathfrak{S}}_{i} \cdot \vec{\mathfrak{S}}_{i+1} + \text{const} \; . 
\end{equation}
Putting $S=1$ in this equation, we recover Eq.~\eqref{eqn:eff-Ham-half-filling} in Sec.~\ref{halffilling_case}.  
This effective Hamiltonian indicates that the spin-$S$ Kondo lattice model with $J_{\text{K}}>0$ in the strong-coupling region 
is an insulator whose spin sector is described by the spin-$(S{-}1/2)$ Heisenberg model;   
according to the Haldane conjecture \cite{Haldane_conjecture,Haldane_conjecture-2}, the spin correlation is qualitatively different 
when $S$ is integer and when $S$ is half-odd-integer.  
If $S$ is integer, then the spin sector exhibits antiferromagnetic quasi-long-range order, 
while the ground state is disordered if $S$ is half-odd.   
This is consistent with the prediction \cite{Tsvelik_kondo} based on field-theory mapping.  

It is straightforward to take the Heisenberg term $J_{\text{H}}$ into account.  
To this end, we follow similar steps to find the projection of the localized spin onto the ground-state subspace:
\begin{equation}
\mathcal{P}_{S-1/2}^{i} \, \vec{S}_{i}\, \mathcal{P}_{S-1/2}^{i}  = \frac{2(S+1)}{2S+1} \vec{\mathfrak{S}}_{i}    \; ,
\label{eqn:localized-spin-projected-gen-S}
\end{equation}
which means that, in the strong-coupling limit, the localized spin $\vec{S}_{i}$ behaves like the effective spin-$(S{-}1/2)$ $\vec{\mathfrak{S}}_{i}$ 
except for the overall normalization factor.   
From this, one immediately sees that the Heisenberg term just gives the same Heisenberg model 
as before, leading to the total effective Hamiltonian:
\begin{equation}
\begin{split}
& H_{\text{eff}}^{(n=1)} \\
& = \left\{ \frac{4t^{2}}{(2S+1)^{2}(S+1) J_{\text{K}} } + \left( \frac{2(S+1)}{2S+1} \right)^{2} J_{\text{H}} \right\} 
\sum_{i} \vec{\mathfrak{S}}_{i}  {\cdot} \vec{\mathfrak{S}}_{i+1}   \; ,
\end{split}
\label{eqn:eff-Ham-half-filling} 
\end{equation} 
which generalizes Eq.~\eqref{effHam}.   
%%%%%%%%%%%%%%%%%%%% TABLE 2 %%%%%%%%%%%%%%%%%%%%%%%%%%%%%%%%
\begin{table}[htb]
\caption{\label{tab:local-states} Local states of spin-$S$ KH model. 
Quantum number of total spin $\vec{T}_{i} =\vec{s}_{i}+\vec{S}_{i}$ is denoted by $T$.}
\begin{ruledtabular}
\begin{tabular}{lccc}
conduction electron & $T$ & Kondo energy & degeneracy \\
\hline
$n_{i} =0$ ($|0\rangle$) & $ S $ & $0$ & $2S+1$  
\\
\hline
$n_{i} =1$  & $ S + 1/2$ & $J_{\text{K}} S/2$ & $2S+2$ \\
($c_{i,\ua}|0\rangle$, $c_{i,\da}|0\rangle$) &  $S - 1/2$  & $- J_{\text{K}} (S+1)/2$ & $2S$ \\  
\hline
$n_{i} =2$ ($c_{i,\ua}c_{i,\da} |0\rangle$) & $S$ & $0$ & $2S+1$ 
\\
\end{tabular}
\end{ruledtabular}
\end{table}
%%%%%%%%%%%%%%%%%%%%%%%%%%%%%%%%%%%%%%%%%%%%%%%%%%%%%%%%
%%%%%%%%%%%%%%%%%%%     FIG       %%%%%%%%%%%%%%%%%%%%%%%%%%%%%%%%%%%%%%%%%%
%%%%%%%%%%%%%%%%%%%%%%%%%%%%%%%%%%%%%%%%%%%%%%%%%%%%%%%%%%%%%%%%%%%
%%%%%%%%%%%%%%%%%%%%%%%%%%%%%%%%%%%%%%%%%%%%%%%%%%%
\section{Derivation of effective Hamiltonian \eqref{hopping1st_effectiveH}}
\label{Appendix_derivation_1st_order_hopping}
%%%%%%%%%%%%%%%%%%%%%%%%%%%%%%%%%%%%%%%%%%%%%%%%%%%
Among the $4 (2S+1)$ possible on-site states listed in Table~\ref{tab:local-states}, 
the $2S$ states $|n {=} 1, T{=}S{-}1/2; T^{z}\rangle$ with a single conduction electron, 
and the $(2S+1)$-fold degenerate states $| n {=}0, T{=}S; T^{z} \rangle$ 
without electron are relevant in the strong-coupling limit (less than half-filling $n <1$).  
As we can distinguish $\ket{n{=}0,T{=}S;T^{z}}$ from $\ket{n{=}1,T{=}S{-}1/2;T^{z}}$ by the value of $T$ ($S$ or $S{-}1/2$) 
in the strong-coupling Hilbert space, 
we can omit $n$ in specifying the states, and we will abbreviate, e.g., $\ket{n=0,T=S; T^z =\mathcal{M}}_{i}$ as $\ket{S;M}_{i}$ 
from now on.   
The low-energy Hilbert space is spanned by the tensor products of $\ket{S; \mathcal{M} }_{i}$ and $\ket{S{-}1/2; \mathcal{M}}_{i}$.  

So the action of electron-creation operators on $\ket{S;\mathcal{M}}_{i}$ is
\begin{subequations}
\begin{align}
& c_{i,\ua}^{\dagger} =   \sum_{\mathcal{M}=-S }^{S} 
\sqrt{\frac{S-\mathcal{M}}{2S+1}} \Ket{S{-}1/2 ;\mathcal{M}+1/2}_{i}\bra{S;\mathcal{M}}_{i} \\
& c_{i,\da}^{\dagger} = - \sum_{\mathcal{M}=-S }^{S} 
\sqrt{\frac{S+\mathcal{M}}{2S+1}} \Ket{S{-}1/2 ;\mathcal{M}- 1/2}_{i} \bra{S;\mathcal{M}}_{i} \;  .
\end{align}
\end{subequations}
\begin{widetext}
Therefore, the action of hopping operators of the conduction electron on $\ket{S;\mathcal{M}_i}\otimes \ket{S-1/2;\mathcal{M}_{i\pm 1} }_{i\pm 1}$ is 
\begin{subequations}
\begin{equation}
\begin{split}
-t c_{i,\ua}^{\dagger} c_{i\pm 1,\ua} =& - \frac{t}{2S+1} \displaystyle{\sum_{\mathcal{M}_{i}= - S }^{S} 
\sum_{\mathcal{M}_{i\pm 1} = -(S{-}1/2) }^{S{-}1/2} } \sqrt{S-\mathcal{M}_{i}} \sqrt{S-\mathcal{M}_{i\pm 1}+1/2 }   \\
&\times \Ket{S{-}1/2 ;\mathcal{M}_{i}+1/2}_{i} \otimes \Ket{S;\mathcal{M}_{i\pm 1}-1/2}_{i\pm 1} 
\left( \bra{S;\mathcal{M}_{i}}_{i} \otimes \Bra{S{-}1/2 ;\mathcal{M}_{i\pm 1} }_{i\pm 1} \right) 
\label{hopping_general_S_1}
\end{split}
\end{equation}
\begin{equation}
\begin{split}
-t c_{i,\da}^{\dagger} c_{i\pm 1,\da} =& - \frac{t}{2S+1} \displaystyle{\sum_{\mathcal{M}_{i} = -S }^{S} 
\sum_{\mathcal{M}_{i\pm 1} = -(S{-}1/2) }^{S{-}1/2} } \sqrt{S+\mathcal{M}_{i}} \sqrt{S+\mathcal{M}_{i\pm 1} +1/2 }   \\
&\times \Ket{S{-}1/2 ;\mathcal{M}_{i}-1/2}_{i} \otimes \Ket{S;\mathcal{M}_{i\pm 1} +1/2}_{i\pm 1} 
\left( \bra{S;\mathcal{M}_{i}}_{i} \otimes \Bra{S{-}1/2 ;\mathcal{M}_{i\pm 1} }_{i\pm 1} \right)  
\label{hopping_general_S_2}
\end{split}
\end{equation}
\end{subequations}
%%%%%%%%%%%%%%%%%%%%%%%%%%%%%%%%%%%%%%%%%%%%%%%%%%%%%%%%%%%%
%\subsection{Case of general $S$}
%%%%%%%%%%%%%%%%%%%%%%%%%%%%%%%%%%%%%%%%%%%%%%%%%%%%%%%%%%%%
If we introduce the exchange operator $X_{i,j}$ as
\eq{X_{i,j} \ket{\psi}_{i} \otimes \ket{\varphi}_j = \ket{\varphi}_{i} \otimes  \ket{\psi}_j \; ,}
then \eqref{hopping_general_S_1} and \eqref{hopping_general_S_2} can be rewritten in the following form:
%\begin{widetext}
\begin{subequations}
\begin{equation}
\begin{split}
-t c_{i,\ua}^{\dagger} c_{i\pm 1,\ua} = &-\frac{t}{2S+1} X_{i,i\pm 1}  
\Biggl\{ \displaystyle{\sum_{\mathcal{M}_{i}=-S }^{S} \sum_{\mathcal{M}_{i\pm 1}= - (S{-}1/2) }^{S{-}1/2} } 
\sqrt{S-\mathcal{M}_{i}} \sqrt{S-\mathcal{M}_{i\pm 1}+1/2}   \\
&\times \Ket{S;\mathcal{M}_{i\pm 1}-1/2}_{i} \otimes \Ket{S{-}1/2;\mathcal{M}_{i}+1/2}_{i\pm 1} 
\left( \bra{S;\mathcal{M}_{i} }_{i} \otimes \Bra{S{-}1/2;\mathcal{M}_{i\pm 1} }_{i\pm 1} \right)  \Biggr\} \label{hopping_general_S_1_2}
\end{split}
\end{equation}
%%%%%%%%%
\begin{equation}
\begin{split}
-t c_{i,\da}^{\dagger} c_{i\pm 1,\da} = &  - \frac{t}{2S+1} X_{i,i\pm 1} 
\Biggl\{ \displaystyle{\sum_{\mathcal{M}_{i} = -S }^{S} \sum_{\mathcal{M}_{i\pm 1} =-(S{-}1/2) }^{S{-}1/2} } 
\sqrt{S+\mathcal{M}_{i}} \sqrt{S+\mathcal{M}_{i\pm 1} + 1/2}  \\
& \times \Ket{S;\mathcal{M}_{i\pm 1}+1/2}_{i} \otimes \Ket{S{-}1/2;\mathcal{M}_{i}-1/2}_{i\pm 1} 
\left( \bra{S;\mathcal{M}_{i}}_{i} \otimes \Bra{S{-}1/2 ;\mathcal{M}_{i\pm 1} }_{i\pm 1} \right) \Biggr\}   \;  .
\end{split}
\label{hopping_general_S_2_2}
\end{equation}
\end{subequations}
If one changes the basis from $\Ket{S;\mathcal{M}_{i} }\otimes \Ket{S {-} 1/2;\mathcal{M}_{i\pm 1}+1/2}$ to the one in which the total spin 
$J\, (=1/2,\ldots, 2S{-}1/2)$ is diagonal:
\begin{equation*}
\Ket{S;\mathcal{M}_i} \otimes  \Ket{S{-}1/2 ;\mathcal{M}_{i\pm 1} } 
=  \sum_{J=1/2}^{2S{-}1/2}
\left(   \langle J; \mathcal{M}_{i}+\mathcal{M}_{i+1} \Ket{S;\mathcal{M}_i} \otimes \Ket{S{-}1/2 ;\mathcal{M}_{i\pm 1} }  \right)   
\Ket{J;\mathcal{M}=\mathcal{M}_{i}+\mathcal{M}_{i+1}}
 \; ,  
\end{equation*}
the above can be further recast as:
\begin{subequations}
\begin{align}
& -t  \left( c_{i,\ua}^{\dagger} c_{i + 1,\ua} + c_{i,\da}^{\dagger} c_{i + 1,\da} \right) 
= - t  X_{i,i+1}\mathcal{P}_{S}^{i} \mathcal{P}_{S-\frac{1}{2}}^{i+ 1} 
 \left[ \frac{1}{2S+1} \sum_{k=0}^{2S-1} (-1)^k (2S-k) P_{i+1\to i} (2S-k- 1/2) \right] \mathcal{P}_{S}^{i} \mathcal{P}_{S-\frac{1}{2}}^{i+ 1} \; , 
 \label{effective_generalS_1}
\\
%& -t \left( c_{i\pm 1,\ua}^{\dagger} c_{i,\ua} + c_{i\pm 1,\da}^{\dagger} c_{i,\da} \right) 
%= -\frac{t}{2S+1} X_{i,i\pm 1} \mathcal{P}_{S-\frac{1}{2}}^{i} \mathcal{P}_{S}^{i \pm1}  
%\left[ \sum_{k=0}^{2S-1} (-1)^k (2S-k) P_{i \to i\pm1} (2S-k- 1/2) \right]  \mathcal{P}_{S-\frac{1}{2}}^{i} \mathcal{P}_{S}^{i\pm1} \; . 
& -t \left( c_{i+ 1,\ua}^{\dagger} c_{i,\ua} + c_{i+ 1,\da}^{\dagger} c_{i,\da} \right) 
= - t X_{i,i+1} \mathcal{P}_{S-\frac{1}{2}}^{i} \mathcal{P}_{S}^{i+1}  
\left[ \frac{1}{2S+1} \sum_{k=0}^{2S-1} (-1)^k (2S-k) P_{i \to i+1} (2S-k- 1/2) \right]  \mathcal{P}_{S-\frac{1}{2}}^{i} \mathcal{P}_{S}^{i+1} \; . 
\label{effective_generalS_2}
\end{align}
\end{subequations}
\end{widetext}
This is the generalization of the so-called double-exchange Hamiltonian \cite{Kubo-82,Muller-Hartmann-D-96} to the case of 
antiferromagnetic $J_{\text{K}}$.  
Here we have defined another operator $P_{i\to i+1}(J)$ [$P_{i+1\to i}(J)$] 
that projects the states of a pair of spins $S{-}1/2$ at site-$i$ [site-$(i+1)$] and $S$ at site-$(i+1)$ (site-$i$) 
onto the subspace with the total spin $J$:
\eq{
P(J) = \sum_{\mathcal{M}=-J}^J \ket{J;\mathcal{M}} \bra{J;\mathcal{M}} \; .
}
In both expressions \eqref{effective_generalS_1} and \eqref{effective_generalS_2} of the electron hopping, 
the projection operator onto $J=2S-1/2=S+(S-1/2)$ (i.e., the maximal value of $\vec{\mathfrak{S}}_{i}{+}\vec{S}_{i+1}$ 
or $\vec{S}_{i}{+}\vec{\mathfrak{S}}_{i+1}$) 
has the largest coefficient suggesting that the ferromagnetic state optimizes the kinetic energy of the conduction electrons 
as in the {\em ferromagnetic} Kondo lattice model.  

It is not difficult to write down the quantity $1/(2S{+}1)\sum_{k=0}^{2S-1} (-1)^k (2S{-}k) P_{i \to i+1} (2S{-}k {-} 1/2)$ 
as a polynomial $f^{(S)} ( \vec{\mathfrak{S}}_i{\cdot} \vec{S}_{i+1} )$ of $\vec{\mathfrak{S}}_{i}{\cdot} \vec{S}_{i+1}$.   
The explicit forms of $f^{(S)}(X )$ are given by:
\begin{equation}
\begin{split}
&  f^{(1/2)}(X) =    1/2     \\
&  f^{(1)} (X) =  \frac{2}{3} X + \frac{1}{3}  \\
&  f^{(3/2)}(X)=  \frac{3}{8} X + \frac{1}{4} X^2 - \frac{3}{8}  \\
& f^{(2)}(X)=  - \frac{1}{3} X +\frac{7}{45} X^{2} + \frac{2}{45} X^{3 }- \frac{4}{5}   \\
& f^{(5/2)}(X)=  \frac{1}{6}\left\{ -\frac{305}{72} X - \frac{79}{144} X^{2} +\frac{7}{36} X^{3} + \frac{1}{36} X^{4} - \frac{85}{48}  \right\} \;  .
\end{split}
\label{eqn:polynomial-spin-S}
\end{equation}
The polynomial for $1/(2S{+}1)\sum_{k=0}^{2S-1} (-1)^k (2S {-} k) P_{i+1 \to i} (2S - k - 1/2)$ is given simply by 
$f^{(S)} ( \vec{S}_i {\cdot} \vec{\mathfrak{S}}_{i+1} )$.  
Setting $S=1$ in Eqs.~\eqref{effective_generalS_1} and \eqref{effective_generalS_2} and 
expressing $X_{i,i+1}$ and $\mathcal{P}$ with $\hat{d}_{i}$ and $\hat{n}^{d}_{i}$ 
reproduce the results \eqref{hopping1st_effectiveH} and \eqref{eqn:eff-int-S-1} in Sec.~\ref{sec:less-than-HF}.
%%%%%%%%%%%%%%%%%%%%%%%%%%%%%%%%%%%%%%%%%%%%%%%%%%%%%%%%%%%%%%

%%%%%%%%%%%%%%%%%%%%%%%%%%%%%%%%%%%%%%%%%%%%%%%%%%%%%%%%%%%%%%
\section{Rigorous proof of the ferromagnetic ground state for effective Hamiltonian (\ref{hopping1st_effectiveH})}
\label{sec:ferro-proof}
%%%%%%%%%%%%%%%%%%%%%%%%%%%%%%%%%%%%%%%%%%%%%%%%%%%%%%%%%%%%%%
In this section, starting from the strong-coupling effective Hamiltonian (\ref{hopping1st_effectiveH}),  
we derive the ferromagnetic ground state of spin-$1$ Kondo lattice model (i.e., $J_{\text{H}} =0$) with filling $0 \lneqq n \lneqq 1$. 
As we can follow almost the same steps to generalize the statement to the arbitrary spin-$S$, we describe the proof 
only for $S=1$ for simplicity.  

For this purpose, let $H^{(l)}$ be the Hamiltonian \eqref{hopping1st_effectiveH} of $l$-site system: 
\begin{equation*}
\begin{split}
H^{(l)}= &
-t\sum_{i=1}^{l-1} \biggl\{  \hat{d}^{\dagger}_{i+1} \hat{d}_{i} \,  f^{(S=1)}_{i\to i+1} ( \vec{D}_i{\cdot} \vec{S}_{i+1} ) \, 
\hat{n}_{\text{d},i} (1-\hat{n}_{\text{d},i+1})  \\
& +  \hat{d}^{\dagger}_{i} \hat{d}_{i+1} \,  f^{(S=1)}_{i+1 \to i} ( \vec{S}_i{\cdot} \vec{D}_{i+1} )  \, 
(1-\hat{n}_{\text{d},i}) \hat{n}_{\text{d},i+1}   \biggr\}  \; , 
\end{split}
\end{equation*}
which is block-diagonal in the number of doublets $N_{\text{d}}(l) =\sum_{i=1}^{l} n_{\text{d},i}$ (which is equal to the number of conduction electrons) 
and the total $S_{\text{tot}}^{z}(l)=\sum_{i=1}^{l} T^{z}_{i}$: 
\eq{
H^{(l)} = \bigoplus_{N_{\text{d}}(l), S_{\text{tot}}^z(l)} H^{(l)}_{N_{\text{d}}(l), S_{\text{tot}}^z(l)}  \; .
}

The first step is to prove that, (A) for $1 \leq N_{\text{d}}(L) \leq L-1$ ($L$: the system size), the matrix representation 
of $H^{(L)}_{N_{\text{d}}(L), S_{\text{tot}}^z(L)}$ is {\em non-positive} and {\em indecomposable} in the standard basis \footnote{%
We do not need to specify the local doublet (i.e., electron) number $n^{d}_{i}$ since $T^{z}_{i}=\pm 1/2$ ($0,\pm 1$) already imply 
$n^{d}_{i}=1$ ($0$).}:
\begin{equation}
\left\{ |T^{z}_{1},\ldots, T^{z}_{L} \rangle = \otimes_{i} \ket{T^z_i } \right\} \quad 
(T^z_i = \pm1/2, 0,\pm 1)  \; .
\label{eqn:def-PF-basis}
\end{equation}
Note that when $T^z_i=\pm 1/2$, an electron exists at site-$i$ forming a Kondo doublet 
($\ket{\Ua}_i$ or $\ket{\Da}_i$), while there is no electron if $T^z_i=0 , \pm 1$.  
Then, we can use the Perron-Frobenius theorem (see, e.g., Ref.~\cite{Tasaki-book-20} for a physicist-friendly exposition of the theorem and 
its applications) to show that the ground state within each sector 
is unique and that the ground-state ``wave function'' in this basis is strictly positive.  
The second step is to show that (B) the above unique ground state has a non-zero overlap with the state of  
maximal total spin: $S_{\text{tot}} = L - N_{\text{d}}/2$, which means that the unique ground state is indeed ferromagnetic.  

The proposition (A) is proven by the mathematical induction in the system size $L$.  
Let us start from the simplest case $L=2$.  In this case, it suffices to consider only $N_{\text{d}}(2)=1$ since $H^{(L=2)}_{N_{\text{d}}, S^z}$ 
is trivially zero for $N_{\text{d}}(2)=0$ and $2(=L)$.  
When $N_{\text{d}}(2)=1$, there are twelve states [six spin states for each of the two possible configurations of $S=1/2$ (electron) and $S=1$ (hole)]:
\[
\begin{split}
& S^{z}_{\text{tot}}(2) =3/2: \; |T^{z}_{1},T_{2}^{z}\rangle
= \ket{1,1/2}  , \, \ket{1/2,1}  \\
& S^{z}_{\text{tot}}(2) =1/2:  \\
& \quad  |T^{z}_{1},T_{2}^{z}\rangle = | 1/2, 0 \rangle  , \,  |-1/2,1\rangle  , \,  |0,1/2\rangle  , \,  |1,-1/2 \rangle  \\ 
& S^{z}_{\text{tot}}(2) = - 1/2:   \\
& \quad |T^{z}_{1},T_{2}^{z}\rangle = | -1/2, 0 \rangle   , \,  |1/2,-1\rangle   , \,  |0,-1/2\rangle  , \,  |-1,1/2 \rangle  \\
& S^{z}_{\text{tot}}(2) = - 3/2: \;  |T^{z}_{1},T_{2}^{z}\rangle
= | -1, -1/2\rangle \, , \; |-1/2, -1\rangle   \; .
\end{split}
\]
The matrix representation of the effective Hamiltonian in the above basis 
can be obtained readily from Eqs.~\eqref{hopping_general_S_1} and \eqref{hopping_general_S_2}.  
For instance, the block Hamiltonian for $S^z_{\text{tot}}(2) =3/2$ and $S^z_{\text{tot}}(2) =1/2$ are respectively given by:
\begin{equation}
H^{(2)}_{1,3/2}=-\frac{t}{3}\left(
\begin{array}{cc}
0 & 2 \\
2 & 0
\end{array}
\right). 
\label{H13/2}
\end{equation}
and
%Here, the basis of the right hand side is $\ket{1,1/2}_{1}\otimes \ket{0,1}_{2}$ and $\ket{0,1}_{1}\otimes \ket{1,1/2}_{2}$. 
%In the case $(n^d, S^z)=(1, 1/2)$,
\begin{equation}
H^{(2)}_{1,1/2}=
-\frac{t}{3}\left(
\begin{array}{cccc}
0 & 0 & 1 & \sqrt{2} \\
0 & 0 & \sqrt{2} & 0 \\
1 & \sqrt{2} & 0 & 0 \\
\sqrt{2} & 0  & 0 & 0\\
\end{array}
\right)
\label{H11/2}
\end{equation}
(the others are: $H^{(2)}_{1,-1/2}=H^{(2)}_{1,1/2}$ and $H^{(2)}_{1,-3/2}=H^{(2)}_{1,3/2}$).   
Clearly, all the off-diagonal elements of these matrices are non-positive.  

\begin{figure}[ht]
\begin{center}
\includegraphics[scale=0.6]{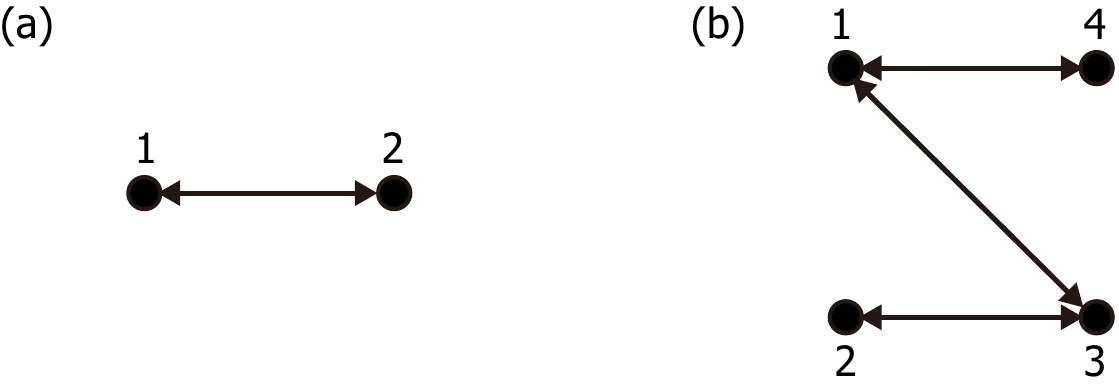}
\caption{Oriented graphs representing the block Hamiltonians(a) $H^{(2)}_{1,3/2}$ [Eq.~\eqref{H13/2}] and 
(b) $H^{(2)}_{1,1/2}$ [Eq.~\eqref{H11/2}].   
The numbers on the vertices denote the matrix indices, and the arrows running from the vertex $j$ to $i$ mean that the $(i,j)$-elements are non-zero. 
As the Hamiltonian is represented by Hermitian matrices, all the arrows are bi-directional.  
\label{connectivity_L2}}
\end{center}
\end{figure}

The connectivity of these matrices can be represented by the connected graph shown in Fig.~\ref{connectivity_L2}, 
in which the vertices represent the basis states and the edges correspond to non-zero matrix elements among them.  
It is easy to see that for any pair of vertices (i.e., basis states) we can go from one to the other by following the edges (i.e., non-zero 
matrix elements); a matrix is said to be indecomposable if the corresponding graph is connected (as in Fig.~\ref{connectivity_L2}).  
Thus, we establish that the block hamiltonians are non-positive and indecomposable for $L =2$.  
%%%%%%% FIG %%%%%%%%%%%%%%%%%%%%%%%%%%%%%%%%%%%%%%%%%%%%%%%%%%%%%%
%%%%%%%%%%%%%%%%%%%%%%%%%%%%%%%%%%%%%%%%%%%%%%%%%%%%%%%%%%%%%%%%

\begin{figure}[tbh]
\begin{center}
\includegraphics[width=\columnwidth,clip]{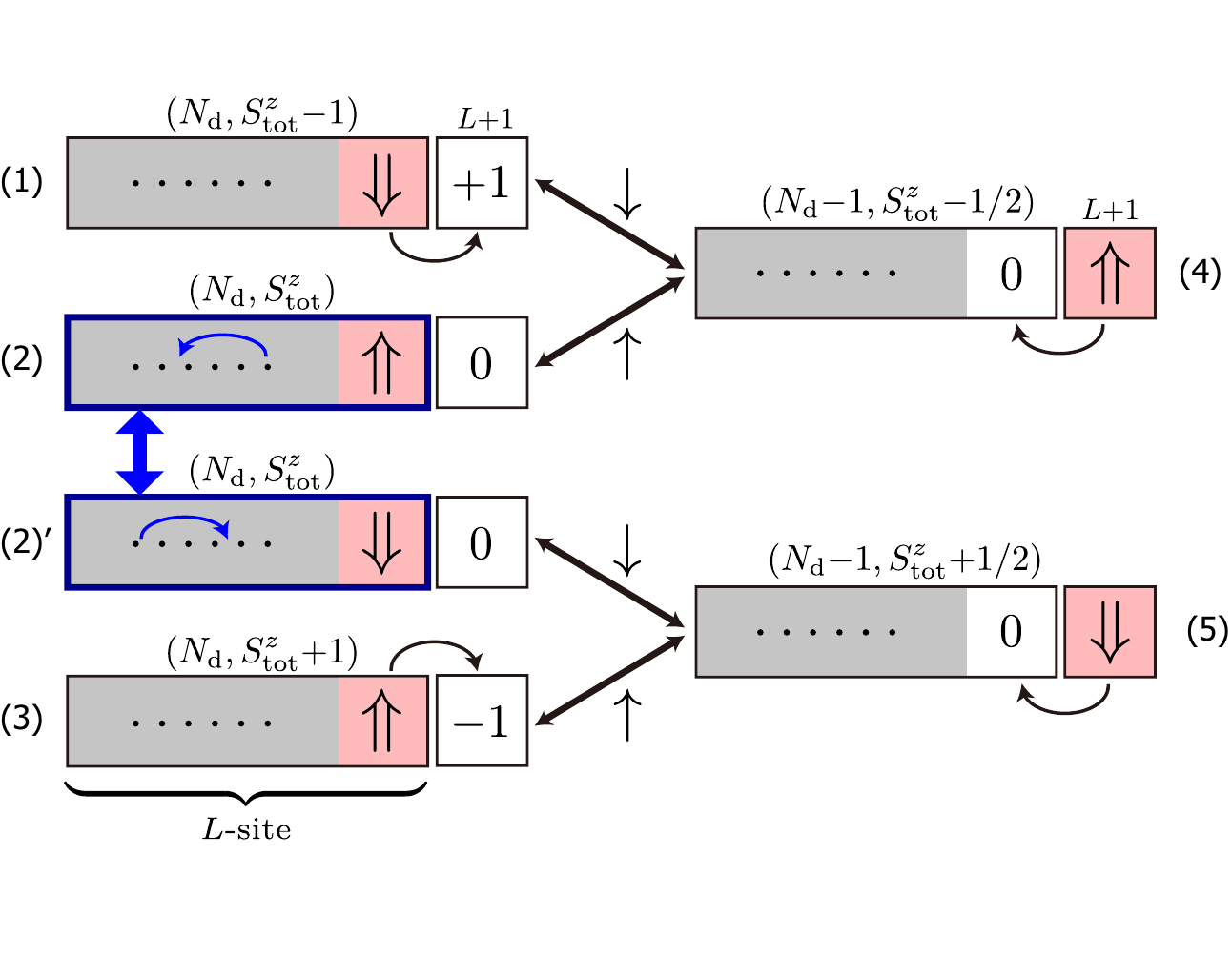}
\caption{Projected hopping \eqref{hopping_general_S_1} and \eqref{hopping_general_S_2} 
between the sites $L$ and $L+1$ connects the five different sectors (1)--(5) with each other [$(1)$--$(5)$ correspond respectively to 
the blocks $M_{1}$--$M_{5}$ in Eq.~\eqref{eqn:MI-block-structure}].  
Note that the hopping Hamiltonian of the size-$L$ sub-system {\em within} each sector is indecomposable by the assumption 
of the induction.  
\label{block-connected}}
\end{center}
\end{figure}

\begin{figure}[tbh]
\begin{center}
\includegraphics[width=\columnwidth,clip]{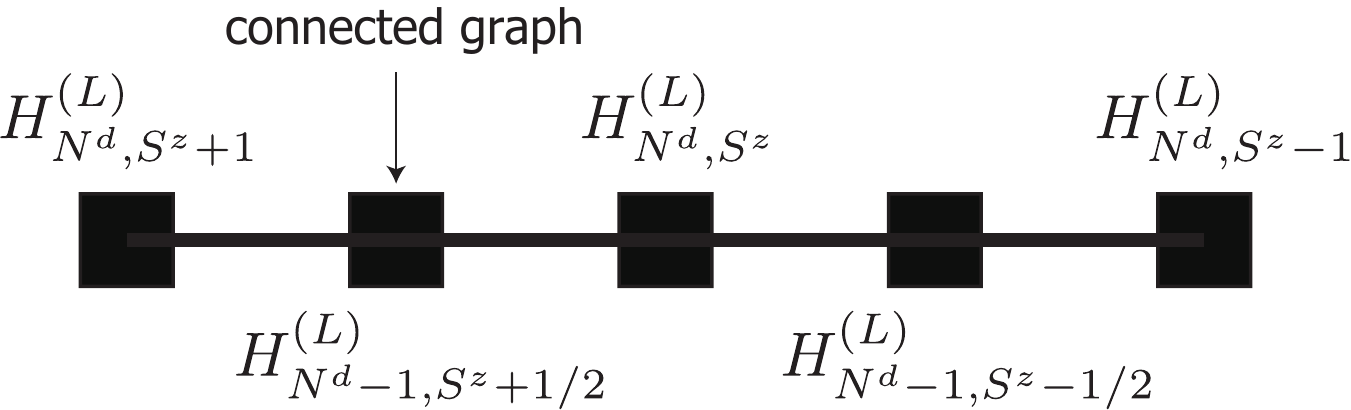}
\caption{Diagrammatic representation of the connectivity structure of the matrix $H^{(L+1)}_{N_{\text{d}}, S^z}$ in Eq.~\eqref{eqn:MI-block-structure}. 
The solid squares and the lines connecting them denote certain connected graphs (representing the diagonal blocks 
in $H^{(L+1)}_{N_{\text{d}}, S^z}$) and the non-zero (non-positive) matrices ``$\ast$''  
in \eqref{eqn:MI-block-structure}, respectively.
\label{connectivity_diagram2}}
\end{center}
\end{figure}

Next, suppose that the statement (A) holds for {\em all} the system sizes $l$ up to $L$, that is, 
all the block Hamiltonians $\{ H^{(l)}_{N_{\text{d}}(l), S_{\text{tot}}^z(l)} \}$ [$2 \leq l \leq L$, $2 \leq N_{\text{d}}(l) \leq l-1$]  
are non-positive and indecomposable in the basis \eqref{eqn:def-PF-basis} to prove the statement for the system size $L+1$.   
Consider the block $H^{(L+1)}_{N_{\text{d}}, S_{\text{tot}}^z}$ of the system with size $L+1$.  
According to the five possible values of $T^{z}_{L+1}\, (=1,0,-1,1/2,-1/2)$, we can decompose the subspace 
with $(N_{\text{d}}, S^z)$ into five different sectors, 
and the hopping between the sites $L$ and $L+1$ connects these five sectors with each other.  
From the explicit expressions \eqref{hopping_general_S_1} and \eqref{hopping_general_S_2} of the hopping term 
(see also Fig.~\ref{block-connected}), 
we see that the Hamiltonian $H^{(L+1)}_{N_{\text{d}}, S_{\text{tot}}^z}$ takes the following block structure: 
\begin{equation}
H^{(L+1)}_{N_{\text{d}}, S_{\text{tot}}^z} = 
\left( 
\begin{array}{ccc|cc}
M_{1}  & \mathbf{0} & \mathbf{0} & \ast & \mathbf{0}   \\
\mathbf{0}   & M_{2}  & \mathbf{0} & \ast & \ast             \\     
\mathbf{0}  &  \mathbf{0}  & M_{3}  & \mathbf{0} & \ast  \\
\hline
\ast & \ast & \mathbf{0} &  M_{4} & \mathbf{0} \\
\mathbf{0} & \ast & \ast &  \mathbf{0} &  M_{5}
\end{array}
\right)
\; , 
\label{eqn:MI-block-structure}
\end{equation}
where the diagonal blocks $M_{1}$, $M_{2}$, $M_{3}$, $M_{4}$, and $M_{5}$ respectively are 
$H^{(L)}_{N_{\text{d}}, S_{\text{tot}}^z-1}$, $H^{(L)}_{N_{\text{d}}, S_{\text{tot}}^z}$, $H^{(L)}_{N_{\text{d}}, S_{\text{tot}}^z+1}$, 
$H^{(L)}_{N_{\text{d}}-1, S_{\text{tot}}^z - \frac{1}{2}}$, 
and $H^{(L)}_{N_{\text{d}}-1, S_{\text{tot}}^z + \frac{1}{2}}$, and $\ast$ denotes non-positive matrices determined 
by \eqref{hopping_general_S_1} and \eqref{hopping_general_S_2}.  
Since $M_{i}$ ($i=1,\ldots, 5$) are all non-positive and indecomposable by the assumption, it is obvious from the corresponding graph 
Fig.~\ref{connectivity_diagram2} that the entire matrix $H^{(L+1)}_{N_{\text{d}}, S_{\text{tot}}^z}$ itself is indecomposable, too.    

A remark is in order about the exceptional cases with $N_{\text{d}}=1$ (one electron in the system) and $L$ 
(one hole in the system).  In these cases, either $(M_{1},M_{2},M_{3})$ (when $N_{\text{d}}=L$) or 
$(M_{4},M_{5})$ (when $N_{\text{d}}=1$) are identically zero and we cannot use the indecomposability of these matrices 
to prove that of $H^{(L+1)}_{N_{\text{d}}, S^z}$.   In fact, we can treat these cases without relying on the mathematical induction.  
First, we note that, in the case of a single electron or hole, we can move it to an arbitrary position 
by the repeated action of the hopping operators (the spin configuration is modified, too). 
Then, we use processes in which the electron/hole moves to a certain site and comes back to the starting point 
to create the spin-flips of the form $T_{i}^{+}T_{j}^{-}$, which connect between any 
two different spin states in the same $(N_{\text{d}},S^{z})$ sector.    
This completes the proof of the statement (A).  Then, by the Perron-Frobenius theorem, there exists a unique 
lowest-energy state $|\psi_{0}; N_{\text{d}}, S_{\text{tot}}^{z}\rangle$ in each of the $(N_{\text{d}},S^{z})$-sectors:
\begin{equation}
H^{(L)}_{N_{\text{d}}, S_{\text{tot}}^z} |\psi_{0}; N_{\text{d}}, S_{\text{tot}}^{z}\rangle 
= E_{\text{g.s.}} (N_{\text{d}}, S_{\text{tot}}^{z})  |\psi_{0}; N_{\text{d}}, S_{\text{tot}}^{z}\rangle \; .
\end{equation}

%%%%%%%%%%%%%%%%%%%%%%%%%%%FIG%%%%%%%%%%%%%%%%%%%%%%%%%%%%%%%%%%%
%%%%%%%%%%%%%%%%%%%%%%%%%%%%%%%%%%%%%%%%%%%%%%%%%%%%%%%%%%%%%%%%
%%%%%%%%%%%%%%%%%%%%%%%%%%%FIG%%%%%%%%%%%%%%%%%%%%%%%%%%%%%%%%%%%
%%%%%%%%%%%%%%%%%%%%%%%%%%%%%%%%%%%%%%%%%%%%%%%%%%%%%%

To prove the statement (B) that the unique ground state found above indeed has the maximal total spin: 
\[
S_{\text{tot}} = S_{\text{max}} = (L-N_{\text{d}}) + (1/2)N_{\text{d}} = L - N_{\text{d}}/2 \; , 
\]
we first apply the Perron-Frobenius theorem to 
the squared total spin: $\left( \sum_{i}^{L} \vec{T }_{i} \right)^{2} = S_{\text{tot}}(S_{\text{tot}}+1)$.  
%%%
As the local spin operator $\vec{s}_{i}+\vec{S}_{i}$ projected onto the ground-state subspace is of the following block-diagonal form:
\begin{equation*}
\vec{T}_{i} 
= \vec{S}_{i} (S=1) \oplus  \vec{D}_{i}(S=1/2)
\; ,
\end{equation*}
the projected total spin $\sum_{i} \vec{T}_{i}$ is block-diagonal with respect to 
the positions of the holes (i.e., spin-$S$) and $S_{\text{tot}}^{z}$, and so is its square \footnote{%
For each $S_{\text{tot}}^{z}$ value, there are $L!/[N_{\text{d}}! (L-N_{\text{d}})!]$ sectors according to different sequences of spin-$S$ and $(S-1/2)$.}.     
Within each block, $\left( \sum_{i}^{L} \vec{T }_{i} \right)^{2}$ is just a squared total spin of a mixed-spin system 
(with a given fixed sequence of spin-$S$ and $S{-}1/2$):
\begin{equation}
\left( \sum_{i}^{L} \vec{T }_{i} \right)^{2}
= \sum_{i,j} \frac{1}{2}( T_{i}^{+}T_{j}^{-} + T_{i}^{-}T_{j}^{+} ) 
+ \sum_{i,j} T_{i}^{z}T_{j}^{z}  \; ,
\label{eqn:squared-total-spin}
\end{equation}
in which $\vec{T}_{i}$ is either spin-$1$ $\vec{S}_{i}$ (hole) or spin-$1/2$ $\vec{D}_{i}$ (electron).   
Then, it is clear that the squared total spin \eqref{eqn:squared-total-spin} within each sector is non-negative and indecomposable 
[in the standard basis \eqref{eqn:def-PF-basis}], 
which allows us to apply the Perron-Frobenius theorem once again to show that the maximum eigenvalue of 
$\left( \sum_{i}^{L} \vec{T }_{i} \right)^{2}$ is unique.   
The corresponding eigenstate is a superposition of {\em all} the basis states allowed for the (given) hole configuration and 
$S_{\text{tot}}^{z}$ with all-positive coefficients.  Since in this $S_{\text{tot}}^{z}$-sector, $S_{\text{tot}}$ can take 
any values in the range $S_{\text{tot}}^{z} \leq S_{\text{tot}} \leq S_{\text{max}}(= L S - N_{\text{e}}/2)$, 
the unique maximal eigenvalue is $S_{\text{max}}(S_{\text{max}}+1)$.   

Due to the peculiar property (guaranteed by the Perron-Frobenius theorem) of the unique lowest-energy state 
$|\psi_{0}; N_{\text{d}},S_{\text{tot}}^{z}\rangle$, its overlap with the above $S_{\text{tot}} = S_{\text{max}}$ state is non-zero, 
i.e. $\mathcal{P} (S_{\text{max}}) |\psi_{0}; N_{\text{d}}, S_{\text{tot}}^{z}\rangle \neq 0$ 
with $\mathcal{P} (S_{\text{max}})$ being the projector onto the space with $S_{\text{tot}} = S_{\text{max}}$ \footnote{%
Since there are $L!/[N_{\text{d}}! (L-N_{\text{d}})!]$ different $S_{\text{tot}} = S_{\text{max}}$ states in the full $S_{\text{tot}}^{z}$-sector, 
$\mathcal{P} (S_{\text{max}})$ is a direct sum of the projectors onto the individual $S_{\text{tot}} = S_{\text{max}}$ states:
\[  \mathcal{P} (S_{\text{max}}) = \oplus_{h \in \text{hole config.}}\mathcal{P}_{h} (S_{\text{max}}) \; . \]}.     
Then, the SU(2)-symmetry of the Hamiltonian implies that the $S_{\text{tot}} = S_{\text{max}}$ state 
$\mathcal{P} (S_{\text{max}}) |\psi_{0}; N_{\text{d}}, S_{\text{tot}}^{z}\rangle$ is another ground state of $H^{(L)}_{N_{\text{d}}, S_{\text{tot}}^z}$:
\begin{equation}
\begin{split}
& H^{(L)}_{N_{\text{d}}, S_{\text{tot}}^z} \mathcal{P} (S_{\text{max}}) |\psi_{0}; N_{\text{d}}, S_{\text{tot}}^{z}\rangle \\
& %= \mathcal{P} (S_{\text{max}})H^{(L)}_{N_{\text{d}}, S_{\text{tot}}^z}  |\psi_{0}; N_{\text{d}}, S_{\text{tot}}^{z}\rangle  
 = E_{\text{g.s.}} (N_{\text{d}}, S_{\text{tot}}^{z}) \mathcal{P} (S_{\text{max}}) |\psi_{0}; N_{\text{d}}, S_{\text{tot}}^{z}\rangle \; .
\end{split}
\end{equation}
The uniqueness of the ground state in the (full) $S_{\text{tot}}^{z}$-sector allows the only possibility 
\[  \mathcal{P} (S_{\text{max}}) |\psi_{0}; N_{\text{d}}, S_{\text{tot}}^{z}\rangle \propto |\psi_{0}; N_{\text{d}}, S_{\text{tot}}^{z}\rangle  \; , \] 
i.e., the ground state $|\psi_{0}; N_{\text{d}}, S_{\text{tot}}^{z}\rangle$ itself is ferromagnetic $S_{\text{tot}} = S_{\text{max}}$ for any values of 
$S_{\text{tot}}^{z}$.  
In particular, the above statement for $S_{\text{tot}}^{z} =0$ (or $1/2$) means that the absolute ground state (i.e., 
the lowest-energy state in the entire Hilbert space) is unique (up to the trivial degeneracy associated with the SU(2)-symmetry) and ferromagnetic.  
It is evident that we can readily generalize the above argument to arbitrary $S \geq 1$ by using Eqs.~\eqref{effective_generalS_1} 
and \eqref{effective_generalS_2} instead of \eqref{hopping1st_effectiveH}.    
%%%%%%%%%%%% BIBLIOGRAPHY %%%%%%%%%%%%%%%%%%%%%%%%%%%%%%%%%%%%%%%
%\bibliographystyle{apsrev4-2}
%\bibliography{bunken,bunken2}

\begin{thebibliography}{63}%
\makeatletter
\providecommand \@ifxundefined [1]{%
 \@ifx{#1\undefined}
}%
\providecommand \@ifnum [1]{%
 \ifnum #1\expandafter \@firstoftwo
 \else \expandafter \@secondoftwo
 \fi
}%
\providecommand \@ifx [1]{%
 \ifx #1\expandafter \@firstoftwo
 \else \expandafter \@secondoftwo
 \fi
}%
\providecommand \natexlab [1]{#1}%
\providecommand \enquote  [1]{``#1''}%
\providecommand \bibnamefont  [1]{#1}%
\providecommand \bibfnamefont [1]{#1}%
\providecommand \citenamefont [1]{#1}%
\providecommand \href@noop [0]{\@secondoftwo}%
\providecommand \href [0]{\begingroup \@sanitize@url \@href}%
\providecommand \@href[1]{\@@startlink{#1}\@@href}%
\providecommand \@@href[1]{\endgroup#1\@@endlink}%
\providecommand \@sanitize@url [0]{\catcode `\\12\catcode `\$12\catcode
  `\&12\catcode `\#12\catcode `\^12\catcode `\_12\catcode `\%12\relax}%
\providecommand \@@startlink[1]{}%
\providecommand \@@endlink[0]{}%
\providecommand \url  [0]{\begingroup\@sanitize@url \@url }%
\providecommand \@url [1]{\endgroup\@href {#1}{\urlprefix }}%
\providecommand \urlprefix  [0]{URL }%
\providecommand \Eprint [0]{\href }%
\providecommand \doibase [0]{https://doi.org/}%
\providecommand \selectlanguage [0]{\@gobble}%
\providecommand \bibinfo  [0]{\@secondoftwo}%
\providecommand \bibfield  [0]{\@secondoftwo}%
\providecommand \translation [1]{[#1]}%
\providecommand \BibitemOpen [0]{}%
\providecommand \bibitemStop [0]{}%
\providecommand \bibitemNoStop [0]{.\EOS\space}%
\providecommand \EOS [0]{\spacefactor3000\relax}%
\providecommand \BibitemShut  [1]{\csname bibitem#1\endcsname}%
\let\auto@bib@innerbib\@empty
%</preamble>
\bibitem [{\citenamefont {Fazekas}(1999)}]{Fazekas-book-99}%
  \BibitemOpen
  \bibfield  {author} {\bibinfo {author} {\bibfnamefont {P.}~\bibnamefont
  {Fazekas}},\ }\href@noop {} {\emph {\bibinfo {title} {Electron Correlation
  and Magnetism}}}\ (\bibinfo  {publisher} {World Scienific},\ \bibinfo {year}
  {1999})\BibitemShut {NoStop}%
\bibitem [{\citenamefont {Coleman}(2015)}]{Coleman-book-15}%
  \BibitemOpen
  \bibfield  {author} {\bibinfo {author} {\bibfnamefont {P.}~\bibnamefont
  {Coleman}},\ }\href@noop {} {\emph {\bibinfo {title} {Introduction to
  Many-Body Physics}}}\ (\bibinfo  {publisher} {Cambridge University Press},\
  \bibinfo {year} {2015})\BibitemShut {NoStop}%
\bibitem [{\citenamefont {Tsunetsugu}\ \emph {et~al.}(1997)\citenamefont
  {Tsunetsugu}, \citenamefont {Sigrist},\ and\ \citenamefont
  {Ueda}}]{Tsunetsugu_review}%
  \BibitemOpen
  \bibfield  {author} {\bibinfo {author} {\bibfnamefont {H.}~\bibnamefont
  {Tsunetsugu}}, \bibinfo {author} {\bibfnamefont {M.}~\bibnamefont
  {Sigrist}},\ and\ \bibinfo {author} {\bibfnamefont {K.}~\bibnamefont
  {Ueda}},\ }\href {https://doi.org/10.1103/RevModPhys.69.809} {\bibfield
  {journal} {\bibinfo  {journal} {Rev. Mod. Phys.}\ }\textbf {\bibinfo {volume}
  {69}},\ \bibinfo {pages} {809} (\bibinfo {year} {1997})}\BibitemShut
  {NoStop}%
\bibitem [{\citenamefont {Gul\'{a}csi}(2004)}]{Gulacsi-review-04}%
  \BibitemOpen
  \bibfield  {author} {\bibinfo {author} {\bibfnamefont {M.}~\bibnamefont
  {Gul\'{a}csi}},\ }\href {https://doi.org/10.1080/00018730412331313997}
  {\bibfield  {journal} {\bibinfo  {journal} {Adv. Phys.}\ }\textbf {\bibinfo
  {volume} {53}},\ \bibinfo {pages} {769} (\bibinfo {year} {2004})}\BibitemShut
  {NoStop}%
\bibitem [{\citenamefont {Yosida}(1966)}]{Yoshida-66}%
  \BibitemOpen
  \bibfield  {author} {\bibinfo {author} {\bibfnamefont {K.}~\bibnamefont
  {Yosida}},\ }\href {https://doi.org/10.1103/PhysRev.147.223} {\bibfield
  {journal} {\bibinfo  {journal} {Phys. Rev.}\ }\textbf {\bibinfo {volume}
  {147}},\ \bibinfo {pages} {223} (\bibinfo {year} {1966})}\BibitemShut
  {NoStop}%
\bibitem [{\citenamefont {Ruderman}\ and\ \citenamefont {Kittel}(1954)}]{RK}%
  \BibitemOpen
  \bibfield  {author} {\bibinfo {author} {\bibfnamefont {M.~A.}\ \bibnamefont
  {Ruderman}}\ and\ \bibinfo {author} {\bibfnamefont {C.}~\bibnamefont
  {Kittel}},\ }\href {https://doi.org/10.1103/PhysRev.96.99} {\bibfield
  {journal} {\bibinfo  {journal} {Phys. Rev.}\ }\textbf {\bibinfo {volume}
  {96}},\ \bibinfo {pages} {99} (\bibinfo {year} {1954})}\BibitemShut {NoStop}%
\bibitem [{\citenamefont {Kasuya}(1956)}]{K}%
  \BibitemOpen
  \bibfield  {author} {\bibinfo {author} {\bibfnamefont {T.}~\bibnamefont
  {Kasuya}},\ }\href {https://doi.org/10.1143/PTP.16.45} {\bibfield  {journal}
  {\bibinfo  {journal} {Progr. Theor. Phys.}\ }\textbf {\bibinfo {volume}
  {16}},\ \bibinfo {pages} {45} (\bibinfo {year} {1956})}\BibitemShut {NoStop}%
\bibitem [{\citenamefont {Yosida}(1957)}]{Y}%
  \BibitemOpen
  \bibfield  {author} {\bibinfo {author} {\bibfnamefont {K.}~\bibnamefont
  {Yosida}},\ }\href {https://doi.org/10.1103/PhysRev.106.893} {\bibfield
  {journal} {\bibinfo  {journal} {Phys. Rev.}\ }\textbf {\bibinfo {volume}
  {106}},\ \bibinfo {pages} {893} (\bibinfo {year} {1957})}\BibitemShut
  {NoStop}%
\bibitem [{\citenamefont {Doniach}(1977)}]{Doniach-77}%
  \BibitemOpen
  \bibfield  {author} {\bibinfo {author} {\bibfnamefont {S.}~\bibnamefont
  {Doniach}},\ }\href {https://doi.org/10.1016/0378-4363(77)90190-5} {\bibfield
   {journal} {\bibinfo  {journal} {Physica B+C}\ }\textbf {\bibinfo {volume}
  {91}},\ \bibinfo {pages} {231} (\bibinfo {year} {1977})}\BibitemShut
  {NoStop}%
\bibitem [{\citenamefont {Coleman}\ \emph {et~al.}(1997)\citenamefont
  {Coleman}, \citenamefont {Georges},\ and\ \citenamefont
  {Tsvelik}}]{Coleman-G-T-97}%
  \BibitemOpen
  \bibfield  {author} {\bibinfo {author} {\bibfnamefont {P.}~\bibnamefont
  {Coleman}}, \bibinfo {author} {\bibfnamefont {A.}~\bibnamefont {Georges}}, \
  and\ \bibinfo {author} {\bibfnamefont {A.~M.}\ \bibnamefont {Tsvelik}},\
  }\href {\doibase 10.1088/0953-8984/9/2/002} {\bibfield  {journal} {\bibinfo
  {journal} {Journal of Physics: Condensed Matter}\ }\textbf {\bibinfo {volume}
  {9}},\ \bibinfo {pages} {345} (\bibinfo {year} {1997})}\BibitemShut {NoStop}%
\bibitem [{\citenamefont {Gorshkov}\ \emph {et~al.}(2010)\citenamefont
  {Gorshkov}, \citenamefont {Hermele}, \citenamefont {Gurarie}, \citenamefont
  {Xu}, \citenamefont {Julienne}, \citenamefont {Ye}, \citenamefont {Zoller},
  \citenamefont {Demler}, \citenamefont {Lukin},\ and\ \citenamefont
  {Rey}}]{Gorshkov-et-al-10}%
  \BibitemOpen
  \bibfield  {author} {\bibinfo {author} {\bibfnamefont {A.~V.}\ \bibnamefont
  {Gorshkov}}, \bibinfo {author} {\bibfnamefont {M.}~\bibnamefont {Hermele}},
  \bibinfo {author} {\bibfnamefont {V.}~\bibnamefont {Gurarie}}, \bibinfo
  {author} {\bibfnamefont {C.}~\bibnamefont {Xu}}, \bibinfo {author}
  {\bibfnamefont {P.~S.}\ \bibnamefont {Julienne}}, \bibinfo {author}
  {\bibfnamefont {J.}~\bibnamefont {Ye}}, \bibinfo {author} {\bibfnamefont
  {P.}~\bibnamefont {Zoller}}, \bibinfo {author} {\bibfnamefont
  {E.}~\bibnamefont {Demler}}, \bibinfo {author} {\bibfnamefont {M.~D.}\
  \bibnamefont {Lukin}},\ and\ \bibinfo {author} {\bibfnamefont {A.~M.}\
  \bibnamefont {Rey}},\ }\href {http://dx.doi.org/10.1038/nphys1535} {\bibfield
   {journal} {\bibinfo  {journal} {Nat Phys}\ }\textbf {\bibinfo {volume}
  {6}},\ \bibinfo {pages} {289} (\bibinfo {year} {2010})}\BibitemShut {NoStop}%
\bibitem [{\citenamefont {Riegger}\ \emph {et~al.}(2018)\citenamefont
  {Riegger}, \citenamefont {Darkwah~Oppong}, \citenamefont {H\"ofer},
  \citenamefont {Fernandes}, \citenamefont {Bloch},\ and\ \citenamefont
  {F\"olling}}]{Riegger-et-al-KLM-18}%
  \BibitemOpen
  \bibfield  {author} {\bibinfo {author} {\bibfnamefont {L.}~\bibnamefont
  {Riegger}}, \bibinfo {author} {\bibfnamefont {N.}~\bibnamefont
  {Darkwah~Oppong}}, \bibinfo {author} {\bibfnamefont {M.}~\bibnamefont
  {H\"ofer}}, \bibinfo {author} {\bibfnamefont {D.~R.}\ \bibnamefont
  {Fernandes}}, \bibinfo {author} {\bibfnamefont {I.}~\bibnamefont {Bloch}},\
  and\ \bibinfo {author} {\bibfnamefont {S.}~\bibnamefont {F\"olling}},\ }\href
  {https://doi.org/10.1103/PhysRevLett.120.143601} {\bibfield  {journal}
  {\bibinfo  {journal} {Phys. Rev. Lett.}\ }\textbf {\bibinfo {volume} {120}},\
  \bibinfo {pages} {143601} (\bibinfo {year} {2018})}\BibitemShut {NoStop}%
\bibitem [{\citenamefont {Suzuki}\ and\ \citenamefont
  {Hattori}(2019)}]{Suzuki_Hattori1}%
  \BibitemOpen
  \bibfield  {author} {\bibinfo {author} {\bibfnamefont {K.}~\bibnamefont
  {Suzuki}}\ and\ \bibinfo {author} {\bibfnamefont {K.}~\bibnamefont
  {Hattori}},\ }\href {https://doi.org/10.7566/JPSJ.88.024707} {\bibfield
  {journal} {\bibinfo  {journal} {J. Phys. Soc. Jpn.}\ }\textbf {\bibinfo
  {volume} {88}},\ \bibinfo {pages} {024707} (\bibinfo {year}
  {2019})}\BibitemShut {NoStop}%
\bibitem [{\citenamefont {Suzuki}\ and\ \citenamefont
  {Hattori}(2020)}]{Suzuki_Hattori2}%
  \BibitemOpen
  \bibfield  {author} {\bibinfo {author} {\bibfnamefont {K.}~\bibnamefont
  {Suzuki}}\ and\ \bibinfo {author} {\bibfnamefont {K.}~\bibnamefont
  {Hattori}},\ }\href {https://doi.org/10.7566/JPSJ.89.034703} {\bibfield
  {journal} {\bibinfo  {journal} {J. Phys. Soc. Jpn.}\ }\textbf {\bibinfo
  {volume} {89}},\ \bibinfo {pages} {034703} (\bibinfo {year}
  {2020})}\BibitemShut {NoStop}%
  %%%%%%%%%%%%%%%%%%%%%%%%%%%%%%%%%%%%%%%%%%%%%%%%%%%%%%%%%
\bibitem [{Note1()}]{Note1}%
  \BibitemOpen
  \bibinfo {note} {%
  Our higher-spin Kondo lattice model \eqref{klH} and its extension \eqref{khH} need some 
  fine-tuning when realized in the heavy-fermion setting.  
  In this respect, the study of the model \eqref{klH} is mainly motivated by theoretical interests, as is discussed in the text}\BibitemShut {NoStop}%
  %%%%%%%%%%%%%%%%%%%%%%%%%%%%%%%%%%%%%%%%%%%%%%%%%%%%%%%%%
\bibitem [{\citenamefont {Sikkema}\ \emph {et~al.}(1997)\citenamefont
  {Sikkema}, \citenamefont {Affleck},\ and\ \citenamefont
  {White}}]{Sikkema-A-W-97}%
  \BibitemOpen
  \bibfield  {author} {\bibinfo {author} {\bibfnamefont {A.~E.}\ \bibnamefont
  {Sikkema}}, \bibinfo {author} {\bibfnamefont {I.}~\bibnamefont {Affleck}},\
  and\ \bibinfo {author} {\bibfnamefont {S.~R.}\ \bibnamefont {White}},\ }\href
  {https://doi.org/10.1103/PhysRevLett.79.929} {\bibfield  {journal} {\bibinfo
  {journal} {Phys. Rev. Lett.}\ }\textbf {\bibinfo {volume} {79}},\ \bibinfo
  {pages} {929} (\bibinfo {year} {1997})}\BibitemShut {NoStop}%
\bibitem [{\citenamefont {Zachar}(2001)}]{Zachar-01}%
  \BibitemOpen
  \bibfield  {author} {\bibinfo {author} {\bibfnamefont {O.}~\bibnamefont
  {Zachar}},\ }\href {https://doi.org/10.1103/PhysRevB.63.205104} {\bibfield
  {journal} {\bibinfo  {journal} {Phys. Rev. B}\ }\textbf {\bibinfo {volume}
  {63}},\ \bibinfo {pages} {205104} (\bibinfo {year} {2001})}\BibitemShut
  {NoStop}%
\bibitem [{\citenamefont {Zachar}\ and\ \citenamefont
  {Tsvelik}(2001)}]{Zachar-T-01}%
  \BibitemOpen
  \bibfield  {author} {\bibinfo {author} {\bibfnamefont {O.}~\bibnamefont
  {Zachar}}\ and\ \bibinfo {author} {\bibfnamefont {A.~M.}\ \bibnamefont
  {Tsvelik}},\ }\href {https://doi.org/10.1103/PhysRevB.64.033103} {\bibfield
  {journal} {\bibinfo  {journal} {Phys. Rev. B}\ }\textbf {\bibinfo {volume}
  {64}},\ \bibinfo {pages} {033103} (\bibinfo {year} {2001})}\BibitemShut
  {NoStop}%
\bibitem [{\citenamefont {Sigrist}\ \emph {et~al.}(1991)\citenamefont
  {Sigrist}, \citenamefont {Tsunetsugu},\ and\ \citenamefont
  {Ueda}}]{kondo_lowdendity_FM}%
  \BibitemOpen
  \bibfield  {author} {\bibinfo {author} {\bibfnamefont {M.}~\bibnamefont
  {Sigrist}}, \bibinfo {author} {\bibfnamefont {H.}~\bibnamefont
  {Tsunetsugu}},\ and\ \bibinfo {author} {\bibfnamefont {K.}~\bibnamefont
  {Ueda}},\ }\href {https://doi.org/10.1103/PhysRevLett.67.2211} {\bibfield
  {journal} {\bibinfo  {journal} {Phys. Rev. Lett.}\ }\textbf {\bibinfo
  {volume} {67}},\ \bibinfo {pages} {2211} (\bibinfo {year}
  {1991})}\BibitemShut {NoStop}%
\bibitem [{\citenamefont {Sigrist}\ \emph {et~al.}(1992)\citenamefont
  {Sigrist}, \citenamefont {Tsunetsugu}, \citenamefont {Ueda},\ and\
  \citenamefont {Rice}}]{Sigrist-T-U-R-92}%
  \BibitemOpen
  \bibfield  {author} {\bibinfo {author} {\bibfnamefont {M.}~\bibnamefont
  {Sigrist}}, \bibinfo {author} {\bibfnamefont {H.}~\bibnamefont {Tsunetsugu}},
  \bibinfo {author} {\bibfnamefont {K.}~\bibnamefont {Ueda}},\ and\ \bibinfo
  {author} {\bibfnamefont {T.~M.}\ \bibnamefont {Rice}},\ }\href
  {https://doi.org/10.1103/PhysRevB.46.13838} {\bibfield  {journal} {\bibinfo
  {journal} {Phys. Rev. B}\ }\textbf {\bibinfo {volume} {46}},\ \bibinfo
  {pages} {13838} (\bibinfo {year} {1992})}\BibitemShut {NoStop}%
\bibitem [{\citenamefont {Moukouri}\ and\ \citenamefont
  {Caron}(1996)}]{Moukouri-C-96}%
  \BibitemOpen
  \bibfield  {author} {\bibinfo {author} {\bibfnamefont {S.}~\bibnamefont
  {Moukouri}}\ and\ \bibinfo {author} {\bibfnamefont {L.~G.}\ \bibnamefont
  {Caron}},\ }\href {https://doi.org/10.1103/PhysRevB.54.12212} {\bibfield
  {journal} {\bibinfo  {journal} {Phys. Rev. B}\ }\textbf {\bibinfo {volume}
  {54}},\ \bibinfo {pages} {12212} (\bibinfo {year} {1996})}\BibitemShut
  {NoStop}%
\bibitem [{\citenamefont {Berg}\ \emph {et~al.}(2010)\citenamefont {Berg},
  \citenamefont {Fradkin},\ and\ \citenamefont {Kivelson}}]{Berg-F-K-10}%
  \BibitemOpen
  \bibfield  {author} {\bibinfo {author} {\bibfnamefont {E.}~\bibnamefont
  {Berg}}, \bibinfo {author} {\bibfnamefont {E.}~\bibnamefont {Fradkin}}, \
  and\ \bibinfo {author} {\bibfnamefont {S.~A.}\ \bibnamefont {Kivelson}},\
  }\href {\doibase 10.1103/PhysRevLett.105.146403} {\bibfield  {journal}
  {\bibinfo  {journal} {Phys. Rev. Lett.}\ }\textbf {\bibinfo {volume} {105}},\
  \bibinfo {pages} {146403} (\bibinfo {year} {2010})}\BibitemShut {NoStop}%
\bibitem [{\citenamefont {Thalmeier}(2002)}]{Thalmeier2002}%
  \BibitemOpen
  \bibfield  {author} {\bibinfo {author} {\bibfnamefont {P.}~\bibnamefont
  {Thalmeier}},\ }\href {https://doi.org/10.1140/epjb/e20020127} {\bibfield
  {journal} {\bibinfo  {journal} {Euro. Phys. J. B - Condensed Matter and
  Complex Systems}\ }\textbf {\bibinfo {volume} {27}},\ \bibinfo {pages} {29}
  (\bibinfo {year} {2002})}\BibitemShut {NoStop}%
\bibitem [{\citenamefont {Alexandrov}\ and\ \citenamefont
  {Coleman}(2014)}]{topological_kondo_insulator}%
  \BibitemOpen
  \bibfield  {author} {\bibinfo {author} {\bibfnamefont {V.}~\bibnamefont
  {Alexandrov}}\ and\ \bibinfo {author} {\bibfnamefont {P.}~\bibnamefont
  {Coleman}},\ }\href {https://doi.org/10.1103/PhysRevB.90.115147} {\bibfield
  {journal} {\bibinfo  {journal} {Phys. Rev. B}\ }\textbf {\bibinfo {volume}
  {90}},\ \bibinfo {pages} {115147} (\bibinfo {year} {2014})}\BibitemShut
  {NoStop}%
\bibitem [{\citenamefont {Mezio}\ \emph {et~al.}(2015)\citenamefont {Mezio},
  \citenamefont {Lobos}, \citenamefont {Dobry},\ and\ \citenamefont
  {Gazza}}]{topological_kondo_insulator2}%
  \BibitemOpen
  \bibfield  {author} {\bibinfo {author} {\bibfnamefont {A.}~\bibnamefont
  {Mezio}}, \bibinfo {author} {\bibfnamefont {A.~M.}\ \bibnamefont {Lobos}},
  \bibinfo {author} {\bibfnamefont {A.~O.}\ \bibnamefont {Dobry}},\ and\
  \bibinfo {author} {\bibfnamefont {C.~J.}\ \bibnamefont {Gazza}},\ }\href
  {https://doi.org/10.1103/PhysRevB.92.205128} {\bibfield  {journal} {\bibinfo
  {journal} {Phys. Rev. B}\ }\textbf {\bibinfo {volume} {92}},\ \bibinfo
  {pages} {205128} (\bibinfo {year} {2015})}\BibitemShut {NoStop}%
\bibitem [{\citenamefont {Hagym\'asi}\ and\ \citenamefont
  {Legeza}(2016)}]{topological_kondo_insulator3}%
  \BibitemOpen
  \bibfield  {author} {\bibinfo {author} {\bibfnamefont {I.}~\bibnamefont
  {Hagym\'asi}}\ and\ \bibinfo {author} {\bibfnamefont {O.}~\bibnamefont
  {Legeza}},\ }\href {https://doi.org/10.1103/PhysRevB.93.165104} {\bibfield
  {journal} {\bibinfo  {journal} {Phys. Rev. B}\ }\textbf {\bibinfo {volume}
  {93}},\ \bibinfo {pages} {165104} (\bibinfo {year} {2016})}\BibitemShut
  {NoStop}%
\bibitem [{\citenamefont {Henriques}\ \emph {et~al.}(1984)\citenamefont
  {Henriques}, \citenamefont {Alcacer}, \citenamefont {Pouget},\ and\
  \citenamefont {Jerome}}]{Henriques-et-al-84}%
  \BibitemOpen
  \bibfield  {author} {\bibinfo {author} {\bibfnamefont {R.~T.}\ \bibnamefont
  {Henriques}}, \bibinfo {author} {\bibfnamefont {L.}~\bibnamefont {Alcacer}},
  \bibinfo {author} {\bibfnamefont {J.~P.}\ \bibnamefont {Pouget}},\ and\
  \bibinfo {author} {\bibfnamefont {D.}~\bibnamefont {Jerome}},\ }\href
  {https://doi.org/10.1088/0022-3719/17/29/019} {\bibfield  {journal} {\bibinfo
   {journal} {J. Phys. C: Solid State Phys.}\ }\textbf {\bibinfo {volume}
  {17}},\ \bibinfo {pages} {5197} (\bibinfo {year} {1984})}\BibitemShut
  {NoStop}%
\bibitem [{\citenamefont {Bourbonnais}\ \emph {et~al.}(1991)\citenamefont
  {Bourbonnais}, \citenamefont {Henriques}, \citenamefont {Wzietek},
  \citenamefont {K\"ongeter}, \citenamefont {Voiron},\ and\ \citenamefont
  {J\'erme}}]{Bourbonnais-et-al-91}%
  \BibitemOpen
  \bibfield  {author} {\bibinfo {author} {\bibfnamefont {C.}~\bibnamefont
  {Bourbonnais}}, \bibinfo {author} {\bibfnamefont {R.~T.}\ \bibnamefont
  {Henriques}}, \bibinfo {author} {\bibfnamefont {P.}~\bibnamefont {Wzietek}},
  \bibinfo {author} {\bibfnamefont {D.}~\bibnamefont {K\"ongeter}}, \bibinfo
  {author} {\bibfnamefont {J.}~\bibnamefont {Voiron}},\ and\ \bibinfo {author}
  {\bibfnamefont {D.}~\bibnamefont {J\'erme}},\ }\href
  {https://doi.org/10.1103/PhysRevB.44.641} {\bibfield  {journal} {\bibinfo
  {journal} {Phys. Rev. B}\ }\textbf {\bibinfo {volume} {44}},\ \bibinfo
  {pages} {641} (\bibinfo {year} {1991})}\BibitemShut {NoStop}%
\bibitem [{\citenamefont {Green}\ \emph {et~al.}(2011)\citenamefont {Green},
  \citenamefont {Brooks}, \citenamefont {Kuhns}, \citenamefont {Reyes},
  \citenamefont {Lumata}, \citenamefont {Almeida}, \citenamefont {Matos},
  \citenamefont {Henriques}, \citenamefont {Wright},\ and\ \citenamefont
  {Brown}}]{Green-et-al-11}%
  \BibitemOpen
  \bibfield  {author} {\bibinfo {author} {\bibfnamefont {E.~L.}\ \bibnamefont
  {Green}}, \bibinfo {author} {\bibfnamefont {J.~S.}\ \bibnamefont {Brooks}},
  \bibinfo {author} {\bibfnamefont {P.~L.}\ \bibnamefont {Kuhns}}, \bibinfo
  {author} {\bibfnamefont {A.~P.}\ \bibnamefont {Reyes}}, \bibinfo {author}
  {\bibfnamefont {L.~L.}\ \bibnamefont {Lumata}}, \bibinfo {author}
  {\bibfnamefont {M.}~\bibnamefont {Almeida}}, \bibinfo {author} {\bibfnamefont
  {M.~J.}\ \bibnamefont {Matos}}, \bibinfo {author} {\bibfnamefont {R.~T.}\
  \bibnamefont {Henriques}}, \bibinfo {author} {\bibfnamefont {J.~A.}\
  \bibnamefont {Wright}},\ and\ \bibinfo {author} {\bibfnamefont {S.~E.}\
  \bibnamefont {Brown}},\ }\href {https://doi.org/10.1103/PhysRevB.84.121101}
  {\bibfield  {journal} {\bibinfo  {journal} {Phys. Rev. B}\ }\textbf {\bibinfo
  {volume} {84}},\ \bibinfo {pages} {121101} (\bibinfo {year}
  {2011})}\BibitemShut {NoStop}%
\bibitem [{\citenamefont {Haldane}(1983{\natexlab{a}})}]{Haldane_conjecture}%
  \BibitemOpen
  \bibfield  {author} {\bibinfo {author} {\bibfnamefont {F.~D.~M.}\
  \bibnamefont {Haldane}},\ }\href
  {https://doi.org/10.1103/PhysRevLett.50.1153} {\bibfield  {journal} {\bibinfo
   {journal} {Phys. Rev. Lett.}\ }\textbf {\bibinfo {volume} {50}},\ \bibinfo
  {pages} {1153} (\bibinfo {year} {1983}{\natexlab{a}})}\BibitemShut {NoStop}%
\bibitem [{\citenamefont {Haldane}(1983{\natexlab{b}})}]{Haldane_conjecture-2}%
  \BibitemOpen
  \bibfield  {author} {\bibinfo {author} {\bibfnamefont {F.}~\bibnamefont
  {Haldane}},\ }\href {https://doi.org/10.1016/0375-9601(83)90631-X} {\bibfield
   {journal} {\bibinfo  {journal} {Phys. Lett. A}\ }\textbf {\bibinfo {volume}
  {93}},\ \bibinfo {pages} {464 } (\bibinfo {year}
  {1983}{\natexlab{b}})}\BibitemShut {NoStop}%
\bibitem [{\citenamefont {Gu}\ and\ \citenamefont {Wen}(2009)}]{Gu-W-09}%
  \BibitemOpen
  \bibfield  {author} {\bibinfo {author} {\bibfnamefont {Z.-C.}\ \bibnamefont
  {Gu}}\ and\ \bibinfo {author} {\bibfnamefont {X.-G.}\ \bibnamefont {Wen}},\
  }\href {http://link.aps.org/abstract/PRB/v80/e155131} {\bibfield  {journal}
  {\bibinfo  {journal} {Phys. Rev. B}\ }\textbf {\bibinfo {volume} {80}},\
  \bibinfo {pages} {155131} (\bibinfo {year} {2009})}\BibitemShut {NoStop}%
\bibitem [{\citenamefont {Chen}\ \emph {et~al.}(2013)\citenamefont {Chen},
  \citenamefont {Gu}, \citenamefont {Liu},\ and\ \citenamefont {Wen}}]{spt}%
  \BibitemOpen
  \bibfield  {author} {\bibinfo {author} {\bibfnamefont {X.}~\bibnamefont
  {Chen}}, \bibinfo {author} {\bibfnamefont {Z.-C.}\ \bibnamefont {Gu}},
  \bibinfo {author} {\bibfnamefont {Z.-X.}\ \bibnamefont {Liu}},\ and\ \bibinfo
  {author} {\bibfnamefont {X.-G.}\ \bibnamefont {Wen}},\ }\href
  {https://doi.org/10.1103/PhysRevB.87.155114} {\bibfield  {journal} {\bibinfo
  {journal} {Phys. Rev. B}\ }\textbf {\bibinfo {volume} {87}},\ \bibinfo
  {pages} {155114} (\bibinfo {year} {2013})}\BibitemShut {NoStop}%
\bibitem [{\citenamefont {Else}\ \emph {et~al.}(2012)\citenamefont {Else},
  \citenamefont {Schwarz}, \citenamefont {Bartlett},\ and\ \citenamefont
  {Doherty}}]{Else-S-B-D-12}%
  \BibitemOpen
  \bibfield  {author} {\bibinfo {author} {\bibfnamefont {D.~V.}\ \bibnamefont
  {Else}}, \bibinfo {author} {\bibfnamefont {I.}~\bibnamefont {Schwarz}},
  \bibinfo {author} {\bibfnamefont {S.~D.}\ \bibnamefont {Bartlett}},\ and\
  \bibinfo {author} {\bibfnamefont {A.~C.}\ \bibnamefont {Doherty}},\ }\href
  {https://doi.org/10.1103/PhysRevLett.108.240505} {\bibfield  {journal}
  {\bibinfo  {journal} {Phys. Rev. Lett.}\ }\textbf {\bibinfo {volume} {108}},\
  \bibinfo {pages} {240505} (\bibinfo {year} {2012})}\BibitemShut {NoStop}%
\bibitem [{\citenamefont {White}(1992)}]{White1}%
  \BibitemOpen
  \bibfield  {author} {\bibinfo {author} {\bibfnamefont {S.~R.}\ \bibnamefont
  {White}},\ }\href {https://doi.org/10.1103/PhysRevLett.69.2863} {\bibfield
  {journal} {\bibinfo  {journal} {Phys. Rev. Lett.}\ }\textbf {\bibinfo
  {volume} {69}},\ \bibinfo {pages} {2863} (\bibinfo {year}
  {1992})}\BibitemShut {NoStop}%
\bibitem [{\citenamefont {White}(1993)}]{White2}%
  \BibitemOpen
  \bibfield  {author} {\bibinfo {author} {\bibfnamefont {S.~R.}\ \bibnamefont
  {White}},\ }\href {https://doi.org/10.1103/PhysRevB.48.10345} {\bibfield
  {journal} {\bibinfo  {journal} {Phys. Rev. B}\ }\textbf {\bibinfo {volume}
  {48}},\ \bibinfo {pages} {10345} (\bibinfo {year} {1993})}\BibitemShut
  {NoStop}%
\bibitem [{\citenamefont {Schollw\"{c}k}(2011)}]{SCHOLLWOCK}%
  \BibitemOpen
  \bibfield  {author} {\bibinfo {author} {\bibfnamefont {U.}~\bibnamefont
  {Schollw\"{c}k}},\ }\href
  {https://doi.org/https://doi.org/10.1016/j.aop.2010.09.012} {\bibfield
  {journal} {\bibinfo  {journal} {Ann. Phys.}\ }\textbf {\bibinfo {volume}
  {326}},\ \bibinfo {pages} {96 } (\bibinfo {year} {2011})}\BibitemShut
  {NoStop}%
\bibitem [{\citenamefont {Shibata}\ and\ \citenamefont {Hotta}(2011)}]{SSD}%
  \BibitemOpen
  \bibfield  {author} {\bibinfo {author} {\bibfnamefont {N.}~\bibnamefont
  {Shibata}}\ and\ \bibinfo {author} {\bibfnamefont {C.}~\bibnamefont
  {Hotta}},\ }\href {https://doi.org/10.1103/PhysRevB.84.115116} {\bibfield
  {journal} {\bibinfo  {journal} {Phys. Rev. B}\ }\textbf {\bibinfo {volume}
  {84}},\ \bibinfo {pages} {115116} (\bibinfo {year} {2011})}\BibitemShut
  {NoStop}%
\bibitem [{\citenamefont {Hotta}\ and\ \citenamefont
  {Shibata}(2012)}]{SSD_chargegap}%
  \BibitemOpen
  \bibfield  {author} {\bibinfo {author} {\bibfnamefont {C.}~\bibnamefont
  {Hotta}}\ and\ \bibinfo {author} {\bibfnamefont {N.}~\bibnamefont
  {Shibata}},\ }\href {https://doi.org/10.1103/PhysRevB.86.041108} {\bibfield
  {journal} {\bibinfo  {journal} {Phys. Rev. B}\ }\textbf {\bibinfo {volume}
  {86}},\ \bibinfo {pages} {041108} (\bibinfo {year} {2012})}\BibitemShut
  {NoStop}%
\bibitem [{\citenamefont {Hikihara}\ and\ \citenamefont
  {Nishino}(2011)}]{SSD_nishino}%
  \BibitemOpen
  \bibfield  {author} {\bibinfo {author} {\bibfnamefont {T.}~\bibnamefont
  {Hikihara}}\ and\ \bibinfo {author} {\bibfnamefont {T.}~\bibnamefont
  {Nishino}},\ }\href {https://doi.org/10.1103/PhysRevB.83.060414} {\bibfield
  {journal} {\bibinfo  {journal} {Phys. Rev. B}\ }\textbf {\bibinfo {volume}
  {83}},\ \bibinfo {pages} {060414} (\bibinfo {year} {2011})}\BibitemShut
  {NoStop}%
\bibitem [{\citenamefont {Giamarchi}(2004)}]{Giamarchi}%
  \BibitemOpen
  \bibfield  {author} {\bibinfo {author} {\bibfnamefont {T.}~\bibnamefont
  {Giamarchi}},\ }\href
  {https://doi.org/10.1093/acprof:oso/9780198525004.001.0001} {\emph {\bibinfo
  {title} {{Quantum physics in one dimension}}}},\ Internat. Ser. Mono. Phys.\
  (\bibinfo  {publisher} {Clarendon Press},\ \bibinfo {address} {Oxford},\
  \bibinfo {year} {2004})\BibitemShut {NoStop}%
\bibitem [{\citenamefont {Tsunetsugu}\ \emph {et~al.}(1992)\citenamefont
  {Tsunetsugu}, \citenamefont {Hatsugai}, \citenamefont {Ueda},\ and\
  \citenamefont {Sigrist}}]{kondo_spingap}%
  \BibitemOpen
  \bibfield  {author} {\bibinfo {author} {\bibfnamefont {H.}~\bibnamefont
  {Tsunetsugu}}, \bibinfo {author} {\bibfnamefont {Y.}~\bibnamefont
  {Hatsugai}}, \bibinfo {author} {\bibfnamefont {K.}~\bibnamefont {Ueda}},\
  and\ \bibinfo {author} {\bibfnamefont {M.}~\bibnamefont {Sigrist}},\ }\href
  {https://doi.org/10.1103/PhysRevB.46.3175} {\bibfield  {journal} {\bibinfo
  {journal} {Phys. Rev. B}\ }\textbf {\bibinfo {volume} {46}},\ \bibinfo
  {pages} {3175} (\bibinfo {year} {1992})}\BibitemShut {NoStop}%
\bibitem [{\citenamefont {Yu}\ and\ \citenamefont {White}(1993)}]{Yu-W-93}%
  \BibitemOpen
  \bibfield  {author} {\bibinfo {author} {\bibfnamefont {C.~C.}\ \bibnamefont
  {Yu}}\ and\ \bibinfo {author} {\bibfnamefont {S.~R.}\ \bibnamefont {White}},\
  }\href {https://doi.org/10.1103/PhysRevLett.71.3866} {\bibfield  {journal}
  {\bibinfo  {journal} {Phys. Rev. Lett.}\ }\textbf {\bibinfo {volume} {71}},\
  \bibinfo {pages} {3866} (\bibinfo {year} {1993})}\BibitemShut {NoStop}%
\bibitem [{\citenamefont {Tsvelik}(1994)}]{Tsvelik_kondo}%
  \BibitemOpen
  \bibfield  {author} {\bibinfo {author} {\bibfnamefont {A.~M.}\ \bibnamefont
  {Tsvelik}},\ }\href {https://doi.org/10.1103/PhysRevLett.72.1048} {\bibfield
  {journal} {\bibinfo  {journal} {Phys. Rev. Lett.}\ }\textbf {\bibinfo
  {volume} {72}},\ \bibinfo {pages} {1048} (\bibinfo {year}
  {1994})}\BibitemShut {NoStop}%
\bibitem [{\citenamefont {Tsvelik}\ and\ \citenamefont
  {Yevtushenko}(2019)}]{Tsvelik-Y-19}%
  \BibitemOpen
  \bibfield  {author} {\bibinfo {author} {\bibfnamefont {A.~M.}\ \bibnamefont
  {Tsvelik}}\ and\ \bibinfo {author} {\bibfnamefont {O.~M.}\ \bibnamefont
  {Yevtushenko}},\ }\href {https://doi.org/10.1103/PhysRevB.100.165110}
  {\bibfield  {journal} {\bibinfo  {journal} {Phys. Rev. B}\ }\textbf {\bibinfo
  {volume} {100}},\ \bibinfo {pages} {165110} (\bibinfo {year}
  {2019})}\BibitemShut {NoStop}%
\bibitem [{\citenamefont {Zener}(1951)}]{Zener-51}%
  \BibitemOpen
  \bibfield  {author} {\bibinfo {author} {\bibfnamefont {C.}~\bibnamefont
  {Zener}},\ }\href {https://doi.org/10.1103/PhysRev.82.403} {\bibfield
  {journal} {\bibinfo  {journal} {Phys. Rev.}\ }\textbf {\bibinfo {volume}
  {82}},\ \bibinfo {pages} {403} (\bibinfo {year} {1951})}\BibitemShut
  {NoStop}%
\bibitem [{\citenamefont {Anderson}\ and\ \citenamefont
  {Hasegawa}(1955)}]{Anderson-H-55}%
  \BibitemOpen
  \bibfield  {author} {\bibinfo {author} {\bibfnamefont {P.~W.}\ \bibnamefont
  {Anderson}}\ and\ \bibinfo {author} {\bibfnamefont {H.}~\bibnamefont
  {Hasegawa}},\ }\href {https://doi.org/10.1103/PhysRev.100.675} {\bibfield
  {journal} {\bibinfo  {journal} {Phys. Rev.}\ }\textbf {\bibinfo {volume}
  {100}},\ \bibinfo {pages} {675} (\bibinfo {year} {1955})}\BibitemShut
  {NoStop}%
\bibitem [{\citenamefont {de~Gennes}(1960)}]{deGennes-60}%
  \BibitemOpen
  \bibfield  {author} {\bibinfo {author} {\bibfnamefont {P.~G.}\ \bibnamefont
  {de~Gennes}},\ }\href {https://doi.org/10.1103/PhysRev.118.141} {\bibfield
  {journal} {\bibinfo  {journal} {Phys. Rev.}\ }\textbf {\bibinfo {volume}
  {118}},\ \bibinfo {pages} {141} (\bibinfo {year} {1960})}\BibitemShut
  {NoStop}%
\bibitem [{\citenamefont {Kubo}(1982)}]{Kubo-82}%
  \BibitemOpen
  \bibfield  {author} {\bibinfo {author} {\bibfnamefont {K.}~\bibnamefont
  {Kubo}},\ }\href {https://doi.org/10.1143/jpsj.51.782} {\bibfield  {journal}
  {\bibinfo  {journal} {J. Phys. Soc. Jpn.}\ }\textbf {\bibinfo {volume}
  {51}},\ \bibinfo {pages} {782} (\bibinfo {year} {1982})}\BibitemShut
  {NoStop}%
\bibitem [{\citenamefont {McCulloch}\ \emph {et~al.}(2002)\citenamefont
  {McCulloch}, \citenamefont {Juozapavicius}, \citenamefont {Rosengren},\ and\
  \citenamefont {Gulacsi}}]{McCulloch-J-R-G-02}%
  \BibitemOpen
  \bibfield  {author} {\bibinfo {author} {\bibfnamefont {I.~P.}\ \bibnamefont
  {McCulloch}}, \bibinfo {author} {\bibfnamefont {A.}~\bibnamefont
  {Juozapavicius}}, \bibinfo {author} {\bibfnamefont {A.}~\bibnamefont
  {Rosengren}},\ and\ \bibinfo {author} {\bibfnamefont {M.}~\bibnamefont
  {Gulacsi}},\ }\href {https://doi.org/10.1103/PhysRevB.65.052410} {\bibfield
  {journal} {\bibinfo  {journal} {Phys. Rev. B}\ }\textbf {\bibinfo {volume}
  {65}},\ \bibinfo {pages} {052410} (\bibinfo {year} {2002})}\BibitemShut
  {NoStop}%
\bibitem [{\citenamefont {Peters}\ and\ \citenamefont
  {Kawakami}(2012)}]{Peters-K-12}%
  \BibitemOpen
  \bibfield  {author} {\bibinfo {author} {\bibfnamefont {R.}~\bibnamefont
  {Peters}}\ and\ \bibinfo {author} {\bibfnamefont {N.}~\bibnamefont
  {Kawakami}},\ }\href {https://doi.org/10.1103/PhysRevB.86.165107} {\bibfield
  {journal} {\bibinfo  {journal} {Phys. Rev. B}\ }\textbf {\bibinfo {volume}
  {86}},\ \bibinfo {pages} {165107} (\bibinfo {year} {2012})}\BibitemShut
  {NoStop}%
\bibitem [{\citenamefont {Tsunetsugu}\ \emph {et~al.}(1993)\citenamefont
  {Tsunetsugu}, \citenamefont {Sigrist},\ and\ \citenamefont
  {Ueda}}]{boundary_FM_PM}%
  \BibitemOpen
  \bibfield  {author} {\bibinfo {author} {\bibfnamefont {H.}~\bibnamefont
  {Tsunetsugu}}, \bibinfo {author} {\bibfnamefont {M.}~\bibnamefont
  {Sigrist}},\ and\ \bibinfo {author} {\bibfnamefont {K.}~\bibnamefont
  {Ueda}},\ }\href {https://doi.org/10.1103/PhysRevB.47.8345} {\bibfield
  {journal} {\bibinfo  {journal} {Phys. Rev. B}\ }\textbf {\bibinfo {volume}
  {47}},\ \bibinfo {pages} {8345} (\bibinfo {year} {1993})}\BibitemShut
  {NoStop}%
\bibitem [{\citenamefont {Fishman}\ \emph {et~al.}(2020)\citenamefont
  {Fishman}, \citenamefont {White},\ and\ \citenamefont
  {Stoudenmire}}]{itensor}%
  \BibitemOpen
  \bibfield  {author} {\bibinfo {author} {\bibfnamefont {M.}~\bibnamefont
  {Fishman}}, \bibinfo {author} {\bibfnamefont {S.~R.}\ \bibnamefont {White}},\
  and\ \bibinfo {author} {\bibfnamefont {E.~M.}\ \bibnamefont {Stoudenmire}},\
  }\href@noop {} {\bibinfo {title} {The \mbox{ITensor} software library for
  tensor network calculations}} (\bibinfo {year} {2020}),\ \Eprint
  {https://arxiv.org/abs/2007.14822} {arXiv:2007.14822} \BibitemShut {NoStop}%
\bibitem [{\citenamefont {Gendiar}\ \emph {et~al.}(2009)\citenamefont
  {Gendiar}, \citenamefont {Krcmar},\ and\ \citenamefont
  {Nishino}}]{GendiarK-N-SSD-09}%
  \BibitemOpen
  \bibfield  {author} {\bibinfo {author} {\bibfnamefont {A.}~\bibnamefont
  {Gendiar}}, \bibinfo {author} {\bibfnamefont {R.}~\bibnamefont {Krcmar}},\
  and\ \bibinfo {author} {\bibfnamefont {T.}~\bibnamefont {Nishino}},\ }\href
  {https://doi.org/10.1143/PTP.122.953} {\bibfield  {journal} {\bibinfo
  {journal} {Prog. Theor. Phys.}\ }\textbf {\bibinfo {volume} {122}},\ \bibinfo
  {pages} {953} (\bibinfo {year} {2009})}\BibitemShut {NoStop}%
\bibitem [{Note2()}]{Note2}%
  \BibitemOpen
  \bibinfo {note} {Almost parallel shifts of the four curves in Fig.~\ref
  {half_filling_spin_corr_long_range_log_log}(b) suggest that $\langle \protect
  \vec {S}_{i} {\cdot } \protect \vec {S}_{j}\rangle $ and $\langle \protect
  \vec {D}_{i} {\cdot } \protect \vec {D}_{j} \rangle $ differ only by
  numerical factors.}\BibitemShut {Stop}%
\bibitem [{Note3()}]{Note3}%
  \BibitemOpen
  \bibinfo {note} {Although in critical isotropic spin systems, the spin-spin
  correlation function is generically expected to behaves like $(\protect
  \qopname \relax o{ln}|i-j|)^{1/2}/|i-j|$ \cite {Giamarchi} except at the
  fine-tuned points, we did not find such logarithmic corrections in our
  simulations. We do not know whether this absence of the logarithmic
  correction is explained by some effective long-range spin-spin interactions
  generated by the electron motion or not.}\BibitemShut {Stop}%
\bibitem [{\citenamefont {Huang}\ \emph {et~al.}(2020)\citenamefont {Huang},
  \citenamefont {Sheng},\ and\ \citenamefont {Ting}}]{dimer2}%
  \BibitemOpen
  \bibfield  {author} {\bibinfo {author} {\bibfnamefont {Y.}~\bibnamefont
  {Huang}}, \bibinfo {author} {\bibfnamefont {D.~N.}\ \bibnamefont {Sheng}},\
  and\ \bibinfo {author} {\bibfnamefont {C.~S.}\ \bibnamefont {Ting}},\ }\href
  {https://doi.org/10.1103/PhysRevB.102.245143} {\bibfield  {journal} {\bibinfo
   {journal} {Phys. Rev. B}\ }\textbf {\bibinfo {volume} {102}},\ \bibinfo
  {pages} {245143} (\bibinfo {year} {2020})}\BibitemShut {NoStop}%
\bibitem [{\citenamefont {Khait}\ \emph {et~al.}(2018)\citenamefont {Khait},
  \citenamefont {Azaria}, \citenamefont {Hubig}, \citenamefont
  {Schollw{\"o}ck},\ and\ \citenamefont {Auerbach}}]{Khait-A-H-S-A-18}%
  \BibitemOpen
  \bibfield  {author} {\bibinfo {author} {\bibfnamefont {I.}~\bibnamefont
  {Khait}}, \bibinfo {author} {\bibfnamefont {P.}~\bibnamefont {Azaria}},
  \bibinfo {author} {\bibfnamefont {C.}~\bibnamefont {Hubig}}, \bibinfo
  {author} {\bibfnamefont {U.}~\bibnamefont {Schollw{\"o}ck}},\ and\ \bibinfo
  {author} {\bibfnamefont {A.}~\bibnamefont {Auerbach}},\ }\href
  {https://doi.org/10.1073/pnas.1719374115} {\bibfield  {journal} {\bibinfo
  {journal} {Proc. Natl. Acad. Sci.}\ }\textbf {\bibinfo {volume} {115}},\
  \bibinfo {pages} {5140} (\bibinfo {year} {2018})}\BibitemShut {NoStop}%
\bibitem [{\citenamefont {Penc}\ and\ \citenamefont
  {Mila}(1994)}]{hubbard_qfill_dimer}%
  \BibitemOpen
  \bibfield  {author} {\bibinfo {author} {\bibfnamefont {K.}~\bibnamefont
  {Penc}}\ and\ \bibinfo {author} {\bibfnamefont {F.}~\bibnamefont {Mila}},\
  }\href {https://doi.org/10.1103/PhysRevB.50.11429} {\bibfield  {journal}
  {\bibinfo  {journal} {Phys. Rev. B}\ }\textbf {\bibinfo {volume} {50}},\
  \bibinfo {pages} {11429} (\bibinfo {year} {1994})}\BibitemShut {NoStop}%
\bibitem [{\citenamefont {Xavier}\ \emph {et~al.}(2003)\citenamefont {Xavier},
  \citenamefont {Pereira}, \citenamefont {Miranda},\ and\ \citenamefont
  {Affleck}}]{Xavier-P-M-A-03}%
  \BibitemOpen
  \bibfield  {author} {\bibinfo {author} {\bibfnamefont {J.~C.}\ \bibnamefont
  {Xavier}}, \bibinfo {author} {\bibfnamefont {R.~G.}\ \bibnamefont {Pereira}},
  \bibinfo {author} {\bibfnamefont {E.}~\bibnamefont {Miranda}},\ and\ \bibinfo
  {author} {\bibfnamefont {I.}~\bibnamefont {Affleck}},\ }\href
  {https://doi.org/10.1103/PhysRevLett.90.247204} {\bibfield  {journal}
  {\bibinfo  {journal} {Phys. Rev. Lett.}\ }\textbf {\bibinfo {volume} {90}},\
  \bibinfo {pages} {247204} (\bibinfo {year} {2003})}\BibitemShut {NoStop}%
\bibitem [{Note4()}]{Note4}%
  \BibitemOpen
  \bibinfo {note} {We follow the convention of Ref.~\cite {Giamarchi}.
  Precisely, we have one more interaction of the form $\protect \qopname \relax
  o{cos}(\protect \sqrt {8}\phi _{\rho })\protect \qopname \relax
  o{cos}(\protect \sqrt {8}\phi _{\sigma })$ with the scaling dimension
  $2+2/K_{\rho } >2$. This is irrelevant and we can safely drop
  it.}\BibitemShut {Stop}%
\bibitem [{\citenamefont {Tsvelik}(2016)}]{Tsvelik-16}%
  \BibitemOpen
  \bibfield  {author} {\bibinfo {author} {\bibfnamefont {A.~M.}\ \bibnamefont
  {Tsvelik}},\ }\href {\doibase 10.1103/PhysRevB.94.165114} {\bibfield
  {journal} {\bibinfo  {journal} {Phys. Rev. B}\ }\textbf {\bibinfo {volume}
  {94}},\ \bibinfo {pages} {165114} (\bibinfo {year} {2016})}\BibitemShut
  {NoStop}%
\bibitem [{\citenamefont {M\"uller-Hartmann}\ and\ \citenamefont
  {Dagotto}(1996)}]{Muller-Hartmann-D-96}%
  \BibitemOpen
  \bibfield  {author} {\bibinfo {author} {\bibfnamefont {E.}~\bibnamefont
  {M\"uller-Hartmann}}\ and\ \bibinfo {author} {\bibfnamefont {E.}~\bibnamefont
  {Dagotto}},\ }\href {https://doi.org/10.1103/PhysRevB.54.R6819} {\bibfield
  {journal} {\bibinfo  {journal} {Phys. Rev. B}\ }\textbf {\bibinfo {volume}
  {54}},\ \bibinfo {pages} {R6819} (\bibinfo {year} {1996})}\BibitemShut
  {NoStop}%
\bibitem [{Note5()}]{Note5}%
  \BibitemOpen
  \bibinfo {note} {We do not need to specify the local doublet (i.e., electron)
  number $n^{d}_{i}$ since $T^{z}_{i}=\pm 1/2$ ($0,\pm 1$) already imply
  $n^{d}_{i}=1$ ($0$).}\BibitemShut {Stop}%
\bibitem [{\citenamefont {Tasaki}(2020)}]{Tasaki-book-20}%
  \BibitemOpen
  \bibfield  {author} {\bibinfo {author} {\bibfnamefont {H.}~\bibnamefont
  {Tasaki}},\ }\href {https://doi.org/10.1007/978-3-030-41265-4} {\emph
  {\bibinfo {title} {Physics and Mathematics of Quantum Many-Body Systems}}}\
  (\bibinfo  {publisher} {Springer},\ \bibinfo {year} {2020})\BibitemShut
  {NoStop}%
\bibitem [{Note6()}]{Note6}%
  \BibitemOpen
  \bibinfo {note} {For each $S_{\protect \text {tot}}^{z}$ value, there are
  $L!/[N_{\protect \text {d}}! (L-N_{\protect \text {d}})!]$ sectors according
  to different sequences of spin-$S$ and $(S-1/2)$.}\BibitemShut {Stop}%
\bibitem [{Note7()}]{Note7}%
  \BibitemOpen
  \bibinfo {note} {Since there are $L!/[N_{\protect \text {d}}! (L-N_{\protect
  \text {d}})!]$ different $S_{\protect \text {tot}} = S_{\protect \text
  {max}}$ states in the full $S_{\protect \text {tot}}^{z}$-sector, $\protect
  \mathcal {P} (S_{\protect \text {max}})$ is a direct sum of the projectors
  onto the individual $S_{\protect \text {tot}} = S_{\protect \text {max}}$
  states: \protect \[ \protect \mathcal {P} (S_{\protect \text {max}}) = \oplus
  _{h \in \protect \text {hole config.}}\protect \mathcal {P}_{h} (S_{\protect
  \text {max}}) \protect \tmspace +\thickmuskip {.2777em} . \protect
  \]}\BibitemShut {NoStop}%
\end{thebibliography}
%%%%%%%%%%%%%%%%%%%%%%%%%%%%%%%%%%%%%%%%%%%%%%%%%%%%%%%%%%%%
%%%%%%%%% REFERENCES %%%%%%%%%%%%%%%%%%%%%%%%%%%%%%%%%%%
%apsrev4-2.bst 2019-01-14 (MD) hand-edited version of apsrev4-1.bst
%Control: key (0)
%Control: author (72) initials jnrlst
%Control: editor formatted (1) identically to author
%Control: production of article title (-1) disabled
%Control: page (0) single
%Control: year (1) truncated
%Control: production of eprint (0) enabled
%
%%%%%%%%%%%%%%%%%%%%%%%%%%%%%%%%%%%%%%%%%%%%%%%%%%%
\end{document}